\documentclass [a4paper, 11pt]{article}

\usepackage[margin=2.0cm]{geometry}
\usepackage{setspace}
\usepackage{amsmath}
\usepackage{amssymb}
\usepackage{amsthm}
\usepackage{dsfont}
\usepackage{gensymb}
\usepackage{bbm}
\usepackage{color}
\usepackage{multirow}
\usepackage[dvipsnames]{xcolor}
\usepackage{graphicx}
\usepackage{caption}
\usepackage{subcaption}
\usepackage{mathrsfs}
\usepackage{bbm}
\usepackage{xfrac}
\usepackage{pgfplots}
\usepackage{geometry}
\usetikzlibrary{intersections}
\usetikzlibrary{patterns}
\usetikzlibrary{arrows}
\usepackage{empheq}
\usepackage[skins,theorems]{tcolorbox}
\tcbset{highlight math style={enhanced,
  colframe=red,colback=white,arc=0pt,boxrule=1pt}}
  
\usepgfplotslibrary{fillbetween}
\usetikzlibrary{shapes,snakes,positioning}
\usetikzlibrary{automata,fit}
\usepackage{todonotes}
\usepackage{newtxmath}
\usepackage[T1]{fontenc}
\usepackage[bookmarksnumbered=true]{hyperref} 
\hypersetup{
     colorlinks = true,
     linkcolor = red,
     anchorcolor = blue,
     citecolor = blue,
     filecolor = blue,
     urlcolor = blue
     }
\usepackage{natbib}

\newcommand{\s}{\mathbbm{S}}
\renewcommand{\i}{\mathbbm{I}}
\renewcommand{\t}{\mathbbm{T}}
\renewcommand{\r}{\mathbbm{R}}
\newcommand{\h}{\mathbbm{H}}

\newcommand{\gas}{\overline{S}}
\newcommand{\gai}{\overline{I}}
\newcommand{\gait}{\overline{IT}}
\newcommand{\gar}{\overline{R}}
\newcommand{\gart}{\overline{RT}}
\newcommand{\gah}{\overline{H}}
\newcommand{\gs}{\overline{\mathbbm{S}}}
\newcommand{\gi}{\overline{\mathbbm{I}}}
\newcommand{\git}{\overline{\mathbbm{IT}}}
\newcommand{\gr}{\overline{\mathbbm{R}}}
\newcommand{\grt}{\overline{\mathbbm{RT}}}
\newcommand{\gh}{\overline{\mathbbm{H}}}
\newcommand{\da}{\delta_A}
\newcommand{\dg}{\delta_G}
\newcommand{\bl}{\lambda}
\newcommand{\bt}{\tau}

\def\underbracex#1#2{\mathop{\vtop{\m@th\ialign{##\crcr
   $\hfil\displaystyle{#2}\hfil$\crcr
   \noalign{\kern3\p@\nointerlineskip}%
   #1\crcr\noalign{\kern3\p@}}}}\limits}

\def\underbracea{\underbracex\upbracefilla}

\def\upbracefilla{$\m@th \setbox\z@\hbox{$\braceld$}%
  \bracelu\leaders\vrule \@height\ht\z@ \@depth\z@\hfill 
\kern\p@\vrule \@width\p@\kern\p@\vrule \@width\p@\kern\p@\vrule \@width\p@
$}

\def\upbracefillb{$\m@th \setbox\z@\hbox{$\braceld$}%
\vrule \@width\p@\kern\p@\vrule \@width\p@\kern\p@\vrule \@width\p@\kern\p@
 \leaders\vrule \@height\ht\z@ \@depth\z@\hfill\bracerd
  \braceld\leaders\vrule \@height\ht\z@ \@depth\z@\hfill
\kern\p@\vrule \@width\p@\kern\p@\vrule \@width\p@\kern\p@\vrule \@width\p@
$}

\def\upbracefillc{$\m@th \setbox\z@\hbox{$\braceld$}%
\vrule \@width\p@\kern\p@\vrule \@width\p@\kern\p@\vrule \@width\p@\kern\p@
\leaders\vrule \@height\ht\z@ \@depth\z@\hfill
\kern\p@\vrule \@width\p@\kern\p@\vrule \@width\p@\kern\p@\vrule \@width\p@
$}

\def\upbracefilld{$\m@th \setbox\z@\hbox{$\braceld$}%
\vrule \@width\p@\kern\p@\vrule \@width\p@\kern\p@\vrule \@width\p@\kern\p@
 \leaders\vrule \@height\ht\z@ \@depth\z@\hfill\braceru$}

\def\underbracebd{\underbracex\upbracefillbd}
\def\upbracefillbd{$\m@th \setbox\z@\hbox{$\braceld$}%
\vrule \@width\p@\kern\p@\vrule \@width\p@\kern\p@\vrule \@width\p@\kern\p@
\bracerd\braceld
 \leaders\vrule \@height\ht\z@ \@depth\z@\hfill\braceru$}

\title{Behavioral epidemiology: An economic model to evaluate \\optimal policy in the midst of a pandemic\footnote{Chakrabarti: \href{mailto: shomak.chakrabarti@manchester.ac.uk}{\texttt{shomak.chakrabarti@manchester.ac.uk}}, University of Manchester; Krasikov: \href{mailto: krasikovis.main@gmail.com}{\texttt{krasikovis.main@gmail.com}}, Higher School of Economics Moscow; Lamba: \href{mailto: rlamba@psu.edu}{\texttt{rlamba@psu.edu}}, Pennsylvania State University. We are grateful to David Argente, Shoumitro Chatterjee, Krishna Dasaratha, Elisa Giannone, Callum Jones, Kei Hirano, Shouyong Shi, Shamim Sinnar, Jakub Steiner and Flavio Toxvaerd for helpful comments and suggestions.}}
\author{Shomak Chakrabarti\hspace{18mm} Ilia Krasikov\hspace{18mm} Rohit Lamba}
\date{February 2022}

\begin{document}

\maketitle

\begin{abstract}
This paper combines a canonical epidemiology model of disease dynamics with government policy of lockdown and testing, and agents' decision to social distance in order to avoid getting infected. The model is calibrated with data on deaths and testing outcomes in the Unites States. 
It is shown that an intermediate but prolonged lockdown is socially optimal when both mortality and GDP are taken into account.
This is because the government wants the economy to keep producing some output and the slack in reducing infection is picked up by social distancing agents. 
Social distancing best responds to the optimal government policy to keep the effective reproductive number at one and avoid multiple waves through the pandemic. 
Calibration shows testing to have been effective, but it could have been even more instrumental if it had been aggressively pursued from the beginning of the pandemic.
Not having any lockdown or shutting down social distancing would have had extreme consequences. Greater centralized control on social activities would have mitigated further the spread of the pandemic. \vspace{10mm}

\end{abstract} 
\begin{small}
\begin{quotation}
 "Coronavirus is Germany's greatest challenge since World War II, says Angela Merkel", \citet{Merkel_speech}
\end{quotation}
\begin{quotation}
"Covid-19 restrictions not affecting social distancing, says ONS: UK statistics agency says level of people still meeting likely to lead to increased hospitalisations and deaths." \citet{FT_UK}
\end{quotation}
\begin{quotation}
 "Vietnam abandons zero-Covid strategy after record drop in GDP: Warnings that lockdowns were crippling businesses heaped pressure on Communist government."  \citet{FT_Vietnam}
\end{quotation}

\end{small}

\section{Introduction}

Three instruments have been salient in the global response to the Covid-19: government policy of lockdown and testing, and people's decision to practice social distancing. The objective for the collective has largely been the minimization of direct mortality while ensuring a steady pace of the economy, and the objective of the individual has been lowering the chance of getting infected while ensuring some normalcy in life.

While the academic literature in epidemiology is primarily concerned with understanding the evolution of the disease and how to control it, economists are attempting to understand the co-evolution of mortality and economic output using some subset of the aforementioned three instruments as choice variables subject to some capacity constraint. In that vein, this paper builds a model of disease dynamics where government tries to manage mortality and the economy through lockdown and testing policies, and agents social distance balancing the chance of getting infected with maintaining some social interactions. 

While lockdown is good at reducing the spread of infection, it is often accompanied with severe economic costs which may further jeopardize livelihoods even after the pandemic recedes. Social distancing as a behavioral response to the spread of the disease influences the government's choice of severity of the lockdown. And, testing spreads more information all around as to who should be participating in social and economic activities. 

We start from an epidemiological framework with five possible states---susceptible, infected, hospitalized, recovered and dead. On it we build a model of economic and social interactions where people are matched and the infection spreads. Economic activities produce output, and can be restricted by the government's lockdown policy. Social activities provide utility to the agents, and are controlled endogenously by their social distancing decisions. Agents suffer disutility from getting infected and infecting others. Both types of interactions-- economic and social-- can be mitigated by tracing and testing and those who are found to be infected are forced to stay at home. The pandemic can end with the arrival of a vaccine at a random time distributed between one and two years since the inception with a mean of about a year and a half. Finally, anticipating the behavioral response of the agents, the government chooses the optimal testing and lockdown policies, by maximizing a weighted sum of total output and mortality. Each agent in turn takes the government's lockdown and testing policy and other agents's social distancing decisions as given and choose their best response of social distancing.\footnote{Section \ref{section model} details the main ingredients of the model. Section \ref{section dynamics} describes the system of equations that quantify the propagation of the disease. Section \ref{section behavior} introduce behavioral response and the solves for the agent's optimization problem. And, Section \ref{govt_opt} states and solves for the government's optimization problem.}

The main contribution of the paper is to model and analyze the three instruments together and explore their pairwise substitutability. This constitutes a technical challenge because we are solving for (i) a planning problem of a forward looking government, (ii) a strategic problem between forward looking agents, and (iii) the fixed point of the government's and the agents' problems as a "stackelberg game". A key difficulty arises from the fact that tracing-testing introduces heterogeneity in the population on who was tested when. Our approach allows us to conclude that time since the last test is a sufficient statistic for the agent's choice of how much to social distance. We then setup an optimal control problem for the agent and another one for the government (Sections \ref{section behavior} and  \ref{govt_opt} respectively) and solve the two simultaneously using the forward-backward sweep algorithm. 

Two key equations summarize the government's lockdown and agents' social distancing choices. The one for the agents pins down the static and dynamic tradeoffs from social distancing. The static trade-off is described as follows: The fraction of social distancers (or the probability of social distancing) is directly proportional to the lump-sum cost of getting infected and infected others normalized by the flow cost of social distancing. Further, the dynamic trade-off adds to this proportionality the net present value of the normalized cost from getting infected while participating in social activities. Analogously, the government implements the lockdown while trading off drop in current output with mortality and net present value of future output. Both these equations also incorporate the heterogeneity of information introduced by testing.

A secondary contribution of the paper is to calibrate the model using data on testing and deaths in the United States due to Covid-19 (Section \ref{section calibration}). It is by now well understood that the standard SIR model is poorly identified with the typical time series on the number of infected and dead (see \citet{jesus_jones} and \citet{sir_unident}). We take a pragmatic approach by fixing the medical parameters using the aggregated wisdom of various studies in medical journals and then use the data to estimate three types of parameters; (i) prevalence or matching of the susceptible with the infected, (ii) efficacy of tracing and testing, and (iii) cost for agents from getting infected and infecting others. These provide a quantitative sense respectively of how rapidly the disease was spreading in the US, how effective was the government in tracing and testing the infected and asymptomatic, and how agents evaluated their decisions to social distance. 

For testing we feed the model with data on daily tests that have been conducted in the US during the pandemic. By tracing we mean the efficacy to identify those in the population who are infected but haven't developed severe symptoms yet--- this is captured by a parameter which is calibrated using data. Our calibrated estimate of it suggests that the US did reasonably well (on average) over the course of the pandemic in identifying and isolating the infected. Its main shortcoming was in the total number of tests available early in the pandemic say in comparison to South Korea. 

How should we systemically think about the role of social distancing? We evaluate it in terms of a daily output produced by the agent, which we take approximately to be her/his daily wage. In that sense, the daily or flow cost of not partaking in any social activities is about 22\% of the daily wage. In addition the agent also suffers a lump-sum disutility from getting infected; this represents the psychological cost of contracting the virus, medical costs, and the potential (probabilistic cost) of death. The calibration exercise pegs this to be approximately worth one and a half year of wages. A further 50 days of wages is the lump-sum (altruistic) disutility from potentially infecting others. 

After calibrating the model, we execute the algorithm to calculate the optimal policy (Section \ref{section optimal policy}). The government shuts down about 40 percent of the economic activity for a prolonged period of time, about 14 months. The lockdown is not complete in that it does not hit the upper bound (of 70 percent) at any point, but it is consistent. For the same time frame, the agents cut back around 50 percent of their social activities. So the government expects the agents to social distance which allows it to impose a less than severe lockdown and since the government locks down some part of the economic activity the agents cut back on some but not all social activities.

The underlying conceptual tack on these policy choices is that they ensure {\it the effective $R_{t}$, i.e. the reproduction number is maintained almost constant throughout the pandemic at one}. This controls the spread of the virus while maintaining some economic and social activities. The constancy of effective reproduction number is also in contrast to what actually transpired in the US between March 2020 and August 2021---multiple waves in infections and deaths. In fact, the total number of deaths predicted by the optimal policy is less than half of actual number for the United States as of September 2021---around 350,000 as opposed to more than 800,000 observed in the data as of writing this draft.  

\begin{figure}
\centering 
\includegraphics[scale=0.35]{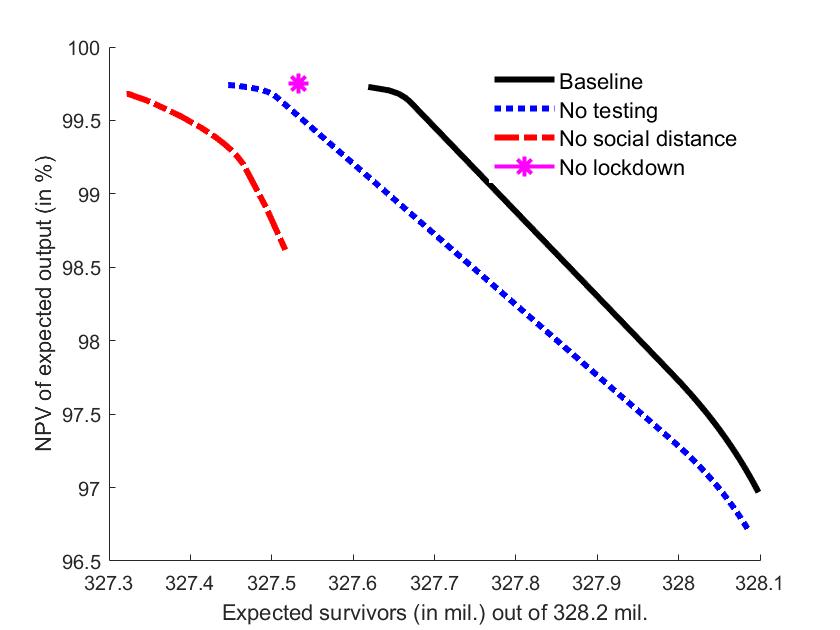}
\caption{The Pareto frontier for different specifications of the model}
\label{fig: PF intro}
\end{figure}

The optimal policy exercise described here is conducted for a specific (widely accepted) Pareto weight on total output and mortality or equivalently for a specific value of life. We then also present the Pareto frontier to illustrate the policy mix of options available with the government. For each possible value of the weight, both optimal control problems have to be solved from scratch yielding distinct value of final output and deaths. This Pareto frontier is given by the solid black line in Figure \ref{fig: PF intro}---net present value of total economic output is on the $y$-axis and total number of survivors is on the $x$-axis. 

The dotted-blue line in Figure \ref{fig: PF intro} shows the Pareto frontier when tracing-testing is shutdown, and lockdown and social distancing are the only instruments for fighting the pandemic, the curve shrinks almost uniformly. In more detailed policy experiments in the paper (in Section \ref{section tt experiment}) we show that if the tracing-testing was more effective and aggressively pursued since the inception of the pandemic, the Pareto frontier shifts out away from the black line and the gains are quite significant. Therefore, for improved tracing-testing technology, keeping the number of survivors fixed, total output expands, and keeping total output fixed, total mortality goes down. 

What if the government imposes no lockdown? The purple star in Figure \ref{fig: PF intro} represents the outcome on total output and mortality. The infection runs through the population very quickly with a single large peak during which agents social distance almost maximally. The total number of deaths is around 750k, which is significantly higher than the optimal policy and in line with the actual realized number as the pandemic has played out in the US. There are of course some economic gains since no part of the economy has been closed. We show however (Section \ref{section ND}) that even with no-lockdown the government can mitigate the extent of the pandemic with more effective and aggressive testing, bringing down the cumulative number to deaths to less than 400k, while maintaining the economics gains. A provocative thought is that even with no lockdown the total number of deaths is in ballpark of the actual realized deaths seen in the data. The main difference is the spread of deaths over time. This raises the question---what is the primary goal of the lockdown policy? One interpretation of our analysis is that lockdown (as eventually practiced in reality, not in the optimal policy of our model) simply slowed down the pandemic so that the institutions fighting it could cope.\footnote{This is echoed somewhat in \citet{atkeson_brookings}: "this model forecast that if efforts to slow transmission were applied early but were only temporary, this dramatic first peak would be delayed but not prevented: cases and deaths would explode again once efforts to slow transmission were relaxed."} 

In the next policy experiment, we ask what happens if we (hypothetically) shut down the channel of social distancing. One way to achieve this is to make the flow cost from social distancing arbitrarily high and the lump sum costs from getting infected to be low. The Pareto frontier from this exercise is given by the dashed-red curve in Figure \ref{fig: PF intro}. The set of feasible output and mortality attainable by the government shrinks considerably. In Section \ref{section ND}, we show that the pandemic runs through the population quickly, leading again to almost 800k deaths by the end of July 2020. The government locks down for a few months but then opens up the economy completely for the pandemic has already run amok. That is why the output doesn't fall much even though the number of deaths is large. This illustrates the significance of incorporating behavioral response in an otherwise mechanical model of disease dynamics.

In the final policy experiment, we allow the government to have greater control over restricting social interactions (Section \ref{section control}). This changes the dynamics of infections and deaths dramatically. The idea being that autocratic governments or communitarian (as opposed to individualist) societies are able to control to social interactions of their citizenry to a much larger degree. Unsurprisingly, the deaths decrease non-linearly as we pass on greater control to a centralized authority in the social realm.\\


\textbf{Related literature}. In the rapidly growing literature on economics of epidemiology various complimentary papers look at some combination of the three forces we seek to model. Table \ref{table_lit} lists some of the studies done by economists and categorizes them on the basis of policy instruments. 

\begin{table}[h!]
\centering
\begin{tabular}{|c|c|c|c|}
\hline
   &  lockdown & tracing-testing & social distancing \\ \hline
\citet*{lockdown_SIR}    &\multirow{2}{*}{\checkmark}  &\multirow{2}{*}{\checkmark}   &  \\ 
\citet*{ dream_team_SIRmodel}     &  &   &  \\ 
\hline
\citet*{testing_SIR}      &   & \multirow{2}{*}{\checkmark}  &  \\ 
\citet*{rishabh_da_paper}      &   &   &  \\ 
\hline 
\citet*{farboodietal_covid}   & \checkmark (exogenous)  &   & \checkmark    \\ 
\hline
\citet{flavio_ier, flavio_covid}, \citet{dasratha_SIR}, \citet{lones_SIR}   & &  &\checkmark    \\ 
\hline 
\end{tabular}
\caption[captionhere]%
{A sample of recent and related economic models of epidemiology} \label{table_lit}
\end{table}

Broadly these papers fall within the realm of augmented SIR models, where one or more instruments from lockdown, tracing-testing and social distancing are added to lend realism to the otherwise mechanical set of equations driving disease dynamics. Lockdown and testing are typically added as a planning problem for the government and social distancing through a strategic or agency model of behavior. To the best of our knowledge, no paper so far has studied a model that incorporates all three instruments together, which is the goal of this paper. As the reader goes further, we hope to convince her/him that this is both technically challenging task and a qualitatively important one, as suggested already in Figure \ref{fig: PF intro}. For more detailed list of references, see \citet{mcadams_survey} for an excellent survey on the recent progress in augmented SIR models. 

The importance of explicitly modeling behavior in SIR models has especially been emphasized. For example, \citet{atkeson_brookings} writes: " "behavior turns what would be a short and extremely sharp epidemic into a long, drawn out one." This also builds on a great body of work in epidemiology, where the importance of behavioral responses to improve the precision of the predictive power of the standard SIR-type model has been emphasized. For example, writing in the Proceedings of the National Academy of Sciences, \citet{behavior_pnas} state:
\begin{quote}
\begin{small}
Results indicate that including adaptive human behavior significantly changes the predicted course of epidemics and that this inclusion has implications for parameter estimation and interpretation and for the development of social distancing policies. Acknowledging adaptive behavior requires a shift in thinking about epidemiological processes and parameters.\footnote{See also \citet{behavior_plos} and \citet{behavior_physics} for other approaches to modeling behavior in epidemiology.} 
\end{small}
\end{quote}

To this discussion in particular, we add the nuance of how a uniform lockdown policy interacts with behavioral response, exploring their substitutability. In addition, tracing-testing introduces (i) heterogeneity in behavioral response because social distancing now is a function of the day of the latest test, and (ii) piecewise substitutability with both lockdown and social distancing in a highly non-linear fashion.

There also has been a rapidly burgeoned literature in the macroeconomics of {COVID-19}. Equilibrium models of the form DSGE-meets-SIR, where representative agents make labor and consumption decisions have been proposed by \citet*{ert_covid}, \citet*{jpv_covid}, and \citet*{kh_covid}, amongst others, and further extended to heterogenous agents by \citet*{moll_SIR} and \citet*{krueger_victor_SIR}, amongst others. Closer to our analysis \citet*{macro_testing_quaran} argue that lockdown type interventions prolong the pandemic to buy time for the health infrastructure, and there are further synergies of this force with testing and quarantining. 

In non-SIR perspectives, \citet*{glsw_covid} explore demand deficiencies created by the pandemic through supply side shortages. Also, \citet{caba_simsek_assets} provide a model of asset prices spiral when the economy is hit by severe supply shock. In contrast to these, we look at a strategic framework to pin down the behavioral response of agents in an augmented SIR-type model where governments incorporate these in determining optimal policy.\footnote{There also has been recent work on understanding the socio-economic implications of the large pandemics, and take lessons that can inform policy decisions in dealing with the health and economic crisis, see \citet*{gabru_ka_paper}, \citet*{spanish_flu} and \citet*{barro_flu}. See the survey \citet*{covid_survey} for more details.}

In terms of data work in the realm of the economics of epidemiology, our work is related to  \citet{jesus_jones} and \citet{sir_unident}---the former fits the standard SIR model to data from various parts of the world to estimate primarily the prevalence parameter and the latter argues that a unique identification of the SIR model is actually impossible. We proceed by fixing medical parameters informed by medical studies and calibrate the prevalence, tracing-testing and behavioral parameters. To the best of our knowledge, this is the first paper that tries to tease out the parameters driving  tracing-testing and behavioral response by fitting the model to the data on deaths and tests.

\section{Model}

\label{section model}

{\color{blue} States and transitions.} There is a continuum of identical agents with mass one who interact in discrete time indexed by $t=1,2,\ldots$.
The agents are forward-looking and discount the future with a factor $\delta_A \in [0,1)$. At any point, a representative agent can find herself in one of five possible health states (or compartments):
\begin{enumerate}
\item[$\bf S$] \emph{Susceptible}, in this state the agent is non-infectious and non-immune; 
\vspace{-2mm}
\item[$\bf I$] \emph{Infected}, in this state the agent is infectious but is either asymptomatic or mildly symptomatic;
\vspace{-2mm}  
\item[$\bf H$] \emph{Hospitalized}, in this state the agent is infectious and clearly symptomatic, thus she is hospitalized or being treated in isolation at home; 
\vspace{-2mm}
\item[$\bf R$] \emph{Recovered}, in this state the agent is non-infectious and immune; it is an absorbing state;
\vspace{-2mm}
\item[$\bf D$] \emph{Dead}, in this state the agent has succumbed to the infection and is dead; again an absorbing state.
\end{enumerate}

The susceptible ($\bf S$) have not had the virus, do not have immunity and may get infected in the future by coming in contact with an infected agent. The agents who get infected ($\bf I$) are at first either asymptomatic or mildly symptomatic. This assumption is especially crucial to study the Covid-19 pandemic since one of the key difficulties has been the seemingly large number of asymptomatic carriers.\footnote{For example, Germany's leading Covid-19 expert informing policy, Christian Drosten, had stated in the summer of 2020: "We now have evidence that almost half of infection events happen before the person passing on the infection develops symptoms -- and people are infectious starting two days prior to that," (\citet{guardian_german_expert})} Without an external intervention such as tracing/testing, agents in state $\bf I$ could potentially be indistinguishable from those in state $\bf S$. 

\begin{figure}
\begin{center}
\resizebox{0.55\textwidth}{!}{
\begin{tikzpicture}[> = stealth, shorten > = 1pt, auto, node distance = 2.5cm, thick]
\tikzstyle{every state}=[draw = black, thick,fill=white, minimum size = 4mm]
\node[state,line width=0.5mm,minimum width=1.2cm,minimum height=1.0cm] (S) {$\bf S$};
\node[state,line width=0.5mm,minimum width=1.2cm,minimum height=1.0cm] (IN) [right=0.7cm and 3.3cm of S] {$\bf I$};
\node[state,line width=0.5mm,minimum width=1.2cm,minimum height=1.0cm] (IT) [below of=IN, yshift=-1cm] {$\bf IT$};
\node[state,line width=0.5mm,minimum width=1.2cm,minimum height=1.0cm] (RN) [left of=IT] {$\bf R$};
\node[state,line width=0.5mm,minimum width=1.2cm,minimum height=1.0cm] (H) [right of=IN] {$\bf H$};
\node[state,line width=0.5mm,minimum width=1.2cm,minimum height=1.0cm] (RT) [below of=H, yshift=-1cm] {$\bf RT$};
\node[state,line width=0.5mm,minimum width=1.2cm,minimum height=1.0cm] (D) [right of=H] {$\bf D$};
\path[->,thick,line width=0.5mm,dotted,color=purple] (S) edge node {\small infection} (IN);
\path[->,thick,line width=0.5mm,color=blue] (IN) edge node[rotate=90,anchor=south] {testing} (IT);
\path[->,line width=0.5mm] (IN) edge node[anchor=south,yshift=6mm] {\small hospitalization} (H);
\path[->,color=red,line width=0.5mm] (IN) edge node {} (RN);
\path[->,line width=0.5mm] (IT) edge node {} (H);
\path[->,color=red,line width=0.5mm] (IT) edge node {} (RT);
\path[->,line width=0.5mm] (H) edge node[anchor=south,yshift=0mm] {\small death} (D);
\path[->,line width=0.5mm,color=blue] (RN) edge[bend right=90] node {\small testing} (RT);
\path[->,thick,line width=0.5mm,color=red] (H) edge node[rotate=90,anchor=south] {\small recovery } (RT);
\end{tikzpicture}
}
\end{center}
\caption{States and transitions in disease dynamics}\label{figure: states and transitions}
\end{figure}
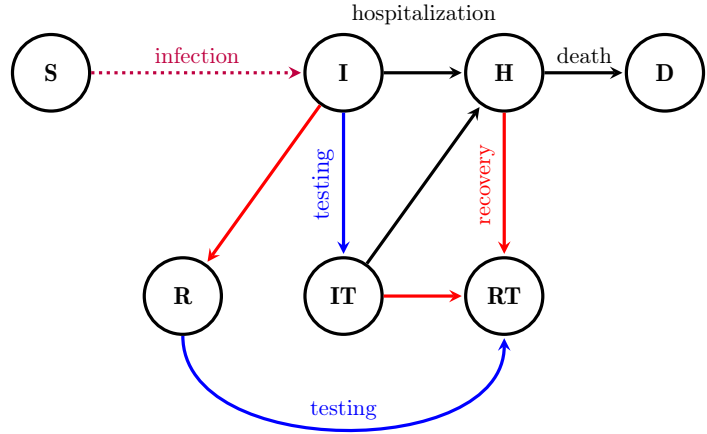

The infected can transition to two states: they either start showing symptoms and be hospitalized ($\bf H$) or they can recover ($\bf R$). The hospitalized $(\bf H)$ can also transition to two states: they can either recover ($\bf R$) or they can die ($\bf D$). Recovery and death are absorbing states. While death is obviously an absorbing state, assuming recovery to be absorbing implies that the agents develop immunity to the virus. Note that we are assuming that being hospitalized is a necessary step before death. This is done for simplicity. We could have added another state for non-hospitalized deaths due to Covid-19. The state $\bf H$ should be broadly interpreted as those showing clear and/or severe symptoms.\\  \vspace{-2mm}

{\color{blue} Tracing and Testing.} In order to separate the infected (and asymptomatic or mildly symptomatic) from the susceptible, we introduce {\it tracing} and {\it testing}. We assume that in each period a fraction of individuals who are in states $\bf S$, $\bf I$ and $\bf R$ are traced and tested. The test perfectly reveals whether the individual is susceptible, infected or recovered. Thus, we are assuming the availability of both the viral and antibody test. This further splits $\bf I$ into two compartments, those \emph{infected and asymptomatic} and those \emph{infected and asymptomatic but known to be so}. These two compartments are denoted by $\bf I$ and $\bf IT$, respectively. Similarly, $\bf R$ is split into two compartments-- $\bf R$ for the \emph{unknown recovered}, and $\bf RT$ for the \emph{known recovered}. Figure \ref{figure: states and transitions} summarizes the various states and transition possibilities. Abusing notation slightly, denote by $S_t$, $I_t$, $IT_{t}$, $R_t$, $RT_t$, $H_{t}$ and $D_t$ the aggregate fractions of agents in respective states at time $t$.

Testing is modeled through the following technology. The count of available tests is exogenous and specified by $\overline{X}_{t}$. At the end of period $t$ a fraction $\tau_t$ is uniformly chosen  amongst the set of ``eligible'' agents, so that exactly $\overline{X}_{t}$ tests are conducted. The set of ``eligible'' agents includes the asymptomatic infected $(\bf I)$ and a fraction $\gamma \in [0,1]$ of the susceptible $(\bf S)$ and unknown recovered $(\bf R)$. The parameter $\gamma$ pins down the efficiency of tracing. If $\gamma=0$, tracing is perfect for only infected people are tested, and if $\gamma=1$, then testing is blind or completely random. Finally, tracing and testing are assumed to be history-independent, so that all agents in states $\left\{\bf S,I, R\right\}$ have the same likelihood of getting traced or tested, irrespective of the past state of tracing and testing. 

This technology is of course simplified and buys us considerable tractability. Most natural extensions, such as making anti-body tests unavailable, making $\overline{X}_t$ a choice variable, and introducing history dependence in testing, can be accommodated.\\ \vspace{-2mm}

{\color{blue} Two worlds.} All agents engage in two types of activities in each period: {\it economic and social}. In each activity, the agents are randomly matched to each other. Matching is bilateral and each pair of agents can get matched independently of matching outcomes of other agents. The matching probabilities are assumed to be $\beta_w$ and $\beta_s$ for work (or economic) and social activities, respectively. Note that $\beta_{w}+\beta_{s}$ is equivalent to the "total prevalence" in the SIR framework, but in the context of our model it refers to the level of interaction an agent has in each period. 

For social interactions, we have in mind activities such as going to a park, or visiting each others' homes that may not directly produce any output. Moreover, the agents can control this rate of matching, as will be formalized later, by practicing social distancing. For economic activities, we have in matches that are generated while engaging in work that directly produces output. These can be controlled be government through lockdown.\\  \vspace{-2mm}

{\color{blue} Economic activity and lockdown.} In addition to introducing tracing and testing, the planner also implements a {\it lockdown}, modeled as the stoppage of a fraction $1-\lambda_t$ of all economic activities, bounded by $\lambda_{t}\geqslant\bar\lambda$.\footnote{We have in mind the necessity of essential services such as hospitals and supply chains for food and gas, and $\bar{\lambda}$ represents that threshold.} 
The total output of the economy during the pandemic, $Y_t$ at time $t$, is assumed to have a simple structure: \\ \vspace{-5mm}

\begin{small}
\[
Y_t := \lambda_t(S_t+I_t+R_t + RT_t).
\]
\end{small}
\vspace{-5mm}

The functional form above states that total output is equal to the total number of agents participating in economic activity, and that at any given point in time only those in states $\left\{\bf S,I,R, RT\right\}$ are productive. We are implicitly assuming all those in states $\left\{\bf IT,H\right\}$ are quarantined and do not participate in economic activities. A selective lockdown policy could allow only those in state $\bf RT$ to work, but we think that is unrealistic, because society still needs essential services to go on, and especially given what we currently see in lockdown policies all over the world. \\  \vspace{-2mm}

{\color{blue} Social activity and behavioral response.} The salient piece of the model is the introduction of behavioral response in the agents' social activities. We allow the agents to voluntarily \emph{social distance} themselves. They have strict preferences to participate in social activities sans the virus. To capture this idea, we suppose that social distancing is costly, its flow costs are denoted by $\frac{c}{2}(1-\alpha)^2$, where $\alpha \in [0,1]$ is a fraction of social activities in which an agent participates. The costs are convex, more specifically quadratic, which captures an intensive margin of the agents' social distancing decision. Since the agents are forward-looking, they account for the current and future costs from social distancing. In addition to $\frac{c}{2}(1-\alpha)^2$, the agents also face two other (lump-sum) costs: a threat of getting infected and altruistic concerns of not infecting others; they suffer disutility of $\phi^+$ and $\phi^-$, respectively, from the two scenarios.\footnote{The idea behind the three parameters $(c,\phi^{+},\phi^{-})$ is that one level we want to differentiate the opportunity cost of not partaking in social activities from the cost of actually getting infected, and at another level we want to decompose the lump-sum cost of getting infected into personal costs and altruistic concerns of infecting others.}\\ \vspace{-2mm}

{\color{blue} Vaccine.} Finally, we also allow for the possibility of discovering a \emph{vaccine} which can cure the virus. For simplicity, upon its arrival all infected agents are assumed to be cured, thus the epidemics effectively stops. Suppose the vaccine arrives at a (random)  time $T$. The total output after the end of epidemics at $t > T$ equals to the measure of the agents who survive, that is $1-D_{T}$.

We choose to model the arrival of vaccine through a negative binomial distribution. This is empirically relevant for at least two reasons. First, the distribution is parametrized by two variables: its mean $\mathbb{E}[T]$ and variance $\mathbb{V}[T]$. So, we can control both aspects of the distribution separately to reflect the reality on vaccine consensus all through the year 2020 before a credible claim on the existence of a vaccine was made. Second, the negative binomial is a sort of "repeated Bernoulli" distribution. Each period there is a probability of success or failure, and only after an exogenously specified number of success do we deem the event "vaccine developed" realized. Thus, there is a minimal time till which no vaccine can be developed, which is precisely given by $\frac{(\mathbb{E}[T])^2}{\mathbb{E}[T]+\mathbb{V}[T]}$. In what follows, we will let $p_{t}$ be the probability that vaccine arrives at time $t$, that is $T=t$, and let $q_{t} := 1-\sum\limits_{s \leqslant t}p_{t}$ be the probability that the vaccine hasn't arrived till time $t$.\footnote{Other leading candidate is the Poisson or geometric arrival, which is stationary, and hence would ensure that likelihood of the arrival of vaccine two months from the start of the pandemic is the same for it to be developed in 14 months at the 12 month mark. This distribution is a limiting case of the negative binomial in which the mean and variance coincide.  This has been used, for example, by \citet{lockdown_SIR} and \citet{farboodietal_covid}.}\\ \vspace{-2mm}

{\color{blue} Tying it all together.} Our main interest lies in understanding how the tradeoff between economic well-being and mortality is shaped by the simultaneous interaction of three forces: social distancing, tracing and testing, and lockdown. At the outset, the planner commits to the testing and lockdown policies which will be in effect until the development of cure. Then, within each period the following sequence of events occurs:
\begin{enumerate}
\item
The agents in $\left\{\bf S,I,R, RT\right\}$ participate in economic and social activities, thus creating matches and spreading the virus.
\item
The agents who were in $\{\bf I,IT,H\}$ may recover or transition to a follow-up compartment.
\item
A fraction $\tau_t\gamma$ of $\left\{\bf S,I,R\right\}$ is tested according to the technology described above, and amongst the tested, those in states $\bf I$ and $\bf R$ transition respectively to $\bf IT$ and $\bf RT$. 
\item
The vaccine maybe discovered at time $T$, which is a random variable.
\end{enumerate}

Finally, the (ex ante) payoff of a (representative) agent is given by $Y_A-C$, where $Y_A$ is the total expected output, which is computed using the agent's discount factor, and $C$ is the total cost. One way to think about $Y_A$ is that the agent directly cares about (or is compensated for) the amount of output produced. The total cost is driven by three parameters: $c$, $\phi^{+}$ and $\phi^{-}$, flow cost from social distancing, and one time disutilities from getting infected and infecting others respectively. The payoff of the government is given by $Y_G + \xi \cdot (1-D)$, where $Y_G$ is the same as $Y_A$ except we will use a (potentially) different discount factor $\delta_G \in [0,1)$ for the government, $1-D$ is the expected number of survivors, and $\xi$ is the relative  Pareto weight the government puts on them. Since the functional forms of these objects require some further notation, we will make them precise later.

\section{Dynamics of infection}

\label{section dynamics}

In this section we define the set of equations that governs the dynamics of transitions amongst $\left\{\bf S,I,IT,R, RT,H,D\right\}$. Importantly, we shut down the behavioral channel for now and try to understand mechanical aspects of the dynamics, bringing in the effect of behavioral response to the dynamics in the next section. Two types of parameters will be introduced here: the average time it takes for an agent to transfer from one state to another, and conditional on the transfer away from one state, the fraction that goes to each of the possible destination states. 

Recollect that the government can control the spread of the disease by uniformly locking down a fraction of $1-\lambda_t$ of productive agents, with the maximal lockdown capped at $1-\bar\lambda$.  Suppose that the susceptible and infected agents always participate in the social activities, then the stock of susceptible evolves as \\ \vspace{-9mm}

\begin{small}
\begin{align}
\label{eq S mech}
\tag{$\Delta S_t$}
S_{t+1} = S_t  - \beta_w \lambda_t^2 S_t I_t - \beta_s  S_t I_t,
\end{align}
\end{small}\vspace{-5mm}

\noindent where we count away the number of new infections due to matches at work and social activities. This is of course {\it one} way of modeling economic and social interactions within the standard epidemiology framework. As a first pass and for tractability we consider this additively separable assumption where the outflow of new infections is "substitutable" with coefficients $\beta_{w}$ and $\beta_{s}$ between the two primary activities for the agents in society. 

At the inception, there is an initial seed of infection $I_{0} = e_{0}$, and all others are susceptible: $S_{0} =1-e_{0}$. Periodic matches at work and in social activities create new infections. We assume that infected agents either develop clear symptoms (and need to be hospitalized) or recover at a fixed rate on average time $t_i$. Taking into an account the attrition from $\bf I$ and addition to $\bf IT$ due to testing, we get that the following equations describe the dynamics of infected cases:\\ \vspace{-9mm}

\begin{small}
\begin{align}
\label{eq I mech}
\tag{$\Delta I_t$}
I_{t+1} &= (1-\tau_t)\bigg(\bigg(1-\frac{1}{t_i}\bigg)I_t + \beta_w\lambda^2_tS_tI_t + \beta_sS_tI_t\bigg),
\\
\label{eq IT mech}
\tag{$\Delta IT_t$}
IT_{t+1} &= \bigg(1-\frac{1}{t_i}\bigg)IT_{t}+ \tau_t\bigg(\bigg(1-\frac{1}{t_i}\bigg)I_{t}+\beta_w \lambda_t^2 S_t I_t + \beta_s  S_t I_t\bigg).
\end{align}
\end{small}
\vspace{-5mm}

Recollect that the agents can transition from being infected, ($\bf I$) or ($\bf IT$), to either developing severe symptoms and being hospitalized ($\bf H$) or recover ($\bf R$). We assume that conditional on transitioning away from $\bf I$ or $\bf IT$, the states $\bf H$ and $\bf R$ are reached with probabilities of transitions given by $m_i$ and $1-m_i$, respectively. Further, we assume that those hospitalized ($\bf H$) transition away from that state on average time  $t_h$. Thus, we have: \\ \vspace{-9mm}

\begin{small}
\begin{align}
\label{eq H}
\tag{$\Delta H_t$}
H_{t+1} = \bigg(1-\frac{1}{t_h}\bigg)H_t + \frac{m_i}{t_i}(I_t+IT_t).
\end{align}
\end{small}
\vspace{-5mm}

Next, the state $\bf D$ is an absorbing state, which keeps count of the number of fatalities amongst those agents that become hospitalized ($\bf H$), and do not recover ($\bf R$). We assume that the hospitalized agents, recover with the probability $1-m_h$ and die with the complimentary probability $m_{h}$. Thus, we have: \\ \vspace{-9mm}

\begin{small}
\begin{align}
\label{eq D}
\tag{$\Delta D_t$}
D_{t+1} & = D_t + \frac{m_h}{t_h}H_t.
\end{align}
\end{small}
\vspace{-5mm}

The state $\bf RT$ is absorbing while the state $\bf R$ is ``almost" absorbing in the sense that the event consists agents that have recovered without ever showing symptoms, and if tested this will be revealed and they would transfer to the state $\bf RT$. There are three sources chipping into $\bf R$ and $\bf RT$: infected and not tested, infected and tested, and hospitalized. These  stocks evolve as follows:\\ \vspace{-7mm}

\begin{small}
\begin{align}
\label{eq R mech}
\tag{$\Delta R_t$}
R_{t+1} &=(1-\tau_t\gamma)\bigg(R_t +\frac{1-m_i}{t_i}I_t\bigg),\\
\label{eq RT}
\tag{$\Delta RT_t$}
RT_{t+1} &= RT_t + \frac{1-m_i}{t_i}IT_t +\frac{1-m_h}{h_t}H_t + \tau_t\gamma\bigg(R_t+\frac{1-m_i}{t_i}I_t\bigg).
\end{align}
\end{small}
\vspace{-5mm}

To sum up, the dynamics of infections and economic activity are jointly determined by the system of equations (\ref{eq S mech}) - (\ref{eq RT}). In the next section we will add behavioral responses to this otherwise mechanical system. Some comments about the set of states we choose to model disease dynamics are instructive. The basic framework here is based on the celebrated SIR model which constitutes three states, namely susceptible, infected and recovered, starting at least from \citet{epidemic_first}, see also \citet{covid_basel} for a recent treatment. We enrich the setup by separating the infected and recovered into two categories on the basis of testing, and creating a separate state for those that show severe symptoms and thus need to be hospitalized. These two choices allow us to (i) introduce testing in a realistic way, (ii) match the widely reported stylized fact that asymptomatic carries are the largest source of contagion, at least for Covid-19, and (iii) generate at least three time series $\left(IT_{t}, H_{t}, D_{t}\right)$ that are observable in the data coming out from various countries.

\section{Behavioral response and the agent's problem}

\label{section behavior}

We augment the epidemiological system of equations described above by giving the agents the agency to social distance themselves. This decision is motivated by two types of parameters: the flow cost of social distancing ($c$) and the (one time) disutility from either getting the infection or infecting some other person ($\phi^+$ and $\phi^-$, respectively). All three are assumed to be fixed {\it behavioral} parameters. Further, the agent chooses a probability of social distancing at each time. Recollect that behavioral response is only relevant for agents in the states $\left\{\bf S,I, R\right\}$ because the others are either quarantined, hospitalized, known recovered, or dead.\footnote{Note that $\phi^{+}$ is the disutility an agent suffers from getting infected, which we take here to include the psychological cost from infecting close friends and family, and the actual perceived cost of dying or the death of these family or friends. On the other hand $\phi^{-}$ refers to the disutility the agent suffers from infecting a person he/she is randomly matched within a workspace, buying groceries or eating out or socializing in crowded places.}

\subsection{Modified dynamics}

In Section \ref{section dynamics} the stock values of all the seven states completely characterized the dynamics. However, testing introduces heterogeneity in the model which is then propagated through the agents' behavioral response. As a consequence, aggregate stock variables are no longer sufficient to keep track of the disease dynamics. In fact, we need to trace the evolution of the susceptible ($\bf S$), infectious ($\bf I$) and recovered ($\bf R$) agents conditional on the \emph {time of last test}. That the time of last test is a sufficient statistic for the dynamics follows from our assumption that testing perfectly reveals an agent's current health state. After taking a test the agent is certain if she is still susceptible, infectious or recovered. In the last two scenarios the agent will transit to the follow-up compartments and never be ``eligible'' for testing in the future. As a result, those who are tested at time $t$ and found to be susceptible have identical (degenerate) beliefs independently of the exact time instance $t$.\footnote{The assumption that testing is history-independent, that is all agents in $\left\{\bf S,R\right\}$ and all agents in $\bf I$ have the same likelihoods of getting tested, can be relaxed to allow for conditioning on the time of last test.}

Define by $S_t^k$, $I_t^k$ and $R_t^k$ the aggregate fractions of agents who are susceptible, infected and recovered, given calendar time $t$ and time of the last test $k \leqslant t$. For example, an agent in state $I_{50}^{35}$ was tested at the 35th day of the pandemic, found to be susceptible, and has since acquired the virus at some point in the last 15 days, but is asymptomatic. Here, $k=1$ tracks the subpopulation of agents who never got any evidence, and $S^{t}_{t}$ is the fraction of people who have just been tested to be susceptible. Further, let $\alpha_t^k$ be the \emph {probability that the agent who received the last test at time $k$ participates in social activities at time $t$}. We note that the total number of susceptible, asymptomatic-infected and recovered satisfy the following:\\ \vspace{-9mm}

\begin{small}
\[
S_t =\sum_{k \leqslant t} S^k_t,\;I_t =\sum_{k \leqslant t} I^k_t,\;R_t=\sum_{k \leqslant t} R^k_t.
\]
\end{small}\\ \vspace{-9mm}

\noindent Similarly, define the total number of susceptible, asymptomatic-infected and recovered
among those who participate in social activities as\\ \vspace{-9mm}

\begin{small}
\[
\widehat S_t := \sum_{k \leqslant t} \alpha_t^k S_t^k, \; \widehat{I}_t := \sum_{k \leqslant t}  \alpha_t^k I_t^k, \;\widehat{R}_t := \sum_{k \leqslant t}  \alpha_t^k R_t^k.
\]
\end{small}\\ \vspace{-9mm}

\noindent Here, $\widehat S_t$, $\widehat {I}_t$ and $\widehat {R}_t$ are equilibrium objects and potentially smaller than the total numbers of people in respective compartments, viz. $S_t$, $I_t$ and $R_t$, respectively. 

When the agents make social distancing decisions, the mechanical system (\ref{eq S mech}) - (\ref{eq RT}) has to be adjusted to incorporate the fact that socially inactive agents cannot spread the virus. Slightly abusing notations, we redefine certain equations taking into an account the agents' behavioral response:\\ \vspace{-9mm}
\begin{small}
\begin{align}
\label{eq S}
\tag{$\Delta S_t^k$}
S^k_{t+1} &= (1-\tau_t\gamma)\bigg(S^k_t-\beta_w\lambda^2_tS^k_tI_t - \beta_s\alpha^k_tS^k_t\widehat{I_t}\bigg)\; \text{for } k=1,\ldots, t,
\\
\tag{$\Delta S_t^{t+1}$}
S^{t+1}_{t+1}&=\tau_t\gamma\bigg(S_t -\beta_w\lambda^2_tS_tI_t - \beta_s\widehat{S_t}\widehat{I_t}\bigg),
\\
\label{eq I}
\tag{$\Delta I_t^k$}
I^k_{t+1} &= (1-\tau_t)\bigg(\bigg(1-\frac{1}{t_i}\bigg)I^k_t + \beta_w\lambda^2_tS^k_tI_t + \beta_s\alpha^k_tS^k_t\widehat{I_t}\bigg)\;\text{for } k=1,\ldots, t, 
\\
\label{eq IT}
\tag{$\Delta IT_t$}
IT_{t+1} &=\bigg(1-\frac{1}{t_i}\bigg)IT_t + \tau_t\bigg(\bigg(1-\frac{1}{t_i}\bigg)I_t + \beta_w\lambda^2_tS_tI_t + \beta_s\widehat{S_t}\widehat{I_t}\bigg),
\\
\label{eq R}
\tag{$\Delta R_t^k$}
R^k_{t+1} &=(1-\tau_t\gamma)\bigg(R^k_t +\frac{1-m_i}{t_i}I^k_t\bigg)\;\text{for } k=1,\ldots, t. 
\end{align}
\end{small}\\ \vspace{-9mm}

\noindent  In addition, of course, $I^{t+1}_{t+1} = 0$, and $R^{t+1}_{t+1}= 0$. The modified dynamics makes it clear that we additively separate the two worlds of economic and social decisions, and allow lockdown policies to affect the former and social distancing to influence the latter. Testing in turn, endogenously affects both decisions by reducing those partaking in both worlds in the state $\bf I$ and potentially increasing those in the state $\bf R$. 

A quick sanity check on the system can be done by noticing that since the total population remains constant (including counting the number of deaths), the sum of all states at all points must equal unity, and in fact we have: $S_t+I_t+R_t+IT_t+RT_t+H_t+D_t  = 1$.

It follows that it is without loss to ignore the equation for the state $\bf D$, i.e., at any point in time $D_t$ can be uniquely determined from the other state variables using the above identity. To save on notations in the future, we stack together the state variables and corresponding equations as\\ \vspace{-9mm}

\begin{small} 
\begin{align}
\Pi_t := &  \Big((S_t^k)_{k \leqslant t},(I_t^k)_{k \leqslant t},IT_t,(R_t^k)_{k \leqslant t},RT_t,H_t\Big),\nonumber
\\ 
 \Delta \Pi_t := & \Big((\Delta S_t^k)_{k \leqslant t+1},(\Delta I_t^k)_{k \leqslant t},\Delta IT_t,(\Delta R_t^k)_{k \leqslant t},\Delta RT_t,\Delta H_t\Big).\nonumber
\label{eqpi} 
\end{align}
\end{small}\\ \vspace{-10mm}

Equations \hyperref[eqpi]{$(\Delta \Pi_t)$} jointly define the disease dynamics in an equilibrium, when the tests follows the process $\overline{X}_{t}$, the government sets the lockdown policy  $\lambda_{t}$ and the agents best-respond by choosing social distancing $(\alpha_{t}^{k})_{k\leqslant t}$ as a function of the last time of being tested. In the parlance of dynamic optimization theory, these state equations are termed {\it forward} because they are pinned down by the initial conditions at the outset, i.e., $S_{1}^1=1-e_{0}$ and $I_{1}^1 = e_{0}$. 

\subsection{Agent's optimization problem}

\label{agent_opt}

We now turn to the agent's problem. Since a representative agent is infinitesimal, she cannot influence the aggregate state variables. Instead, the agent takes the dynamics of aggregate states as well as the government's lockdown and testing policies as given, and maximizes her own expected payoff by choosing the rate of social distancing $(\alpha_t^k)_{k \leqslant t}$. In this quest the agent internalizes how her current social distancing decision will affect likelihood of getting infected in the future. It turns out that the agent's problem is a dynamic control problem with the set of constraints which resembles Equations \hyperref[eqpi]{$(\Delta \Pi_t)$} modulo the fact that certain aggregate variables are taken to be fixed.

To formally define the agent's optimization problem, we should distinguish between an agent's state vector $\pi_t:= \big((s_t^k)_{k \leqslant t},(i_t^k)_{k \leqslant t},it_t,(r_t^k)_{k \leqslant t},rt_t,h_t\big)$ and its aggregate counterpart $\Pi_t$. For example, $s_t^k$ is the ex-ante probability that a representative agent is susceptible at time $t$ when she last time got tested at time $k$. The remaining variables in $\pi_t$ are interpreted similarly. We will make use the following shorthand notations: $y_t := \lambda_t (s_t+i_t+r_t+rt_t)$, and\\ \vspace{-9mm}

\begin{small}
\begin{align*}
&s_t:=\sum_{k \leqslant t}s_t^k,\;i_t:=\sum_{k \leqslant t}i_t^k,\;r_t:=\sum_{k \leqslant t}r_t^k,\;\widehat s_t:=\sum_{k \leqslant t}\alpha_t^ks_t^k,\;\widehat i_t:=\sum_{k \leqslant t}\alpha_t^k i_t^k,\;\widehat r_t:=\sum_{k \leqslant t}\alpha_t^k r_t^k.
\end{align*}
\end{small}\\ \vspace{-9mm}

\noindent In addition, note that the probability of being dead $d_t$ at time $t$ follows from the accounting identity: $s_t+i_t+r_t+it_t+rt_t+h_t+d_t=1$.
\vspace{2mm}

We now have all the concepts and notation in place to define the agent’s preferences. The (ex ante) expected payoff of the representative agent is given by $Y_A-C$, where  \\ \vspace{-9mm}

\begin{small}
\[
Y_A :=(1-\da)\sum_{t =1}^{\infty} \da^{t-1}\bigg(y_tq_t + \frac{\da}{1-\da}(1-d_{t+1}) p_t\bigg),
\]
\end{small}
and
\begin{small}
\begin{align*}
C := (1-\da)\frac{c}{2}\sum_{t =1}^{\infty}\da^{t-1}\bigg((it_t+h_t+d_t)q_t+\frac{\da}{1-\da}d_{t+1}p_t + \sum_{k\leqslant t}(1-\alpha_t^k)^2(s_t^k+i_t^k+r_t^k)q_t \bigg) + \\
+(1-\da)\sum_{t =1}^{\infty}\da^{t-1}\bigg(\phi^+\left(\beta_w\bl_t^2s_t{I}_t+\beta_s\widehat s_t{\widehat I}_t\right)+\phi^-\left(\beta_w\bl_t^2{S}_ti_t+\beta_s{\widehat S}_t \widehat i_t\right)\bigg)q_t.
\end{align*}
\end{small}\\ \vspace{-9mm}

\noindent The functional form of $Y_A$ is straightforward, it simply aggregates the total (expected) output produced by the agent using her individual discount factor. One way to think about it is that the agent directly cares (or is compensated for) the amount of output produced.

The total cost is driven by three parameters: $c$, $\phi^{+}$ and $\phi^{-}$. The agents incur a flow cost $\frac{c}{2}(1-\alpha^k_t)^2$ from social distancing-- these arise in the states $\left\{\bf IT, H, D\right\}$ at their maximal level of $\frac{c}{2}$,  because complete  social distancing here is a necessity, and in the states $\left\{\bf S, I, R\right\}$, because social distancing here is a choice based on information of last test.

In addition, the agent suffers the one-shot disutility from getting infected $\phi^{+}$ and another one from infecting others $\phi^{-}$. The former represents the probabilistic cost of potentially getting very sick, or even facing death, and infecting members of family, while latter is an altruism parameters that captures the cost of infecting a random person at work or in the social meetings. We want the reader (relatively speaking) to the think of $c$ as being small, $\phi^{+}$ as being large, and $\phi^{-}$ as being somewhat intermediate.\footnote{There is a subtlety in how to interpret the altruistic parameter $\phi^{-}$. Each agent is non-atomistic, so they have a negligible impact on the aggregate. How should we then interpret the altruism parameter? What we are doing here is plugging in the beliefs of the individual into the preferences. So, it is an “as if” component of their utility. We want the reader to think of it as a psychological cost from the prospect of infecting others and since we are imposing a rational expectations equilibrium this cost is realized ex post. It can also be regarded as a ``warm glow effect" (in the sense of \citet{warm_glow}) from controlling the likelihood of infecting others.}

We now describe the laws of motion of the individual state vector. These differ from the laws of motion of the aggregate state vector, because each agent individually cannot influence behavior of the others on the matching markets. We have the following system:\\ \vspace{-9mm}

\begin{small}
\begin{align}
\label{eq S A}
\tag{$\Delta s_t^{k}$}
s^k_{t+1} &= (1-\bt_t\gamma)\bigg(s^k_t-\beta_w\bl^2_ts^k_tI_t - \beta_s\alpha^k_ts^k_t\widehat{I_t}\bigg)\;\text{for } k=1,\ldots, t,
\\
\tag{$\Delta s_t^{t+1}$}
s^{t+1}_{t+1}&=\bt_t\gamma\bigg(s_t -\beta_w\bl^2_ts_tI_t - \beta_s\widehat{s_t}\widehat{I_t}\bigg),
\\
\label{eq I A}
\tag{$\Delta i_t^k$}
i^k_{t+1} &= (1-\bt_t)\bigg(\bigg(1-\frac{1}{t_i}\bigg)i^k_t + \beta_w\bl^2_ts^k_tI_t + \beta_s\alpha^k_ts^k_t\widehat{I_t}\bigg)\;\text{for } k=1,\ldots, t, 
\\
\label{eq IT A}
\tag{$\Delta it_t$}
it_{t+1} &=\bigg(1-\frac{1}{t_i}\bigg)it_t + \bt_t\bigg(\bigg(1-\frac{1}{t_i}\bigg)i_t + \beta_w\bl^2_ts_tI_t + \beta_s\widehat{s_t}\widehat{I_t}\bigg),
\\
\label{eq R A}
\tag{$\Delta r_t^k$}
r^k_{t+1} &=(1-\bt_t\gamma)\bigg(r^k_t +\frac{1-m_i}{t_i}i^k_t\bigg)\;\;\text{for } k=1,\ldots, t,
\\
\label{eq RT A}
\tag{$\Delta rt_t$}
rt_{t+1} &= rt_t + \frac{1-m_i}{t_i}it_t +\frac{1-m_h}{h_t}h_t + \bt_t\gamma\bigg(r_t+\frac{1-m_i}{t_i}i_t\bigg),
\\
\label{eq H A}
\tag{$\Delta h_t$}
h_{t+1} &= \bigg(1-\frac{1}{t_h}\bigg)h_t + \frac{m_i}{t_i}(i_t+it_t).
\end{align}
\end{small}
In addition, $i_{t+1}^{t+1} = r_{t+1}^{t+1} = 0$ at all dates. Stack the above equations in a vector $\Delta\pi_t$, that is
\begin{small}
\[
\Delta\pi_t:=\Big((\Delta s_t^k)_{k \leqslant t+1},(\Delta i_t^k)_{k \leqslant t},\Delta it_t,(\Delta r_t^k)_{k \leqslant t},\Delta rt_t,\Delta h_t\Big) \label{eqpi A}  
\]
\end{small}\\ \vspace{-10mm}

\noindent The reader can verify that Equations \hyperref[eqpi A]{$(\Delta \pi_t)$} are almost identical to \hyperref[eqpi]{$(\Delta \Pi_t)$}; the only difference is that we have $I_t$ and $\widehat I_t$ instead of $i_t$ and $\widehat i_t$ in the first three equations. 

Finally, the agents' optimization problem is given by the choice of the social distancing $(\alpha_t^k)_{k \leqslant t}$ and individual state variables $\pi_t$ at each date to maximize expected output net of costs subject to the family of state equations \hyperref[eqpi A]{$(\Delta \pi_t)$}: 
\begin{empheq}[box=\tcbhighmath]{align*}
\max\limits_{(\alpha_t^k)_{k \leqslant t},\pi_t} \quad Y_A-C + (1-\da)c\kappa P \quad \text{subject to}\quad \text{\hyperref[eqpi A]{$(\Delta \pi_t)$}}; \; \text{given }\big(S_t,I_t,\widehat S_t,\widehat I_t\big)\;\text{and}\;(\lambda_t,\;\tau_t).
\end{empheq}
In the objective, $\kappa > 0$ is a small number and $P$ is a {\it punishment} term to avoid boundary solutions,\footnote{Interior solutions for $(\alpha_t^k)_{k \leqslant t}$ make the execution of the algorithm that finds the optimal solution to the government's problem  tractable. For context, we actually assume $\kappa$ to be $10^{-12}$ in our numerical calculations.} 
that is\\ \vspace{-9mm}

\begin{small}
\[
P:=\sum_{t=1}^{\infty}\da^{t-1}\sum_{k \leqslant t}\Big(\ln \alpha^k_t+\ln \big(1-\alpha^k_t\big)\Big)q_t.
\]
\end{small}\\ \vspace{-9mm}

The model admits as a special case the scenario where the agents are myopic: this will constitute substituting $\delta_A=0$. The government, as we will model later, will always be assumed to be forward looking. This assumption of myopia for the agents dramatically simplifies all the calculations and some have argued generates a model with nonetheless reasonable predictions. As mentioned in the introduction, we do not take a call on the agent's forward looking capacity and allow the analyst the flexibility of varying the agents' discount factors as she/he deems fit to the situation being studied. To provide intuition and to point out the implications of the myopic model, we will refer sometimes to the special case as we go along. 

\subsection{Equilibrium response and adjoint equations}

\label{eqm cond}

Solving for the (representative) agent's optimal social distancing rule involves setting up the Lagrangian function. Let the Lagrange multipliers corresponding to Equations \hyperref[eqpi A]{$(\Delta \pi_t)$} at time $t$ be $\Upsilon_t$, i.e., \\ \vspace{-9mm}

\begin{small} 
\[
\Upsilon_t:=\Big((\s^k_t)_{k \leqslant t+1},(\i_t^k)_{k \leqslant t},\i\t_t,(\r^k_t)_{k \leqslant t},\r\t_t,\h_t\Big).
\]  
\end{small}
\vspace{-5mm}

\noindent We calculate the agent's first order-condition with respect to the control variable $(\alpha_t^k)_{k \leqslant t}$ and then take the first-order condition with respect to each state variable in the optimization problem to get the agent's adjoint equations. 

The agent's optimal social distancing decision is thus characterized as a unique solution to the necessary first-order condition of the Lagrangian with respect to $\alpha^{k}_{t}$, which is of course sufficient for optimality whenever $\kappa$ is small enough: \\ \vspace{-9mm}

\begin{small}
\begin{align*}
\label{agent_foc}
\tag{$FOC-\alpha_t^k$}
1-\alpha^k_t + \frac{\kappa}{s_t^k+i_t^k+r_t^k}\bigg(\frac{1}{\alpha^k_t}-\frac{1}{1-\alpha^k_t}\bigg)
= \frac{\phi^+}{c}\frac{s^k_t}{s^k_t+i^k_t+r^k_t}\beta_s{\widehat I_t} + \frac{\phi^-}{c}\frac{i^k_t}{s^k_t+i^k_t+r^k_t}\beta_s{\widehat S_t} +
\\
+\frac{(1-\bt_t\gamma)\s^k_t +\s^{t+1}_t\bt_t\gamma -(1-\bt_t)\i^k_t -\bt_t\i\t_t}{c(1-\da)\da^{t-1}q_t}\frac{s_t^k}{s_t^k+i_t^k+r_t^k}\beta_s {\widehat I_t}.
\nonumber
\end{align*}
\end{small}
\vspace{-5mm}
  
\noindent  One intuitive way of thinking about this condition is to first consider the myopic case. Suppose $\da=0$, then the second line in Equation (\ref{agent_foc}) disappears.\footnote{This is because when $\da = 0$, the dual variables are zero at all dates.} Now, for this myopic case, take an agent who got evidence of being susceptible at time $k$, then at date $t$ she assigns probabilities \\ \vspace{-7mm}

\begin{small}
\[
f_{t}^{k} := \frac{s_t^k}{s_t^k+i_t^k+ r^{k}_{t}} \;\; \text{and} \;\; g_{t}^{k} :=  \frac{i_t^k}{s_t^k+i_t^k+r^{k}_{t}}
\]
\end{small}\\ \vspace{-7mm}

\noindent  to the events that she is susceptible and infectious, respectively.\footnote{If the agent has been already discovered to be infected/recovered, then the agent's social distancing decision at $t$ will be mechanical.} Thus, for $\kappa$ close to zero,  Equation (\ref{agent_foc}) can be re-written as follows (whenever the solution for $\alpha_t^k$ is interior):\footnote{In general, as $\kappa \downarrow 0$, the solution for $1-\alpha_t^k$ converges to $\max\{0,\min\{1,\cdot\}\}$, where ``$\cdot$" stays for the right-hand side of Equation (\ref{agent_foc}). The same is true in the more general context with $\da > 0$.} \vspace{-5mm}

\begin{small}
\begin{align*}
\underbrace{1-\alpha_t^k}_{\color{blue}\substack{\text{fraction of }\\\text{social distancers}}} \approx \underbrace{\frac{\phi^+}{c} f_{t}^{k}\beta_s{\widehat I_t}}_{\color{blue}\substack{\text{cost from }\\\text{getting infected}}} + \underbrace{\frac{\phi^-}{c} g_{t}^{k}\beta_s{\widehat S_t}}_{\color{blue}\substack{\text{cost from }\\\text{infecting others}}},
\end{align*}
\end{small} \vspace{-5mm}

\noindent The agent experiences disutility from getting infected and infecting others which are added up. Given the aggregate behavior, each susceptible agent gets infected with a probability $\beta_s {\widehat I_t}$, and each infectious agent infects another agent with a probability $\beta_s {\widehat S_t}$. Moreover, $\frac{\phi^{+}}{c}$ and $\frac{\phi^{-}}{c}$ represent the costs of getting infected and infecting others, respectively, normalized by the cost of not being able to socialize for one period. 

Having motivated the static component of Equation (\ref{agent_foc}), allow now for $\da>0$. The agents are forward looking, and thus internalize the fact that they will transition between states in the future. Yet, they do not internalize that their social distancing decision will affect the aggregate dynamics. 

Let $\Delta_t^k$ be the expected drop in the agent's continuation value due to an infection, that is $\Delta_t^k := (1-\bt_t\gamma)\s^k_t +\bt_t\gamma\s^{t+1}_t -(1-\bt_t)\i^k_t -\bt_t\i\t_t$. Re-writing Equation (\ref{agent_foc}) for small $\kappa$ and assuming that the social distancing decision is interior, we get: \vspace{-5mm}

\begin{small}
\begin{align*}
1-\alpha_t^k \approx \underbrace{\ldots\ldots}_{\text{\color{blue} static, $\delta_A = 0$}} +  \underbrace{\frac{\Delta_t^k}{c(1-\da)\da^{t-1}q_t}f_{t}^{k}\beta_s{\widehat I_t}}_{\color{blue}\substack{\text{dynamic cost of}\\ \text{an infection}}},
\end{align*}
\end{small} \vspace{-5mm}

\noindent A higher value of $\alpha_t^k$ increases the likelihood that the susceptible agent gets infected by $\beta_s \widehat I_t$. In this case the agent will transit to $it_t$ with the probability $\bt_t$ and $i_t^k$ with the probability $1-\bt_t$, thus the agent's expected future value increases in a proportion to $(1-\bt_t)\i_t^k+\bt_t \i\t_t$. At the same time, the likelihood that in the next period the agent will still be susceptible decreases; by the similar logic as above the agent's expected future value decreases in a proportion to $(1-\bt_t\gamma)\s_t^k+\bt_t\gamma \s_t^{t+1}$. Finally, as in the static case, the dynamic (lump-sum) cost is deflated by the net present value of the flow cost of social distancing. 

In addition to the first-order condition with respect to the control vector $(\alpha_t^k)_{k \leq t}$, we have the system of seven partial difference equations (\hyperref[equps A]{$\Delta \Upsilon_t$}) which describes the dynamics of the adjoint variables $\Upsilon_t$. This system is similar in spirit to Equations  \hyperref[eqpi A]{$(\Delta \pi_t)$}, which define the dynamics of individual state variables $\pi_t$. The main difference though is that that the system is \emph{backward}, i.e., it specifies $\Upsilon_t$ as a function of $\Upsilon_{t+1}$ and other variables at time $t+1$. The ``initial'' conditions are given at ``infinity'': \\ \vspace{-5mm}

\begin{small}
\[
\lim_{t \to \infty} \frac{\s^k_t}{p_t\da^{t-1}} = \lim_{t \to \infty} \frac{\i^k_t}{p_t\da^{t-1}} = \lim_{t \to \infty} \frac{\r^k_t}{p_t\da^{t-1}} = \lim_{t \to \infty} \frac{\i\t_t}{p_t\da^{t-1}} = \lim_{t \to \infty} \frac{\r\t_t}{p_t\da^{t-1}} = \lim_{t \to \infty} \frac{\h_t}{p_t\da^{t-1}} = \da\left(1+\frac{c}{2}\right) \; \forall k.
\]
\end{small} \\ \vspace{-7mm}

\noindent  As a result, the adjoint variables $\Upsilon_t$ have to be solved backwards. We present Equations \hyperref[equps A]{$(\Delta \Upsilon_t)$} for $\Upsilon_t$ and their derivations in the appendix (see Section \ref{agentapp}). 

Finally, equilibrium dictates that individual behavior must be consistent with aggregate behavior at every point in time, i.e., $\pi_t = \Pi_t$, $\widehat{s}_t = \widehat{S}_t$ and $\widehat{i}_t = \widehat{I}_t$. For the rest of the analysis we will utilize these equilibrium conditions to \emph{express all the state variables in capital fonts.}

\section{The government's problem}

\label{govt_opt}

The government's objective function is given by a weighted sum of total output, which is computed using the social discount factor, and the number of survivors: $Y_G + \xi \cdot (1-D)$, where \\ \vspace{-6mm}

\begin{small}
\[
Y_G :=(1-\dg)\sum_{t=1}^{\infty}\dg^{t-1}\bigg(Y_tq_t + \frac{\dg}{1-\dg}(1-D_{t+1}) p_t\bigg),\; 1-D :=\sum_{t=1}^{\infty}(1-D_{t+1})p_t,
\]
\end{small} \vspace{-5mm}

\noindent  and the number $\xi$ is the relative Pareto weight on $1-D$. 

The government has three control variables, these are the rates of lockdown $\bl_t$, testing $\bt_t$ and social distancing as a social planner $(\alpha^k_t)_{k \leqslant t}$. Moreover, the government also incorporates the agents' optimal behavior in its decision making. Thus, the Lagrangian of the government's optimal control problem takes as constraints the state equations (forward system \text{\hyperref[eqpi A]{$(\Delta \pi_t)$}}, now written as \hyperref[eqpi]{$(\Delta \Pi_t)$}), the agent's adjoint equations (backward system, \hyperref[equps A]{$(\Delta \Upsilon_t)$}) and the agent's first-order conditions with respect to the social distancing vector (\ref{agent_foc}).

As the last constraint, the government is assumed to be exogenously restricted in its testing capacity to $\overline X_{t}$. We use the actual test data to feed this time series. Making the rate of testing a choice variable, $\tau_{t}$, then gives us the resource constraint:  \vspace{-5mm}

\begin{small}
\begin{align}
\label{eq testing capacity} 
\overline X_t=\bt_t\bigg(\underbrace{\gamma\bigg(S_t-\beta_w\bl^2_tS_tI_t-\beta_s\widehat{S}_t\widehat{I}_t\bigg)}_{\color{blue}\text{``eligible'' ($S$)}}+\underbrace{\gamma\bigg(R_t+\frac{1-m_i}{t_i}I_t\bigg)}_{\color{blue}\text{``eligible'' ($R$)}}+\underbrace{\bigg(1-\frac{1}{t_i}\bigg)I_t + \beta_w\bl^2_tS_tI_t+\beta_s\widehat{S}_t\widehat{I}_t}_{\color{blue}\text{``eligible'' ($I$)}}\bigg).
\tag{$F-\tau_t$}
\end{align}
\end{small} \vspace{-5mm}

\noindent Recall that the agents are tested at the end of period $t$ after new matches, and hence infections, are created. Then, the fraction $\gamma$ of the susceptible $(\bf{S})$ and recovered $(\bf{R})$ added to the pool of the infected $(\bf{I})$, and each agent in this set is tested at the same rate $\bt_t$.\footnote{A comment on modeling choice is in order here: We could have let $\overline X_{t} $ be a choice variable by introducing say a convex cost of tracing and testing, but the varied pathologies of why different countries ended up with their own trajectory of total tests is deeply enmeshed in political economy which requires a separate study of its own. It seems to us more reasonable, given the complexity of the model in other dimensions, to choose the time series of tests that actually materialized.} 

To sum up, the government seeks to maximize $Y_G + \xi \cdot (1-D)$ by choosing the policies $\bl_t$, $\{\alpha_t^k\}_{k \leqslant t}$ and $\tau_t$ which jointly control the dynamics of the state variables $\Pi_t$ and agent's adjoint variables $\Upsilon_t$. The government is further constrained by the resource constraint on testing and the fact that the agent's social distancing decision must be a best-response. Thus, the government's problem can be stated in a consolidated way as follows: 
\begin{empheq}[box=\tcbhighmath]{align*}
&\max\limits_{\bl_t,(\alpha_t^k)_{k \leqslant t},\bt_t,\Pi_t,\Upsilon_t} \quad Y_G + \xi \cdot (1-D)  \quad \text{subject to}\quad \text{\hyperref[eqpi]{$(\Delta \Pi_t)$}},\;\text{\hyperref[equps A]{$(\Delta \Upsilon_t)$}},\;\text{(\ref{agent_foc})}\;\text{and}\;\text{(\ref{eq testing capacity})}.
\end{empheq}

It should be noted that the government is not utilitarian in the usual sense of the word. It takes a specific stance that its main job is to ensure maximal economic output while keeping number of fatalities in check. The agents' utilities enter the government's problem as a constraint since it internalizes their social distancing decision. At a conceptual level, our analysis can thus be regarded as theory of second-best.\footnote{\citet{flavio_paper3} explores implications of the SIS model for different objectives of the ``planner", terming them controlled and uncontrolled decentralized equilibrium and social optimum. In the context of that framework, we are closest to a controlled decentralized equilibrium.}

We next describe the set of optimality conditions associated with the planner's problem and the numerical algorithm to solve the model.

\subsection{Optimality conditions}

\label{opt cond}

We use the Lagrangian approach to solve the government's problem. We now describe the system of constraints as well as the family of Lagrange multipliers which we will attach to them.
\begin{itemize}
\item 
Recollect $\Pi_t = \big((S_t^k)_{k \leqslant t},(I_t^k)_{k \leqslant t},IT_t,(R_t^k)_{k \leqslant t},RT_t,H_t\big)$ satisfies Equations \hyperref[eqpi]{$(\Delta \Pi_t)$}. Define the government's adjoint variables corresponding to these equations to be $\overline \Upsilon_t$, i.e., \\ \vspace{-9mm}

\begin{small}
\[
\overline \Upsilon_t:=\Big((\gs^k_t)_{k \leqslant t+1},(\gi^k_t)_{k \leqslant t},\git_t,(\gr^k_t)_{k \leqslant t},\grt_t,\gh_t\Big).  
\]
\end{small}
\vspace{-8mm}

Similar to the agent's problem, we denote by $\overline{\Delta}_t^k$ the expected drop in the government's continuation value due to an infection, that is\\ \vspace{-9mm}

\begin{small}
\[
\overline{\Delta}_t^k := (1-\bt_t\gamma)\gs^k_t +\bt_t\gamma\gs^{t+1}_t -(1-\bt_t)\gi^k_t -\bt_t\git_t.
\]
\end{small}
\vspace{-8mm}

\item
The government internalizes that each agent will best-respond. This is incorporated in the government's problem by taking the rate of social distancing $(\alpha_t^k)_{k \leqslant t}$ and the agent's adjoint variables $\Upsilon_t$ to be choice variables on their own. Of course, the government is constrained by the fact that the agent's adjoint variables $\Upsilon_t$ must respect the adjoint equations \hyperref[equps A]{$(\Delta \Upsilon_t)$}, which are formally derived in the appendix. Denote the vector of dual variables associated with this system by $\overline{\Pi}_t$:  \\ \vspace{-9mm}

\begin{small}
\[
\overline{\Pi}_t := \Big((\gas_t^k)_{k \leqslant t},(\gai_t^k)_{k \leqslant t},\gait_t,(\gar_t^k)_{k \leqslant t},\gart_t,\gah_t\Big).
\] 
\end{small}
\vspace{-8mm}

\item
Let $\chi_t$ be the vector of Lagrange multipliers associated with the testing constraint, viz. Equation (\ref{eq testing capacity}). 
\item
Finally, we define $\{\eta_t^k\}_{k \leqslant t}$ to be the vector of Lagrange multipliers associated with the agent's first-order condition with respect to the rate of social distancing $(\alpha_t^k)_{k \leqslant t}$, viz. Equation (\ref{agent_foc}). Again, this is needed to guarantee that the rate of social distancing constitutes the agent's best-reply.
\end{itemize}

The next step is to solve for the Lagrange variables $\overline{\Pi}_t$. The reader can verify that these variables satisfy system which resembles Equations \hyperref[eqpi]{$(\Delta \Pi_t)$}, yet there are several noticeable differences: \\ \vspace{-9mm}

\begin{small}
\begin{align}
\label{eq S bar}
\tag{$\Delta \gas_t^k$}
\overline{S}^k_{t+1} &= (1-\bt_t\gamma)\bigg(\overline{S}^k_t-\beta_w\bl^2_t\overline{S}^k_tI_t - \beta_s\bigg(\alpha^k_t\overline{S}^k_t-\eta^k_t{S}^k_t\bigg)\widehat{I}_t\bigg)\;\text{for } k=1,\ldots, t, 
\\
\tag{$\Delta \gas_t^{t+1}$}
\overline{S}^{t+1}_{t+1}&=\bt_t\gamma\bigg(\overline{S}_t -\beta_w\bl^2_t\overline{S}_tI_t - \beta_s\bigg(\widehat{\overline{S}}_t-\sum_{k \leqslant t}\eta_t^kS_t^k\bigg)\widehat{I}_t\bigg),
\\
\label{eq I bar}
\tag{$\Delta \gai_t^k$}
\overline{I}^k_{t+1} &= (1-\bt_t)\bigg(\bigg(1-\frac{1}{t_i}\bigg)\overline{I}^k_t + \beta_w\bl^2_t\overline{S}^k_tI_t +\beta_s\bigg(\alpha^k_t\overline{S}^k_t-\eta^k_t{S}^k_t\bigg)\widehat{I}_t\bigg)\;\text{for } k=1,\ldots, t, 
\\
\label{eq IT bar}
\tag{$\Delta \gait_t$}
\overline{IT}_{t+1} &=\bigg(1-\frac{1}{t_i}\bigg)\overline{IT}_t + \bt_t\bigg(\bigg(1-\frac{1}{t_i}\bigg)\overline{I}_t + \beta_w\bl^2_t\overline{S}_tI_t +\beta_s\bigg(\widehat{\overline{S}}_t-\sum_{k \leqslant t}\eta_t^kS_t^k\bigg)\widehat{I}_t\bigg),
\\
\label{eq R bar}
\tag{$\Delta \gar_t^k$}
\overline{R}^k_{t+1} &=(1-\bt_t\gamma)\bigg(\overline{R}^k_t +\frac{1-m_i}{t_i}\overline{I}^k_t\bigg),\;\;\;\text{for } k=1,\ldots, t, 
\\
\label{eq RT bar}
\tag{$\Delta \gart_t$}
\overline{RT}_{t+1} &= \overline{RT}_t + \frac{1-m_i}{t_i}\overline{IT}_t +\frac{1-m_h}{h_t}\overline{H}_t + \bt_t\gamma\bigg(\overline{R}_t+\frac{1-m_i}{t_i}\overline{I}_t\bigg),
\\
\label{eq H bar}
\tag{$\Delta \gah_t$}
\overline{H}_{t+1} &= \bigg(1-\frac{1}{t_h}\bigg)\overline{H}_t + \frac{m_i}{t_i}(\overline{I}_t+\overline{IT}_t),
\end{align}
\end{small} \vspace{-4mm} 

where $\overline{I}_{t+1}^{t+1} = \overline{R}_{t+1}^{t+1} = 0$, $\overline{S}_t:= \sum_{k \leqslant t} \overline{S}_t^k$, $\overline{I}_t:= \sum_{k \leqslant t} \overline{I}_t^k$, $\overline{R}_t:= \sum_{k \leqslant t} \overline{R}_t^k$ and $\widehat{\overline S}_t:= \sum_{k \leqslant t} \alpha_t^k\overline{S}_t^k$, $\widehat {\overline I}_t:= \sum_{k \leqslant t} \alpha_t^k\overline{I}_t^k$. As usual, we stack together the aforementioned equations as 

\begin{small} 
\[
\Delta \overline{\Pi}_t := \Big((\Delta \gas_t^k)_{k \leqslant t},(\Delta \gai_t^k)_{k \leqslant t},\Delta \gait_t,(\Delta \gar_t^k)_{k \leqslant t},\Delta \gart_t,\Delta \gah_t\Big) \label{eqpi bar}.  
\] 
\end{small}

We note that holding the path of state variables fixed, the Lagrange variables $\overline{\Pi}_t$ follow the same law of motion as the state variables $\Pi_t$, but adjusted proportionally to $(\eta_t^k)_{k \leqslant t}$. In particular, Equations \hyperref[eqpi bar]{$(\Delta \overline{\Pi}_t)$} are forward in the sense that $\overline{\Pi}_{t+1}$ is a function of $\overline{\Pi}_t$ and other variables at time $t$. The initial condition for this system is $\overline{\Pi}_1 = 0$, which should be contrasted with the state equations where we have $\Pi_1 \neq 0$, i.e., $S_1^1 = 1-e_0>0$ and $I_1^1 = e_0>0$.

The variables $\overline{\Pi}_t$ have a natural interpretation. These variables are shadow prices that convert changes in the agent's continuation payoff to changes in the government's continuation payoff. For example, consider the first period. Recall that the agent's social distancing decision at this date depends on two static costs as well as the dynamic component, which is proportional to the expected change in the agent's value due to an infection, i.e., $\Delta_1^k = (1-\bt_1\gamma)\mathbb{S}_1^k+\bt_1\gamma \s_1^{2} - (1-\tau_1)\i_1^k-\bt_1\i\t_1$. Each of these adjoint variables can be thought as a sensitivity of the agent's first period continuation payoff with respect to a certain state at date $t=2$. For instance, $\s_1^1$ is the sensitivity related to $S_2^1$, and if the agent's payoff becomes marginally more sensitive to $S_2^1$, then the agent will social distance more. As a result, the planner's payoff will change in a proportion to $\eta_1^1$, that is the multiplier on the agent's first-order condition with respect to $\alpha_1^1$. In fact, the total adjustment of the government's profit due to the marginal change in $\s_1^1$ is exactly $\gas_1^1 = -\eta_1^1 (1-\bt_1\gamma)\beta_s S_1^1\widehat I_1$.

We now look at the problem of choosing the optimal lockdown. It turns out that the optimal choice of $\bl_t$ depends on the state variables and all Lagrange multipliers through the following:\\ \vspace{-9mm} 

\begin{small}
\begin{align*}
\label{lambda_foc}
\tag{OPT-$\bl_t$}
\lambda_t 
&= \arg\max\limits_{\bl \in [\overline{\bl},1]} \bigg\{\bigg((1-\dg)\dg^{t-1}Y_tq_t+(1-\da)\da^{t-1} \overline{Y}_tq_t\bigg)\bl - \\
&- \bigg(\beta_w\sum_{k \leqslant t} \bigg(\overline{\Delta}_t^k S_t^k I_t+\Delta_t^k\gas_t^k I_t\bigg)+\beta_w\chi_t\bt_t(1-\gamma)S_tI_t+\beta_w(1-\da)\da^{t-1}\bigg(\phi^+\gas_tI_t+\phi^-S_t\gai_t\bigg)q_t
\bigg)\bl^2\bigg\}.\nonumber
\end{align*}
\end{small}
\vspace{-5mm}

To understand Equation (\ref{lambda_foc}), suppose for a moment that the agents are myopic, that is $\da = 0$. Then, the agent's adjoint variables $\Upsilon_t$, and the dual variables $\overline{\Pi}_t$ attached to them, are identically equal to zero. Equation (\ref{lambda_foc}) reduces to the following:\\ \vspace{-9mm} 

\begin{small}
\begin{align*}
\bl_t &= \arg\max\limits_{\bl \in [\overline{\bl},1]} \bigg\{\underbrace{(1-\dg)\dg^{t-1}Y_tq_t}_{\color{blue}\text{direct output}}\bl - \bigg(\underbrace{\beta_w \sum_{k \leqslant t} \overline{\Delta}_t^k  S_t^k I_t}_{\color{blue}\substack{\text{direct dynamic cost}\\ \text{of an infection}}}+\underbrace{\beta_w \chi_t\bt_t(1-\gamma) S_tI_tq_t}_{\color{blue}\substack{\text{direct dynamic cost}\\ \text{propagated through testing}}}+\underbrace{\beta_w (1-\da)\bigg(\phi^+\gas_tI_t+\phi^- S_t\gai_t\bigg)}_{\color{blue}\substack{\text{indirect static lump-sum cost}}}\bigg)\bl^2\bigg\}.
\end{align*}
\end{small}\vspace{-5mm}

The first term is the discounted marginal product at date $t$ multiplied by the probability of having the pandemic going at this date. Consider the second term and recall that $\overline{\Delta}_t^k$ is the net expected change of the planner's continuation payoff due to an additional infection at state $k \leqslant t$. This is then multiplied by the likelihoods of transitions from the susceptible to infected, i.e., $\beta_w S_t^k I_t$, and aggregated across all testing states. As for the third term, note that an increase in economic activities creates $\beta_w S_tI_t$ new infections in total, and these then reduce the testing probability and decrease the planner's continuation payoff in a proportion to $\chi_t$. Finally, consider the last term and recall that the agent suffers disutility $\phi^+$ from getting infected and disutility $\phi^-$ from infecting others. These disutilities are weighted by the probabilities of respective events, pre-multiplied by their shadow prices $\overline{\Pi}_t$ and aggregated over the time of the last time. The first three terms in the above equation can be thought as direct costs, because they are unrelated to the agent's best reply.

In the general case with $\da > 0$, there are two additional terms:\\ \vspace{-9mm} 

\begin{small}
\begin{align*}
\bl_t &= \arg\max\limits_{\bl \in [\overline{\bl},1]} \quad \underbrace{\ldots\ldots}_{\text{\color{blue} static, $\delta_A = 0$}} + \underbrace{(1-\da)\da^{t-1}\overline{Y}_tq_t}_{\color{blue}\text{indirect output}}\bl - \underbrace{\beta_w \sum_{k \leqslant t} {\Delta}_t^k \gas_t^k I_t}_{\color{blue}\substack{\text{indirect dynamic cost}\\ \text{of an infection}}}\bl^2.
\end{align*}
\end{small}\vspace{-5mm}

\noindent We note that the rate of economic activities at time $t$ affects the agent's adjoint process $\Upsilon_t$ in two ways. First, it changes each of $\s_{t-1}^k$, $\i_{t-1}^k$, $\r_{t-1}^k$, $\r\t_{t-1}$ symmetrically and linearly, because the agent's next period output is linear in $\bl_t$. This gives the first new term above. Second, we have the quadratic term that captures certain indirect dynamic cost. Note that $\Delta_t^k \beta_w I_t$ is the expected change in the agent's payoff multiplied by the probability of an infection at $k \leqslant t$. These values are pre-multiplied by $\gas_t^k$ in order to convert them to the planner's payoff, and then they are aggregated across testing states. 

The government's first order condition with respect to $(\alpha_t^k)_{k\leqslant t}$ and $\bt_t$ can be found in the appendix (see Equations (\ref{govt_foc}), (\ref{tau_foc})). In short, the first set of conditions allows us to precisely pin down the set of Lagrange variables $(\eta_{t}^k)_{k \leqslant t}$ by solving a certain linear system, whereas the first-order condition with respect to $\bt_t$ determines the multiplier on the resource constraint.\footnote{A comment on optimization over testing is in order here. Although, the objective is linear in $\bt_t$, we must have $\bt_t \in (0,1)$ at all dates, because the process $\overline{X}_t$, which we infer from the actual testing data, is strictly positive (and not too large). To guarantee that $\bt_t \in (0,1)$ is indeed optimal we set the term in the objective in front of it to zero.}

Finally, we have the system that describes the dynamics of government adjoint variables $\overline{\Upsilon}_t$. We relegate this system to the appendix, because it is rather complicated (see Equations \text{\hyperref[equps A]{$(\Delta \overline{\Upsilon}_t)$}}). We note that this system, as one for $\Upsilon_t$, is backward with the following ``initial" condition: \vspace{-5mm}

\begin{small}
\begin{align*}
\lim_{t \to \infty} \frac{\gs^k_t}{p_t}-\dg^{t} = \lim_{t \to \infty} \frac{\gi^k_t}{p_t}-\dg^{t} = \lim_{t \to \infty} \frac{\gr^k_t}{p_t}-\dg^{t} = \lim_{t \to \infty}
\frac{\git_t}{p_t}-\dg^{t} = \lim_{t \to \infty} \frac{\grt_t}{p_t}-\dg^{t} = \lim_{t \to \infty} \frac{\gh_t}{p_t}-\dg^{t} = \xi \; \forall k.
\end{align*}
\end{small}
\vspace{-5mm}

\subsection{Numerical algorithm} \label{algorithm}

We now briefly outline the algorithm which we use to solve for the optimal policies. First of all, we truncate the problem by fixing a certain large terminal time $\overline{t}$ such that $p_{\overline t}$, the probability that the vaccine has not arrived by time $\overline{t}$, is sufficiently small. The solution to the truncated problem is characterized by exactly the same equations as above, but the terminal conditions for two backward systems have to hold exactly at $t = \overline{t}$.

Our numerical approach is a variation of the so-called {\it forward-backward sweep algorithm}. The forward-backward sweep is a standard iterative method to solve an optimal control problem with control dynamics in the form of partial difference/differential equations. The idea is to start with a policy, use it to simulate state variables forward and then solve for adjoint variables backward. These two sets of variables jointly produce a new policy. If it sufficiently close to the original one, then the algorithm stops. Otherwise, it updates the initial policy as a convex combination of the old and new ones with the weights of  $1-\varepsilon>0$ and $\varepsilon>0$, respectively.

We adapt the forward-backward sweep to two nested control problems: one for the agent and the other is for the government. Thus, we will have two loops. In the inner loop we solve the agent's problem with a fixed lockdown policy, whereas in the outer loop we update the lockdown policy and Lagrange multipliers on the resource constraint and agent's first-order condition with respect to the rate of social distancing. 

Let $\lambda_t$ for $t=1,2,...,\overline{t}$ be given. Then, the inner loop is as follows:
\begin{enumerate}
\item[Step 0] Select the rate of social distancing $(\alpha^k_t)_{k \leqslant t}$.
\item[Step 1] Solve for the state variables $\Pi_t$ and rate of testing $\tau_t$ using Equations \hyperref[equps A]{$(\Delta \Pi_t)$} and (\ref{eq testing capacity}).
\item[Step 2] Solve for the agent's adjoint variables $\Upsilon_t$ using Equations \text{\hyperref[equps A]{$(\Delta \Upsilon_t)$}}.
\item[Step 3] Find the rate of social distancing $(\widetilde\alpha^k_t)_{k \leqslant t}$ using Equation (\ref{agent_foc}).
\item[Step 4] If the distance between $(\alpha^k_t)_{k \leqslant t}$ and $(\widetilde\alpha^k_t)_{k \leqslant t}$ is small enough, then stop the inner loop. Otherwise, update $(\alpha_k^t)_{k \leqslant t}$ to  $(1-\varepsilon) \cdot (\alpha^k_t)_{k \leqslant t}+\varepsilon \cdot (\widetilde\alpha^k_t)_{k \leqslant t}$ and go to Step 1.
\end{enumerate}

The outer loop is as follows: 
\begin{enumerate}
\item[Step 0] Select the rate of lockdown $\lambda_t$, the Lagrange multipliers on the agent's first-order condition with respect to the rate of social distancing $(\eta_t^k)_{k \leqslant t}$ and the Lagrange multipliers on the resource constraint $\chi_t$.
\item[Step 1] Use the inner loop to obtain the rate of social distancing $(\alpha^k_t)_{k \leqslant t}$.
\item[Step 2] Solve for the state variables $\Pi_t$, shadow cost variables $\overline{\Pi}_t$ and rate of testing $\tau_t$ using Equations \text{\hyperref[eqpi]{$(\Delta \Pi_t)$}}, \text{\hyperref[eqpi bar]{$(\Delta \overline{\Pi}_t)$}} and (\ref{eq testing capacity}).
\item[Step 3] Solve for both sets of adjoint variables $\Upsilon_t$ and $\overline{\Upsilon}_t$ using Equations \text{\hyperref[equps A]{$(\Delta \Upsilon_t)$}} and \text{\hyperref[equps G]{$(\Delta \overline{\Upsilon_t})$}}.
\item[Step 4] Find the rate of lockdown $\widetilde\lambda_t$ using Equation (\ref{lambda_foc}), compute the Lagrange multipliers $(\widetilde\eta_t^k)_{k \leqslant t}$ and $\widetilde\chi_t$ using Equations (\ref{govt_foc}) and (\ref{tau_foc}), respectively.
\item[Step 5] If the distance between $\big(\lambda_t,\{\eta_t^k\}_{k \leqslant t},\tau_t\big)$ and $\big(\widetilde\lambda_t,\{\widetilde\eta_t^k\}_{k \leqslant t},\widetilde\tau_t\big)$ is small enough, then stop the inner loop. Otherwise, update $\big(\lambda_t,(\eta_t^k)_{k \leqslant t},\tau_t\big)$ to  $(1-\varepsilon) \cdot \big(\lambda_t,(\eta_t^k)_{k \leqslant t},\tau_t\big)+\varepsilon \cdot \big(\widetilde\lambda_t,(\widetilde\eta_t^k)_{k \leqslant t},\widetilde\tau_t\big)$ and go to Step 1.
\end{enumerate}

The inner loop takes a few seconds to terminate, whereas the outer loop needs around ten minutes to produce the solution. Several comments are in order. First, it is well-known that the forward-backward sweep algorithm might run into cycles, especially, when $\varepsilon$ is large. We control for this by fine-tuning this parameter, in fact, in our simulations with $\varepsilon=0.01$ the algorithm converges monotonically. Second, we validate global optimality by multi-starting the algorithm with different initial values chosen at random. We are not aware of alternative procedures which can be used to solve for the global optima in the case of two nested control problems.

\section{Calibration}
\label{section calibration}
Predicting disease dynamics, calculating optimal policy and evaluating policy experiments all require reliable parametrization of the model. In principle we could let the entire set of parameters be identified using time-series of data on state variables. This unfortunately runs into problems for even an ideal dataset will leave any standard SIR model unidentified (see \citet{sir_unident} and  \citet{jesus_jones}). 

We take the following pragmatic approach. We fix government policies, medical and other miscellaneous parameters using the aggregated wisdom of research from medicine and social science, and calibrate behavioral parameters to match the time series of data available on the state variables. The underlying conceptual thought here is to borrow liberally from other scientific studies, but try our best to tease out the (i) {\it initial prevalence}---the building block of SIR framework, (ii) {\it social distancing costs}, and (iii) {\it efficacy of tracing-testing} by fitting the data on the progress of the pandemic thus far to the predictions of our model. 

Let the vector of parameters that will be estimated be given by $\theta := (\beta_w,\beta_s,c,\phi_+,\phi_-,\gamma)$. In what follows we briefly describe our dataset, the set of fixed parameters and the estimation process to pin down $\theta$. All additional details can be found in the appendix.

\subsection{Method}
\label{section: estimation method}

We combine data from two sources for our estimation exercise. Time series data for daily deaths and positive cases are obtained from "\href{https://github.com/nytimes/covid-19-data}{The New York Times Repository}", while time series data for daily hospitalizations and daily tests are sourced from "\href{https://covidtracking.com/}{COVID Tracking Project}". Our analysis covers data between February 29, 2020 $(t=1)$, when the first death was reported in the US, till February 14, 2021 $(t=352)$. \footnote{The collected data suffers from the ``weekend'' effects: new observations are added in chunks with lags of several days contributing to the high volatility. To mitigate this issue, we smoothen the data by constructing moving averages in the window of seven days for each variable. This smoothening of Covid-19 data is standard and has also been employed by other studies, see for example, \citet{jesus_jones}.}

\begin{figure}[h]
\centering
\begin{subfigure}[b]{0.48\textwidth}    
\includegraphics[scale=0.28]{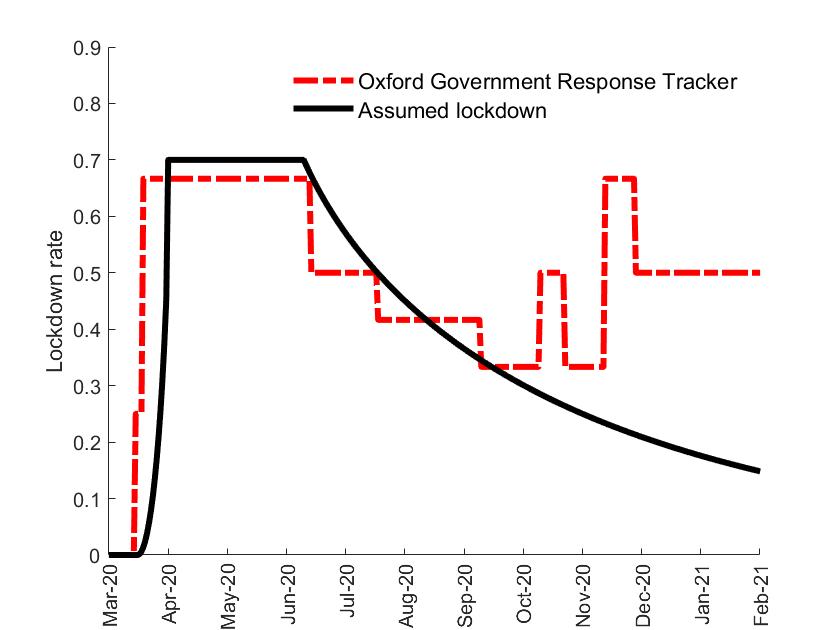}
\caption{Lockdown vs Oxford GRT index}
\label{fig: lock}
\end{subfigure}
\begin{subfigure}[b]{0.48\textwidth}    
\includegraphics[scale=0.28]{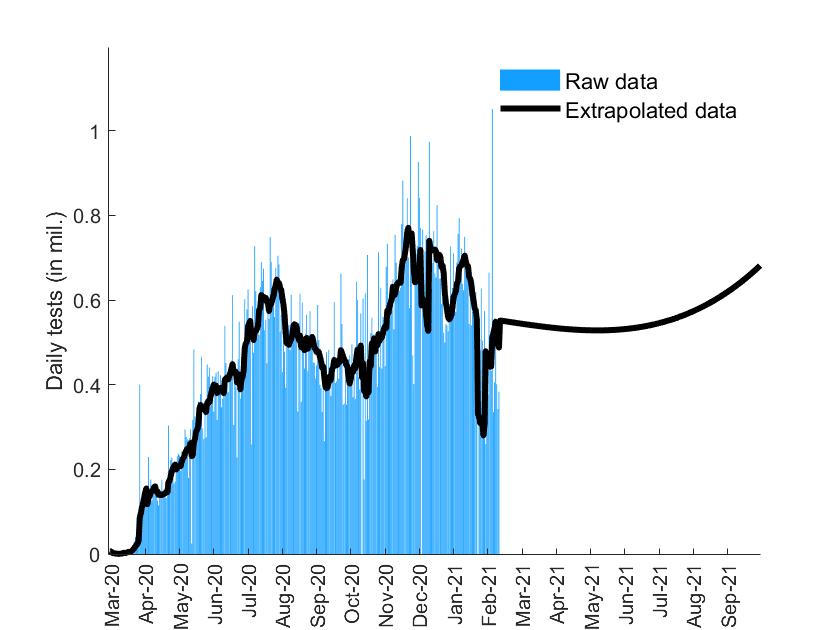}
\caption{Test count}
\label{fig: tests fit}
\end{subfigure}
\caption{Assumed government policies} 
\label{fig: govt pol}
\end{figure}

The primary objective of the estimation exercise is to parametrize our model so that it rationalizes the US Covid-19 data obtained from the above mentioned sources. This requires us to solve the agent's optimization problem given the \textit{actual} lockdown and testing policies followed by the government. Since there is no reliable source to estimate the actual lockdown function implemented by the US government, we use an approximation which is roughly consistent with the government response index published by \citet{hale2021global} (see \href{https://covidtracker.bsg.ox.ac.uk/}{Oxford COVID-19 Government Response Tracker} for the data). Figure \ref{fig: lock} compares our approximation of lockdown function and the Oxford GRT index.\footnote{The Oxford GRT reports a variety of indices to capture the policy responses undertaken by different countries. These include policies for economic reform, stringency measures to reduce transmission, healthcare reform and so on. To construct the index illustrated in Figure \ref{fig: lock}, we use two stringency measures used by the US government to curb economic activities, namely workplace closure and stay-at-home requirements.} 
 
As for the testing capacity, we compute the test count from the state level data as opposed to simply using the publicly available aggregate data, for example, see "\href{https://covidtracking.com/}{COVID Tracking Project}". The reason is that data is collected at the state level in the US, and the states have practiced distinct ways of reporting negative tests. Some states report people tested at least once (unique people), others report either testing encounters or samples collected. To the best of our knowledge, neither "\href{https://covidtracking.com/}{COVID Tracking Project}" nor other data providers, i.e.,  "\href{https://github.com/nytimes/covid-19-data}{The New York Times Repository}", explicitly have corrected for these differences in reporting. 

Ideally, we would like to use testing encounters for our calibration exercise but since only a few states (mainly less populated) reported this metric imputing it might result in imprecise estimates. Another reason for confusion is that more than one test is often conducted on a patient in a short span, especially if she/he tests positive. So, we construct $\overline{X}_t$ by imputing and aggregating the state level testing data on the number of unique people who got tested, which can be thought as a lower bound on the total US testing capacity, in-sample till February 14, 2021. Then, we extrapolate the count of available tests out-of-sample by fitting a 2nd degree polynomial function of time. Figure \ref{fig: tests fit} plots the actual number of conducted tests in US and its out-of-sample extrapolation.

Having set the lockdown and testing policies, we now present the set of medical and other auxiliary parameters that are taken to be fixed to calibrate the other parameters of interest.  Table \ref{table parameters} summarizes these fixed parameters. In short, we rely here on the accumulated knowledge thus far, and we discuss each parameter in details in the appendix.
\begin{table}[h]
\begin{small}
\begin{center}
\begin{tabular}{||c|c|c|c||}
\hline
{\bf Parameter} & {\bf Value} & {\bf Definition} & {\bf Source/Target}\\\hline
$t_i$ & 10 & Average infectious period (incl. latency)  & \\
$t_h$ & 7  & Average hospitalization period & \href{https://www.cdc.gov/coronavirus/2019-ncov/hcp/planning-scenarios.html}{CDC Planning Scenario}\\    \hline
$m_i$ & 0.0176 & Hospitalized as \% of infected & $m_i = IFR/m_h$, where $IFR = 0.3\%$ (\citet{ioannidis2021infection})\\
\hline
$m_h$ & 0.1705 & Daily deaths as \% of hospitalized & Estimated by OLS\\ \hline
$\mathbb{E}T$ & 540 & Expected day the pandemic ends & \\  
$\mathbb{V}T$ & 180 & Variance in the day pandemic ends & \citet{vaccine_nyt}\\
\hline
$\overline{\lambda}$ & 0.3 &  Share of essential services &\\
$\delta_G$ & 0.9999 &  Government's discount factor &  Used by multiple studies incl. \citet{lockdown_SIR}, \\
$\delta_{A}$ & 0.9999  &  Agent's discount factor & \citet{dream_team_SIRmodel} and \citet{farboodietal_covid}\\
\hline 
$e_{0} $ & $0.007\%$ & Share of infected at the start& 23,000 infected agents given the population of 328.2 mil. \\\hline
\end{tabular}
\end{center}
\caption{Fixed parameters}
\label{table parameters}
\end{small}
\end{table}

Given the paucity and limited reliability of the Covid-19 data mentioned above, only the flow data on deaths and positive cases are used to match the model parameters. We employ the following loss function for this purpose:
\[
\omega \cdot \sum_{t=1}^{352}\Big(\Delta D_t(\theta) - \Delta \widetilde D_t\Big)^2 + (1-\omega) \cdot \sum_{t=1}^{352}\Big(X_t^+(\theta) - \widetilde X_t^+\Big)^2  
\]
where $\theta = (\beta_w,\beta_s,c,\phi_+,\phi_-,\gamma)$ is the vector of parameters being estimated, $\Delta D_t(\theta)$ is the predicted flow of deaths at time $t$, $X_t^+(\theta)$ is the predicted flow of positive tests, i.e., the sum of newly hospitalized and known infected, at date $t$, $\Delta \widetilde D_t$ and $\widetilde X_t^+$ are their empirical counterparts. The weights $\omega$ and $1-\omega$ ensure that both variables get the same importance in the computation of the loss function.\footnote{This is required since the daily number of positive cases vastly outnumbers the number of daily deaths in the US. For instance, the mean  daily deaths in the US was 1284 compared to the mean daily positive cases of 70982.}$^{,}$\footnote{It is worthwhile to note that we fix the test capacity $\overline {X}_t$ at the actual empirical level. Hence fitting $X^{+}_t(\theta)$ to the data ensures that the predicted flow negative cases fit the data well too.} 

To avoid over-fitting the model we impose two additional restrictions. First, we require $\beta_w = \beta_s$, which is broadly in agreement with the American Time Use Survey that employed Americans (on average) spend about half time at work and half time in social activities (see \citet{ert_covid}). Second, we fix $\phi^{-}= \frac{\phi^+}{10}$, i.e., we are assuming the one-time cost of infecting another person outside of immediately friends and family to be 10 percent of the cost that we would suffer from getting infected ourself and potentially risking infection for our loved ones as well. This is broadly subjective and can be adjusted in light of more evidence.

\subsection{Estimation} 
\label{section: estimation results}

\begin{figure*}[h]
\centering
\begin{subfigure}[b]{0.48\textwidth}    
\includegraphics[scale=0.28]{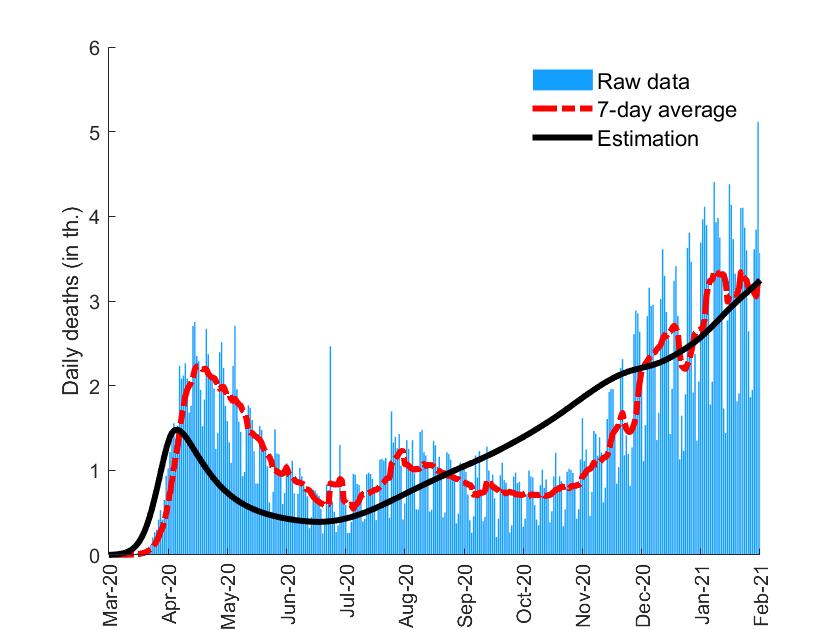}
\caption{Deaths}
\label{fig: deaths fir}
\end{subfigure}
\begin{subfigure}[b]{0.48\textwidth}    
\includegraphics[scale=0.28]{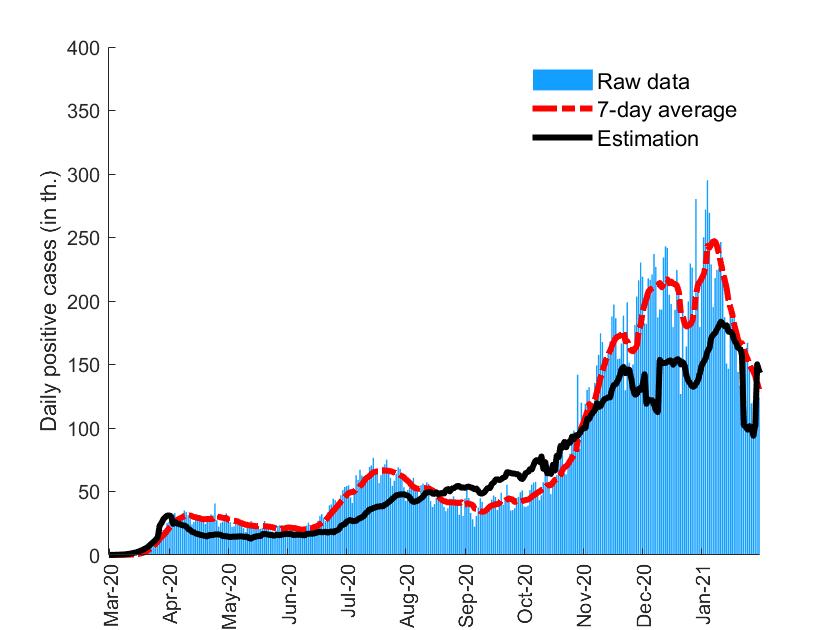}
\caption{Positive confirmed}
\label{fig: positive fit}
\end{subfigure}
\caption{Model fit vs data} 
\label{fig: estimation}
\end{figure*}

Recollect that we are trying to fit the following two time series: the number of daily deaths and the number of daily positive cases. The inner loop of the numerical algorithm described in Section \ref{algorithm} is able to generate the fixed point. The robustness of the fixed point is verified by multi-starting the algorithm. Figures \ref{fig: deaths fir} and \ref{fig: positive fit} compares the model fit and data for daily deaths and daily positive cases, and Table \ref{table parameters 2} reports the calibrated parameters.

\begin{table}[h]
\begin{small}
\begin{center}
\begin{tabular}{||c|c|c||}
\hline
{\bf Parameter} & {\bf Value} & {\bf Definition} \\ 
\hline\hline
$\beta_w$ & $0.2046$ & Prevalence in economic activities\\  \hline
$\beta_s$ & $0.2046$ & Prevalence in social activities\\  \hline  
$\gamma$ & $0.0976$ & Tracing-testing efficacy \\  
\hline
$c$ & $0.4403$ & Daily opportunity cost of infection  \\   \hline
$\phi_{+}$ & $523.932$ & Disutility of getting infected \\ \hline
$\phi_{-}$ & $52.393$ & Disutility of infected someone else \\  
\hline\hline
\end{tabular}
\end{center}
\caption{Calibrated parameters.}
\label{table parameters 2}
\end{small}
\end{table}
Overall, the model with just 4 free parameters fits the data reasonably well. It turns out that our estimates match almost perfectly the cumulative numbers of deaths and positive cases by February, 14. However, they do not pick up magnitudes of some waves smoothing them over time. This can be partially attributed to factors not present in the model such as overload of the healthcare system, information percolation about the pandemic, seasonality in contact rates, i.e., academic year and holidays, etc.\footnote{Note that we could substantially ``improve the fit" visibly by freeing up some more parameters from Table \ref{table parameters}. However, this would surely result in an overfitting of the model.}

Our estimate of the total prevalence $\beta_w+\beta_s$ turns out to be $0.4092$ that implies that the initial rate of growth of infection is slightly more than $0.3$. In other words, the number of infected agents doubles in $3$ days at the beginning of our analysis. This is in line with previous studies. For example,  \citet{farboodietal_covid} calibrated the total prevalence parameter to $0.43$, and \citet{ert_covid} estimated the transmission rate in social activities to be around $0.16$, which is close to our estimate of $0.5 \times 0.4092 = 0.2046$.

The efficacy of trace-testing $\gamma$ is estimated to be $0.0976$. Recall that a fraction $1-\gamma$ of the susceptible ($\mathbf{S}$) and unknown recovered ($\mathbf{R}$) is excluded from testing based on tracing, so here $\gamma=0$ is perfect tracing technology and $\gamma=1$ denotes completely random testing. The estimate of efficacy of trace-testing is novel to the literature, and it essentially tells us that conditional on the number of tests, the tracing and testing infrastructure in the United States has been reasonably effective in targeting the infected population. However, as we will argue later, the delay in scaling up testing had significant consequences. 

For the behavioral parameters, we estimate three types of costs which each agent incurs. The daily opportunity cost of missing work and social activities due to being quarantined, measured by $c$, is estimated to be $0.4403$. We find $\phi^+$ to be estimated to be around $523.9$.  This represents the psychological cost from knowing that one has been infected plus the cost of facing a potential death from the virus or risking the health of a close family member. Then,  $\phi^{-}$ is set to $53.4$, which is the (ex ante) cost of getting someone else infected and contributing indirectly to all their direct costs. As mentioned in the introduction, one way to interpret these numbers is in terms of daily wage as the unit of measurement. 

\section{The optimal policy}

\label{section optimal policy}

In this section we report the results from executing the numerical algorithm described in Section \ref{algorithm} to calculate the optimal lockdown and testing policies for parameters specified in the previous section and the value of life $\xi$ set to $20$. This specific value of $\xi$ is in line with \citet*{jones_pareto} and has been chosen in other papers that have followed. We use this number for the simulation exercise and reporting of some of the results, but then we also map out the Pareto frontier to exposit the full set of policy options available to the government.

\begin{figure*}[h]
\centering
\begin{subfigure}[b]{0.48\textwidth}    
\includegraphics[scale=0.28]{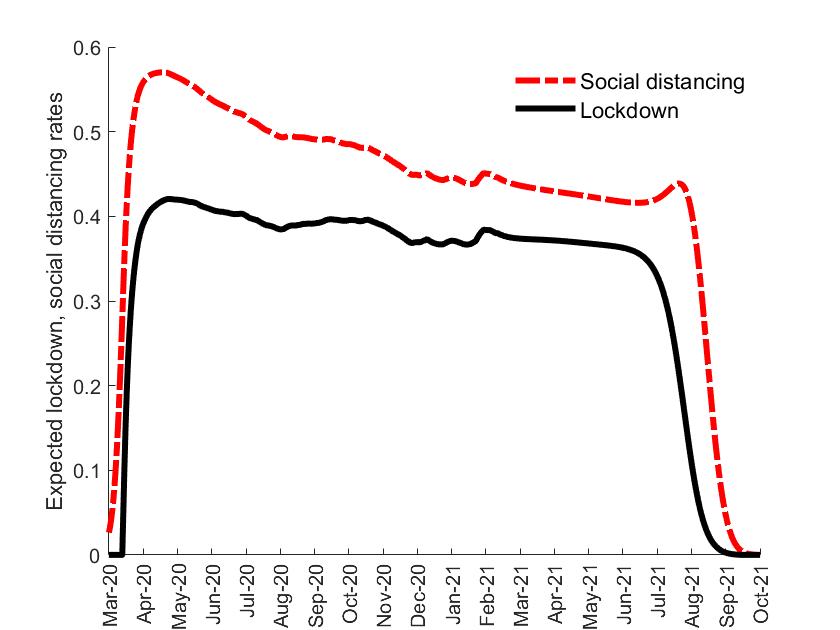}
\caption{Lockdown and social distancing}
\label{fig: lockdown and soc distancing}
\end{subfigure}
\begin{subfigure}[b]{0.48\textwidth}    
\includegraphics[scale=0.28]{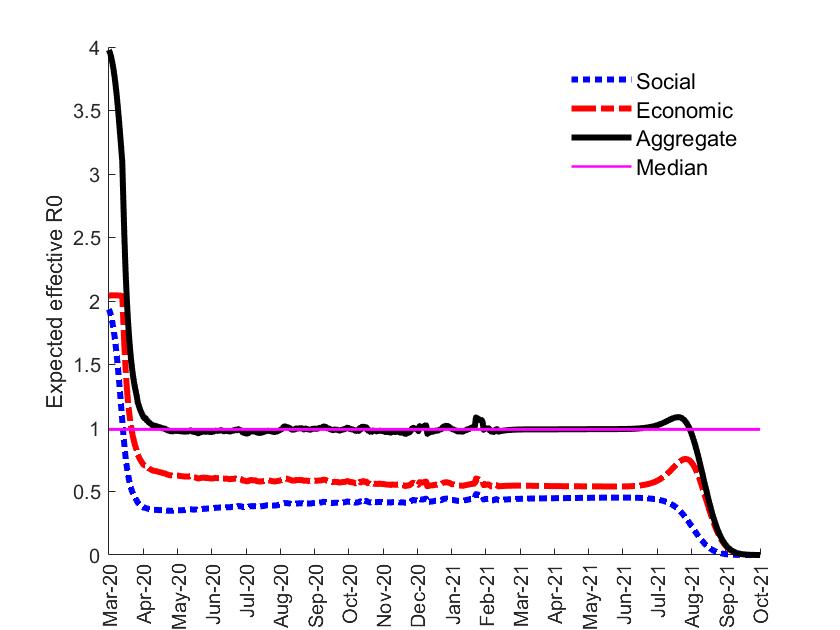}    
\caption{Effective reproductive number}
\label{fig: r0}
\end{subfigure}
\caption{Optimal lockdown and social distancing} 
\label{fig: optimal policy}
\end{figure*}

Figure \ref{fig: lockdown and soc distancing} plots the optimal lockdown and average fraction of social distancers at each date over the course of 20 months. The first thing to notice is that government internalizes that individuals will voluntarily social distance at moderate levels in order to avoid getting infected. Towards the beginning of the pandemic in March 2020, agents are willing to give up around 55\% of their social activities (dashed red line). Interestingly this is the highest level of social distancing that agents engage in throughout the pandemic. Over time the extent of social distancing declines very gradually till August 2021 to a 50\%. As the possibility of the pandemic ending increases, the optimal social distancing levels decline rather steeply to 0 between the period of August 2021 and October 2021.\footnote{The small spike in social distancing before the sharp decline is due to the uncertainty in vaccine arrival or modeling end of the pandemic. Since the government starts to ease the lockdown, if the vaccine is very likely to arrive but hasn't quite arrived yet, agents may social distance more to control the infection rate.}

Given this moderate but persistent level of social distancing, the government imposes an early lockdown of around 40\% which then remains relatively stable (solid black line). It rationally internalizes that the pandemic is going to spread, and under the expectation of vaccine arriving in Summer/Fall 2021 it chooses an intermediate but persistent lockdown to avoid a large outbreak of infections. The rate of lockdown declines gradually to around 35\% till July 2021, followed by a rapid decline as the pandemic is expected to end. However, the government never imposes anywhere close to the maximal lockdown levels in the economy throughout the period of the pandemic.
\begin{figure*}[h]
\centering
\begin{subfigure}[b]{0.48\textwidth}    
\includegraphics[scale=0.28]{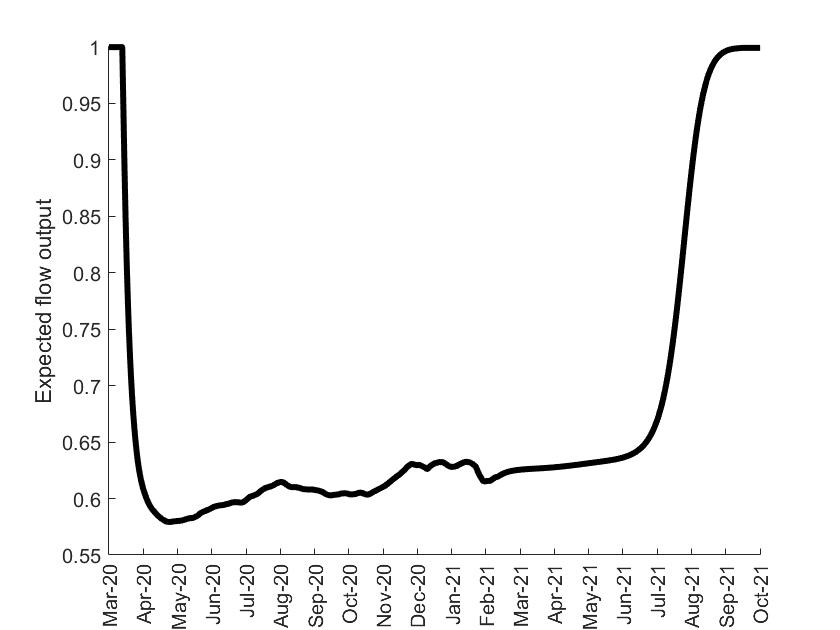}    
\caption{Expected flow output}
\label{fig: output opt}
\end{subfigure}
\begin{subfigure}[b]{0.48\textwidth}    
\includegraphics[scale=0.28]{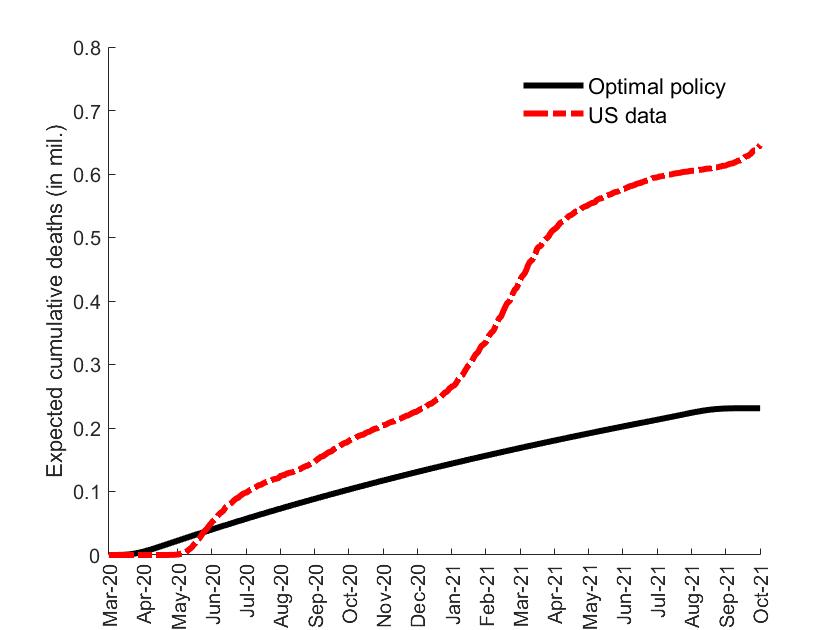}    
\caption{Expected cumulative deaths}
\label{fig: deaths opt}
\end{subfigure}
\caption{Flow output and cumulative deaths}
\label{fig: output and deaths}
\end{figure*}

The rationale behind the optimal lockdown can be seen by examining the famed reproductive number $R_t$. It is an approximation of the expected number of secondary cases produced by a single infected agent at a specific date $t$:\footnote{A variant of this expression without testing is reported in studies like \citet{farboodietal_covid} and \citet{jones_pareto}.} \\ \vspace{-7mm}

\begin{small}
\[
R_t:= \underbrace{\frac{1}{1-(1-\tau_t)(1-\sfrac{1}{t_i})}}_{\text{``avg. duration'' of state $\mathbf{I}$}} \cdot \underbrace{(1-\tau_t) \bigg(\beta_w\lambda^2_tS_t+\beta_s\widehat{S}_t\frac{\widehat{I}_t}{I_t}\bigg)}_{\text{\# of new infections}}. 
\]
\end{small} \\ \vspace{-9mm}

\noindent Recall that an infected agent can transmit the disease through two channels: economic activities, in which case she/he infects $\beta_w\lambda^2_tS_t$, and social activities, in which case she/he infects $\beta_s\widehat{S}_t\frac{\widehat{I}_t}{I_t}$ agents. Taking into an account that $\tau_t$ of these new secondary cases will be test positive immediately at the end of date $t$, the total number of secondary cases produced is given by $(1-\tau_t) \left(\beta_w\lambda^2_tS_t+\beta_s\widehat{S}_t\frac{\widehat{I}_t}{I_t}\right)$. Then, we note that this infected agent stays in $\mathbf{I}$ if and only if she/he has not moved to $\mathbf{IT}$ (tested positive), $\mathbf{R}$ (recovered) or $\mathbf{H}$ (hospitalized). Thus, if testing rate is close to be being a constant, she/he is expected to spend approximately $\frac{1}{1-(1-\tau_t)(1-\sfrac{1}{t_i})}$ days in the compartment of unknown infected agents. We call $R_{t}$ the {\it effective reproductive number}. 

Figure \ref{fig: r0} plots the effective reproductive number implied by the optimal policy and agents' social distancing decision. The government chooses the lockdown function internalizing the agent's best response in a way that $R_t$ is approximately $1$ for almost 17 months. This in turn means that modulo the variation in total available tests the number of active cases is targeted to be approximately constant over this time frame, which can be seen by examining Equation (\ref{eq I mech}). Importance of reducing the effective reproductive number to one has been extensively addressed in public discourse, and we believe that our results provide arguments in favor of keeping this as the target of policy. 

The two outcomes that define the objective function of the government are depicted in Figure \ref{fig: output and deaths}. Recollect that the total number of survivors is weighted by the $\xi=20$ for these calculations. Figure \ref{fig: output opt} shows the flow of output as the pandemic evolves. Total output falls almost by 45\% percent initially according to the optimal policy and then gradually claws its way back to its full potential as the government eases lockdown. This graph of course reflects the lockdown policy in Figure \ref{fig: lockdown and soc distancing} because total output in our model is simply equal to the labor supply.

Figure \ref{fig: deaths opt} plots the cumulative number of deaths predicted by the optimal policy (solid black line) and the actual realized number (dashed red line). The two series depart fairly quickly around the two-three month mark, and while the slope of the actual number gets a bump as winter 2020 approaches, the optimal policy ensures the slope of the number of deaths remains more or less constant rising up to about three hundred thousand deaths, less than half of the actual number of deaths seen in the data, which has already passed seven hundred and fifty thousand as of writing this draft of the paper. 
\begin{figure}
\centering 
\includegraphics[scale=0.35]{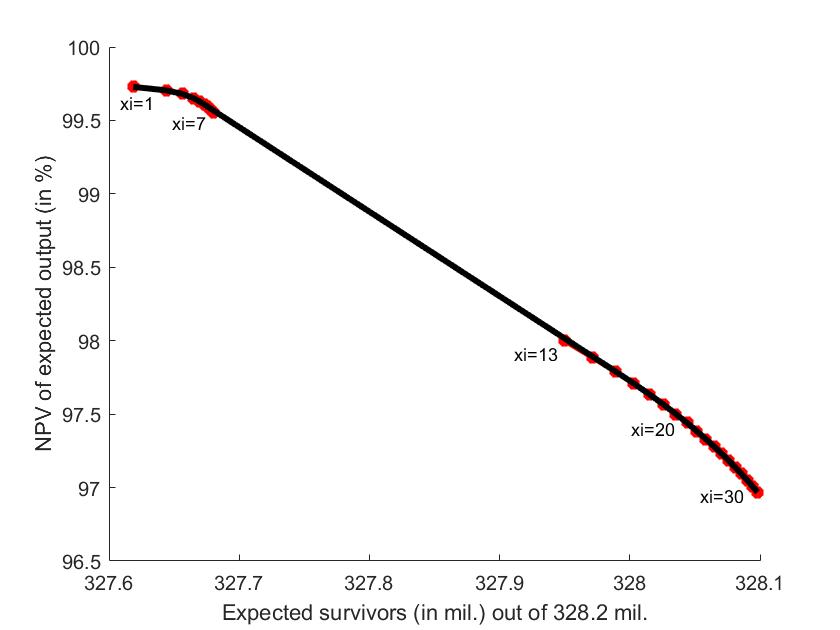}   
\caption{Mapping the Pareto frontier on output and lives saved}
\label{fig: PF basic}
\end{figure}

Finally, going beyond the specific value of $\xi$, in Figure \ref{fig: PF basic} the entire Pareto frontier is presented. This graph is produced by varying $\xi$ on a grid of thirty points ($\xi=0$ to 30), and then solving the government's problem for each possible value. Thus, the curve traces all efficient combinations of total output and mortality by changing the value of the Pareto weight $\xi$. That the curve is downward sloping is obvious. Interestingly, it first has an approximately constant slope followed by a concave transition. Mathematically speaking, the optimization problem here is mapping out the concave closure of the {\it actual} Pareto frontier. So, the initial straight part of the frontier is representative of a weakly convex frontier---hence increasing returns in terms of lives saved for every unit drop in output. Eventually the concave part diminishes these returns quite substantially.

For low values of $\xi$, especially $1\leq\xi\leq10$, the trade-off is stark: the first one percent permanent drop in output increases the expected number of survivors by 281 thousand. For larger values of $\xi$, for example $13\leq\xi\leq30$, a one percent drop in output increases the expected number of survivors by 158 thousand. Thus, even a moderate emphasis on mortality in the objective function produces significant gains in lives saved with a concomitant linear (or even lesser) decline in output. However, as the Pareto weight is increased the trade-off becomes less rewarding---the number of lives saved per unit drop in output decreases marked by the concave rate of transition.

\section{Policy experiments}

In this section we explore policy experiments by changing two parameters--- efficiency of tracing and testing and the fraction of prevalence the government can directly control---and also removing lockdown and social distancing piecemeal from the optimization problem. 

\subsection{Improving tracing and testing}

\label{section tt experiment}

There has been substantial variation in the way different countries have approached large scale testing to fight the pandemic. Some seem to have delayed in making testing widely available and others in making contact tracing particularly effective. Here we do two experiments on the baseline. 

\begin{figure*}[h]
\centering
\begin{subfigure}[b]{0.48\textwidth}    
\includegraphics[scale=0.3]{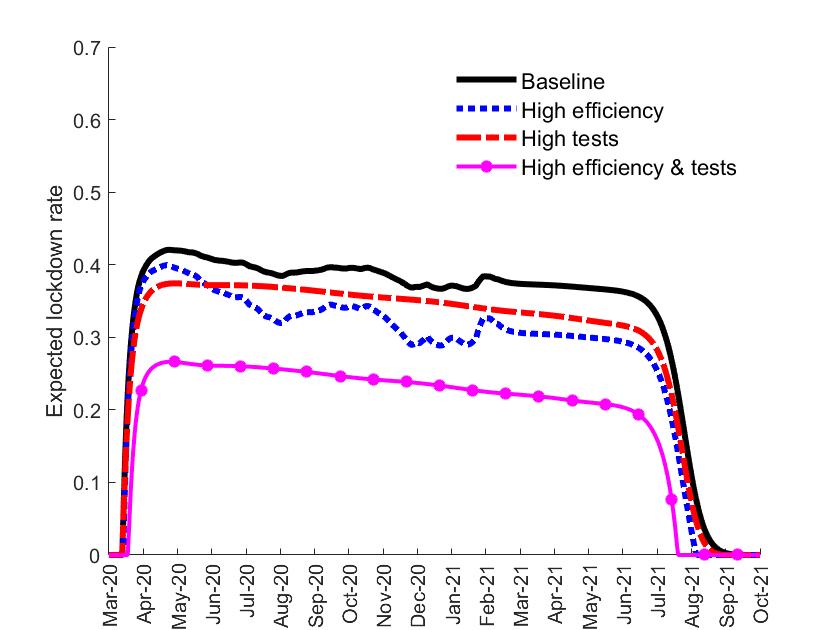}
\caption{Lockdown}
\label{fig: lockdown testing}
\end{subfigure}
\begin{subfigure}[b]{0.48\textwidth}    
\includegraphics[scale=0.3]{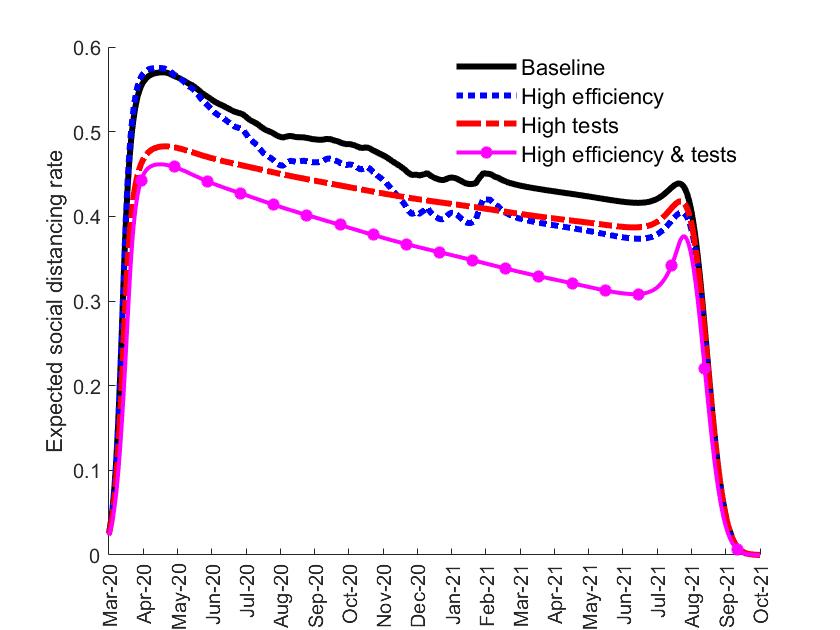}
\caption{Social Distancing}
\label{fig: sd testing}
\end{subfigure}\\
  \begin{subfigure}[b]{0.48\textwidth}    
\includegraphics[scale=0.3]{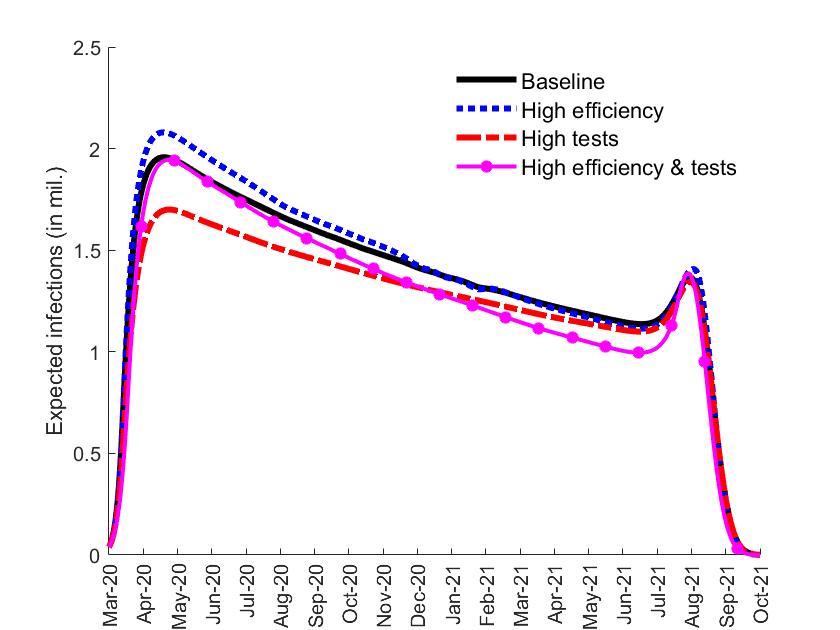}
\caption{Flow of infections}
\label{fig: infections testing}
\end{subfigure}
\begin{subfigure}[b]{0.48\textwidth}    
\includegraphics[scale=0.3]{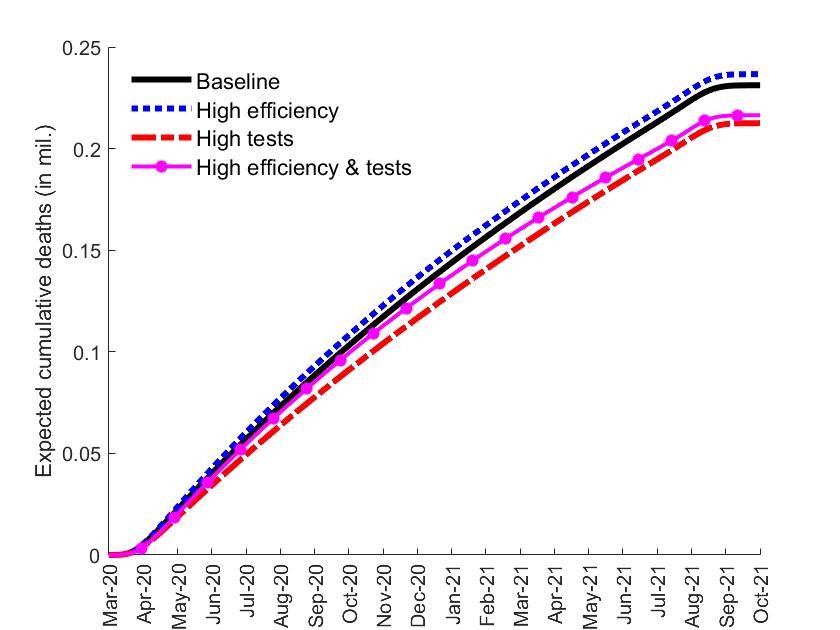}
\caption{Cumulative deaths}
\label{fig: deaths testing}
\end{subfigure}
\caption{Counterfactuals for early testing, and effective tracing and testing} 
\label{fig: counterfactuals testing}
\end{figure*}

First, we allow for early testing. As can be seen in Figure \ref{fig: tests fit}, testing starts late in the pandemic, almost at the two and half month mark. We expand the total set of test capacity at each date to be its highest attained value in the data, which is one million tests a day. We call this the high testing counterfactual. Second, we allow the tracing and testing to be much more efficient. We do that by reducing $\gamma$ by half---from 0.09 to 0.045. This ensures a larger proportion of infected are now being tested and quarantined. Figure \ref{fig: counterfactuals testing} plots the key variables for both counterfactuals and also for the combined scenario of high and efficient testing. 

The pairwise substitutability of better testing with lockdown and social distancing is immediately clear from Figures \ref{fig: lockdown testing} and \ref{fig: sd testing} respectively. When testing is scaled earlier and is more efficient the optimal extent of lockdown and social distancing diminish significantly. On average, lockdown drops by almost 40 percent, and social distancing by about 20 percent. On the social distancing dimension, if an agent is not selected for testing, she/he is relatively more confident of not being infected and thus is willing to engage more in social activities. For lockdown too, since the government is more certain that the pool of agents who are not quarantined or recovered are not infected, it needs to lockdown less. 
\begin{figure}
\centering 
\includegraphics[scale=0.35]{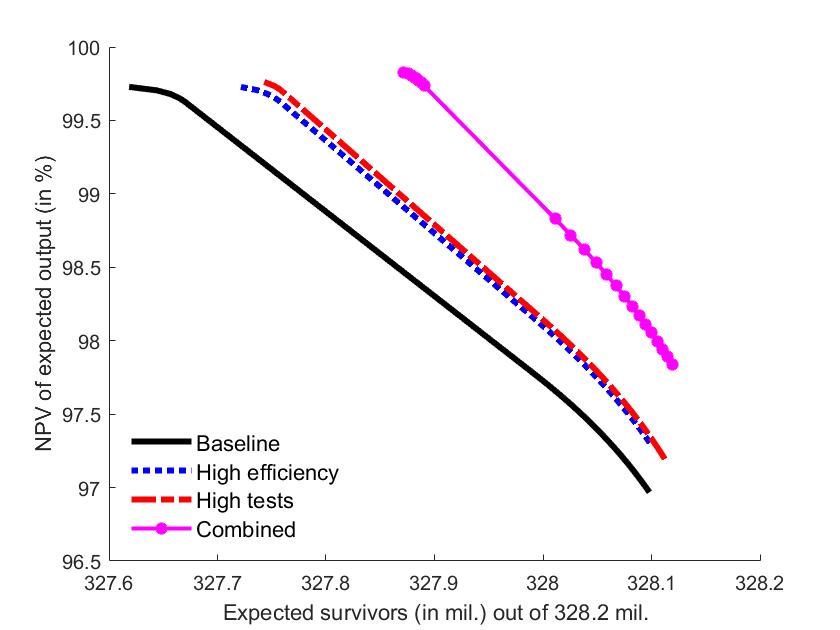}
\caption{The Pareto frontier for high and improved testing}
\label{fig: PF testing}
\end{figure}

Improved tracing-testing leads to less total infections; however, lower lockdown and social distancing lead to more infections. So the aggregate effect of improved tracing-testing on mortality is ambiguous. In fact in Figures \ref{fig: infections testing} and \ref{fig: deaths testing}, we can see that for fixed level of testing, introducing greater efficiency keeps the number of deaths to almost the same level, in fact slightly increases it. Recall these numbers are being determined endogenously through the solutions to the government and the agents' problems for a fixed Pareto weight. It turns out that the government is trading off some mortality for a much larger increase in output. 

Figure \ref{fig: PF testing} plots the Pareto frontier for four different specifications of tracing-testing. Introducing either high number of tests or higher efficiency of testing pushes out the Pareto frontier. And, introducing both considerably expands it. A simple way to think about this expansion is that for a fixed level of mortality, the government now achieves a much higher output and similarly for a fixed level of output it can now achieve considerably more survivors (or less deaths). These findings are analogous to \citet{ dream_team_SIRmodel}---they feature testing levels tagged to age, we have a uniform testing policy; however, agents in our model choose their level of social interactions, in theirs they are mechanical.

\newpage
\subsection{Greater control of the government}

\label{section control}

There has been intense discussion through the pandemic that autocracies have been able to control the diseases better than democracies (see \citet{narita_democracy}). While the moral and institutional intricacies of that questions are outside the scope of this paper, we can shed some light on it through a simple policy experiment. How do the basic results change by increasing the control of the government over social interactions? 
\begin{figure*}[h]
\centering
\begin{subfigure}[b]{0.48\textwidth}    
\includegraphics[scale=0.28]{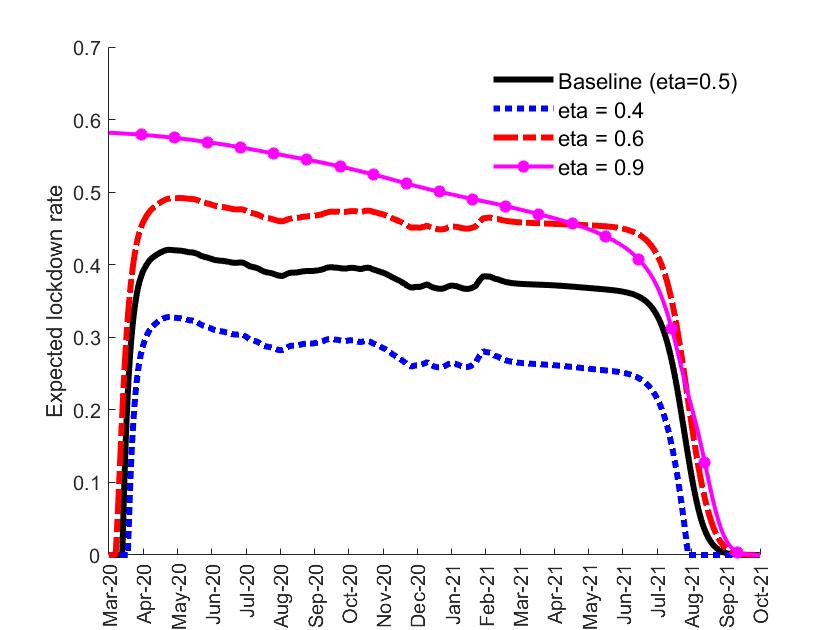}
\caption{Lockdown}
\label{fig: lockdown eta}
\end{subfigure}
\begin{subfigure}[b]{0.48\textwidth}    
\includegraphics[scale=0.28]{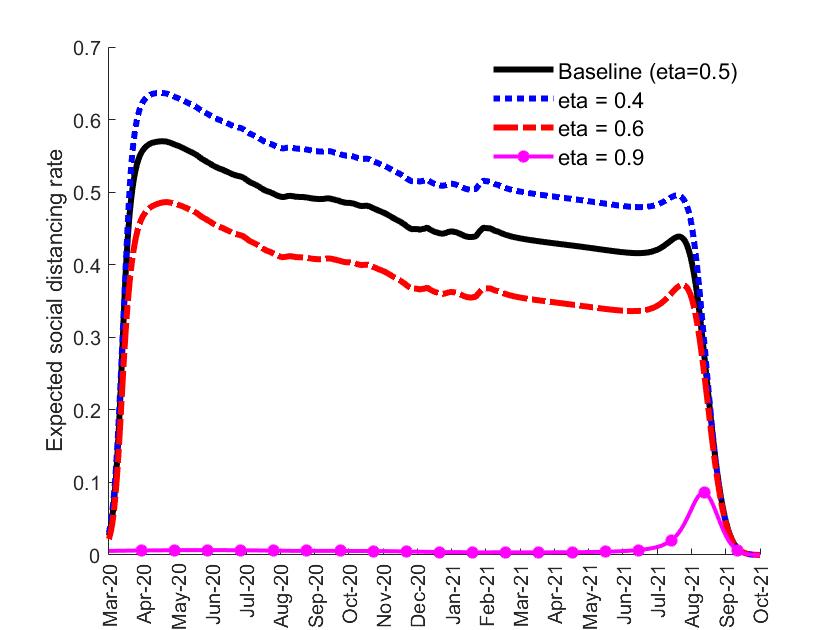}
\caption{Social Distancing}
\label{fig: sd eta}
\end{subfigure}\\
\begin{subfigure}[b]{0.48\textwidth}    
\includegraphics[scale=0.28]{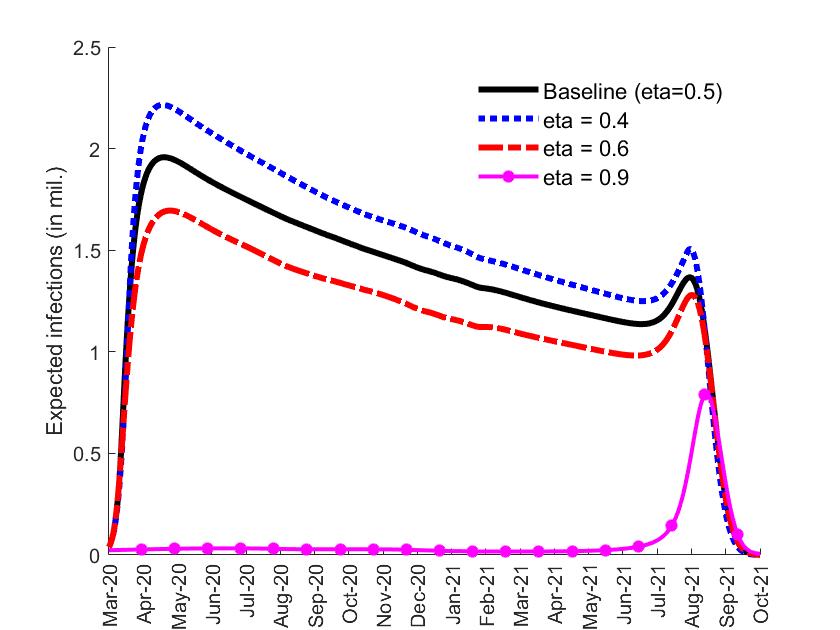}
\caption{Flow of infections}
\label{fig: infections eta}
\end{subfigure}
\begin{subfigure}[b]{0.48\textwidth}    
\includegraphics[scale=0.28]{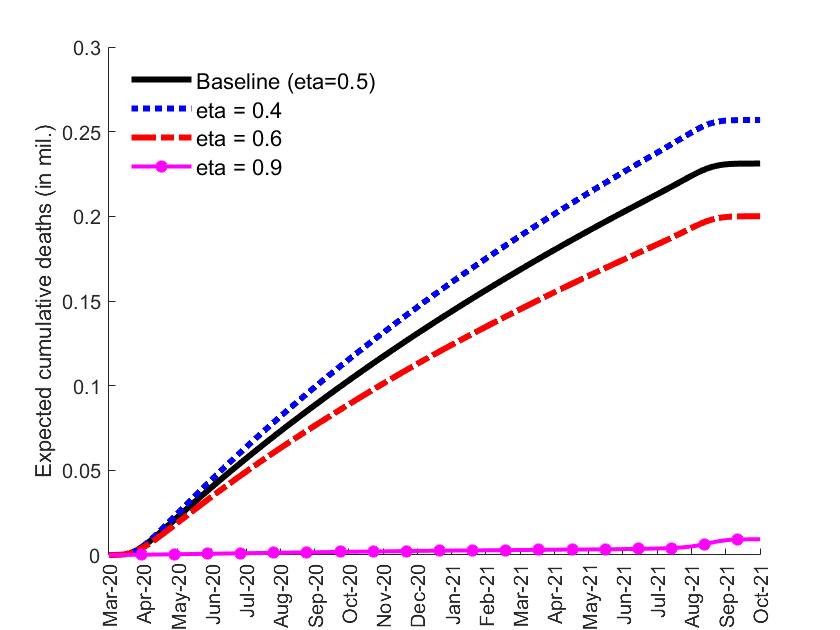}
\caption{Cumulative deaths}
\label{fig: deaths eta}
\end{subfigure}
\caption{Counterfactuals for different levels of $\eta$, where $\beta_{w}=\eta\beta$ and $\beta_{s}=(1-\eta)\beta$} 
\label{fig: counterfactuals eta}
\end{figure*}

We do this through the following reduced form experiment. Recollect that $\beta_{w} = \eta \beta$ and $\beta_{s} = (1-\eta)\beta$, where $\beta$ can be considered as the aggregate prevalence parameter which sets the contagiousness of this disease, and $\eta$ is the fraction of infections that arise at work and $1-\eta$ the fraction that arise in social interactions, respectively. We started out by setting $\eta=0.5$, which is motivated by the American Time Use Survey (see \citet*{ert_covid}). We now present the results for $\eta=0.4, 0.6$ and 0.9. The idea is to vary the matching intensity in the social interactions part of Equation \eqref{eq S mech} and its subsequent avatars.  So, we interpret $\eta=0.4$ to mean that the fraction of matches within the direct control of the government decreases. And, the way we interpret $\eta=0.6, 0.9$ is that government has been able to enforce masking and other other such policies (which decreases matches) more effectively. The agents must follow these policies to a varying degree, as $\eta$ increases. 

As $\eta$ increases the government essentially tries to kill the pandemic through a more extensive lockdown for a larger fraction of infections matches are under its control. Why does the government do this? Remember that saving lives has a direct value in the government's objective function and every life saved produces more output in the future. Even though the flow of output decreases, its net present value is higher because of lives saved through the stricter lockdown. And since the government has greater control on matches the stricter lockdown has a greater bang for buck. The agents on the other hand social distance less because given the strict lockdown in place, the probability of them contracting the infection in social interactions is lower---they try to take advantage of whatever social interactions are permitted. Overall, greater governmental control reduces infections and total deaths. See Figure \ref{fig: counterfactuals eta} for the results. 

\subsection{No lockdown and social distancing}

\label{section ND}

Several governments have faced the dilemma of whether to lockdown economic activities and by how much. Sweden is perhaps the most widely discussed case, it went in for little or almost no lockdown over the early part of the pandemic (\citet{FT_Sweden}). In this section we first evaluate the policy experiment where lockdown is exogenously fixed at zero. We analyze this scenario for both the benchmark constellation of parameters and for the case where testing-tracing is more extensive and efficient. 
\begin{figure*}[h]
\centering
\begin{subfigure}[b]{0.48\textwidth}    
\includegraphics[scale=0.28]{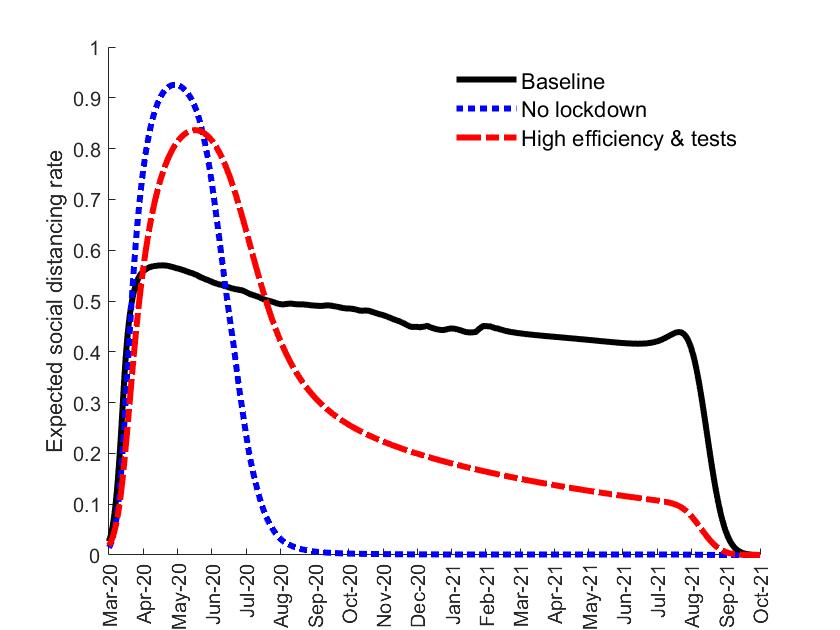}
\caption{Social distancing}
\label{fig: sd ND}
\end{subfigure}
\begin{subfigure}[b]{0.48\textwidth}    
\includegraphics[scale=0.28]{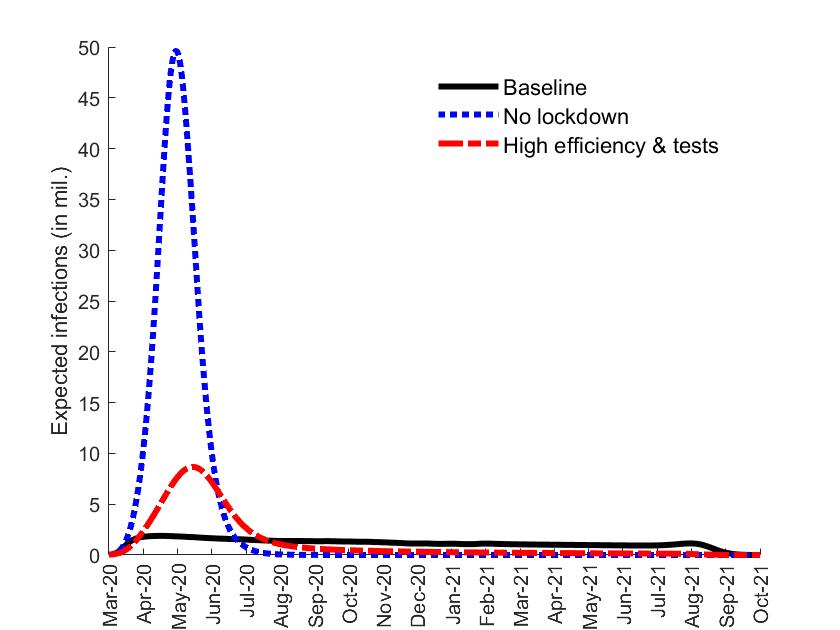}
\caption{Flow of infections}
\label{fig: infections ND}
\end{subfigure}\\
\begin{subfigure}[b]{0.48\textwidth}    
\includegraphics[scale=0.28]{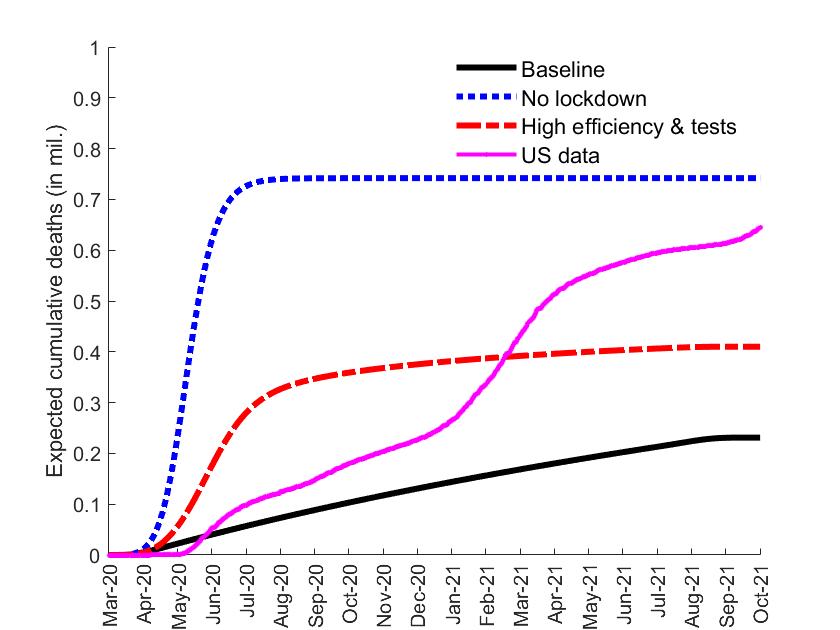}
\caption{Cumulative deaths}
\label{fig: deaths ND}
\end{subfigure}
\caption{Counterfactuals for no lockdown}
\label{fig: counterfactuals ND}
\end{figure*}

Figures \ref{fig: sd ND} and \ref{fig: infections ND} show that pandemic speeds through the population with one large peak of infections at the inception in response to which the agents social distance to a peak level of shutting down 90\% of all social activities (red lines in both graphs). Correspondingly, the number of deaths rises rapidly to about 750 thousands by July 2020, and then stays constant as the population reaches herd immunity (see red line in Figure \ref{fig: deaths ND}). This makes one ponder because the {\it actual} number of deaths in the US as of writing this draft is also in the range of 750 thousands. A provocative thought then is that with its current capacity in testing, the US could have ended up with the same numbers of total deaths even without a lockdown---the difference would be in the distribution of fatalities over time.\footnote{There are obvious caveats here: Agents need to have understood the implications of the pandemic at the outset and were forward looking in their social distancing decisions. Moreover, hospitals all over the country could have handled the extreme peak in infections shown in Figure \ref{fig: infections ND}. Subject to these caveats, this point about the equivalence of deaths is made to highlight that one of the goals of optimal policy is in-fact to spread out the pandemic to allow the government and society to manage it better.} 

The solid purple line in Figure \ref{fig: counterfactuals ND} plots social distancing, infection and deaths for the no lockdown case when it is accompanied with more extensive and effective tracing-testing. It is clear that if testing-tracing capacity is ramped up then even with no lockdown the peak of infections is much smaller, and the total number of deaths is reduced significantly, the number being lower than what we see actually realized in the data. 
\begin{figure*}[h]
\centering
\begin{subfigure}[b]{0.48\textwidth}    
\includegraphics[scale=0.28]{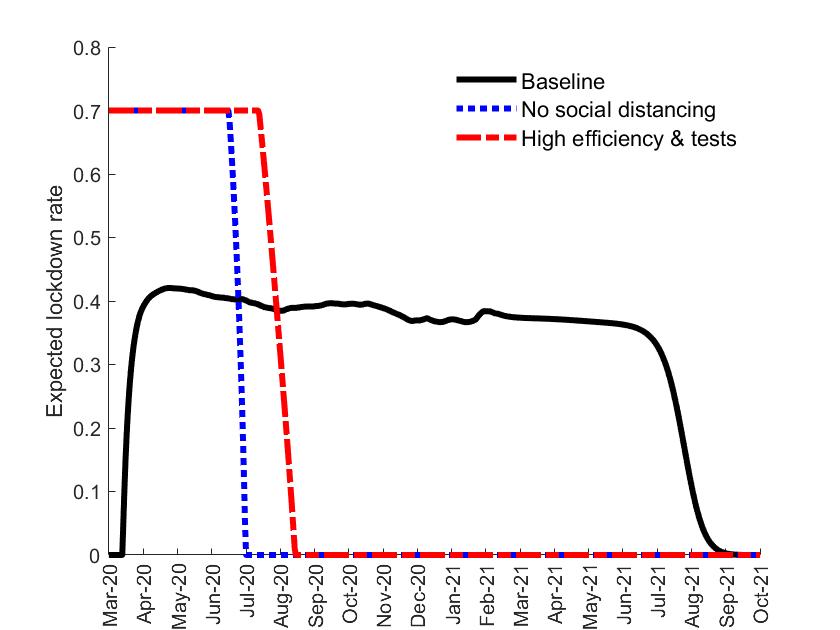}
\caption{Lockdowm}
\label{fig: sd NSD}
\end{subfigure}
\begin{subfigure}[b]{0.48\textwidth}    
\includegraphics[scale=0.28]{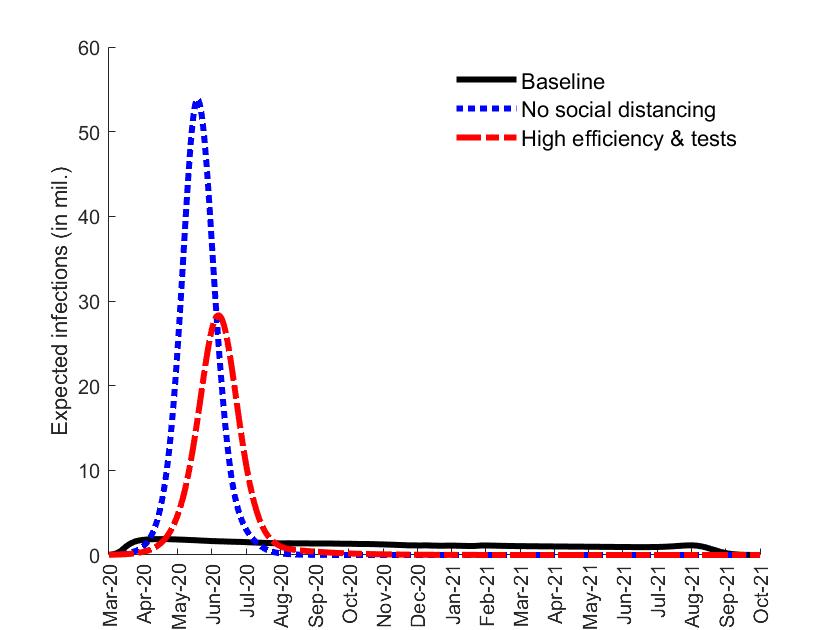}
\caption{Flow of infections}
\label{fig: infections NSD}
\end{subfigure}\\
\begin{subfigure}[b]{0.48\textwidth}    
\includegraphics[scale=0.28]{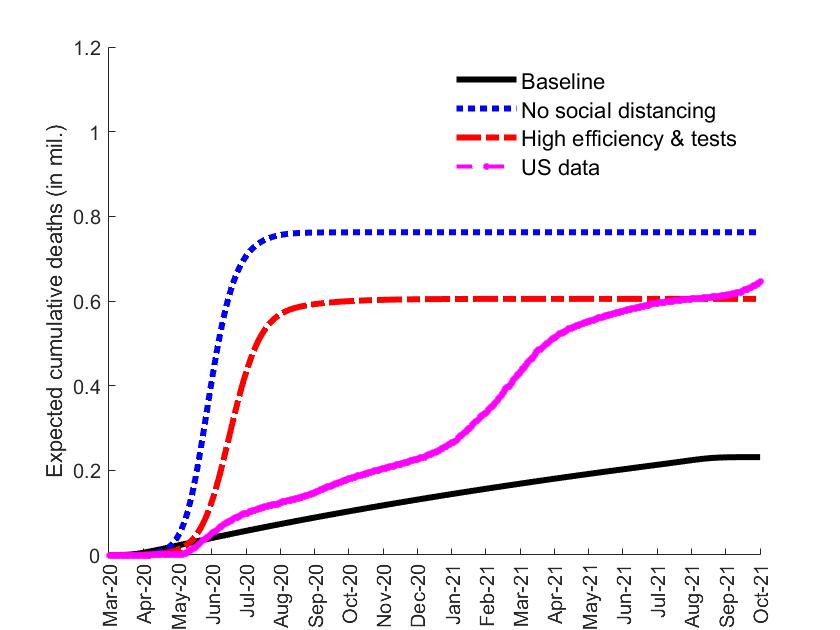}
\caption{Cumulative deaths}
\label{fig: deaths NSD}
\end{subfigure}
\caption{Counterfactuals for no social distancing}
\label{fig: counterfactuals NSD}
\end{figure*}

In the next policy experiment we ask the question: what if social distancing is shutdown in the model? This scenario is not meant to be taken literally, it is simply a theoretical exercise to understand what the analyst or the modeler may miss if agency or behavioral response is not directly modeled in an SIR framework. Figure \ref{fig: counterfactuals NSD} plots the results for this exercise. A pattern similar to the no lockdown case emerges, which highlights the complementarity of the two forces. However, things are more severe when social distancing is shut down in the sense that government locks down maximally while the infections peak (dotted blue in Figures \ref{fig: sd NSD} and \ref{fig: infections NSD} respectively). The total number of deaths rises quickly to more than 750 thousands (dotted blue line in Figure \ref{fig: deaths NSD}). Again increasing efficacy of testing-tracing reduces the peak of infections and total number of deaths (dashed red lines in Figures \ref{fig: infections NSD} and \ref{fig: deaths NSD}). The reduction in the peak is much more extreme in the no lockdown case in comparison to the no social distancing case. This is because better testing technology is used to draw out the pandemic longer  through social distancing; whereas is in the no social distancing case,  because of the concern for total output, the pandemic is made to run through faster. This contrast can be seen by comparing the shape of the cumulative deaths curve (dashed red line) in Figures \ref{fig: infections ND} and \ref{fig: deaths NSD}).

\section{Final remarks}

In this paper we set out to incorporate two policy instruments, namely lockdown and tracing-testing, into a standard epidemiology framework augmented by behavioral response. The theoretical analysis tackles the challenge of capturing the pairwise substitutability of each of the three forces and the heterogeneity introduced by testing. The model is calibrated to the data on the time series of deaths and tests conducted in the US (from Feb 2020 to Feb 2021) to calculate the three key parameters---prevalence of the virus, efficiency of tracing-testing, and costs from social distancing. The testing capacity is fixed by the available data on tests conducted in the US during the same time frame. Under the optimal policy, the government locks down at a moderate but persistent rate and agents social distance too at a moderate but persistent rate, each picking up the slack of the other. These two together contribute in keeping the effective reproductive number at one throughout the pandemic. Pareto frontier maps the feasible set of economic output and total fatalities. Three different policy experiments are conducted to evaluate how the predictions of the model would change under some commonly debated policy prescriptions. 

In this final section, we now briefly discuss some limitations of our analysis, and extensions of the model that could be useful in addressing these concerns.\\ \vspace{-2mm}

{\color{blue} Modeling behavioral response.} We model behavioral response by introducing costs from social distancing that constitute a flow component as well as a lumps-sum component from getting infected and infecting others. This seems like a natural way to incorporate the incentives for social distancing, as has been done, for example by  \citet{flavio_covid} and \citet{farboodietal_covid}. However, we depart from existing work by assuming that those who have recovered without showing severe symptoms cannot know so unless they are tested. There are potentially interesting micro-foundations to these decisions of social distancing worth thinking about carefully. What role do authorities and peer effects play? This requires a better modeling of information percolation.\footnote{See \citet*{kalyan_da_paper} for a model of optimal government policy where agents learn about the pandemic from the intensity of the lockdown.} Also, are individualistic societies less or more likely to internalize social distancing into their daily realities than those that are community driven? \\ \vspace{-2mm}

{\color{blue} Modeling testing}. We introduced a very simple tracing and testing technology wherein a fraction of susceptible, and infectious and recovered without symptoms can be tested every period. The efficacy of testing is controlled by a single parameter. Testing technology can be enriched by first separating rapid tests that only discover infection from antibody tests that also discover whether the person was infected in the past. Restricting testing to only the former, which is more realistic, would change the effectiveness of testing, but by how much is an open question, at least in the context of our model. Another avenue of future research is to focus tracing, testing and quarantining more towards the elderly, who are significantly more vulnerable.\footnote{In response to the question, should all countries be testing uniformly, Germany's leading Covid expert informing policy, Christian Drosten, had stated early in the pandemic: "I'm not sure. Even in Germany, with our huge testing capacity, and most of it directed to people reporting symptoms, we have not had a positivity rate above 8\%. So I think targeted testing might be best, for people who are really vulnerable... This is not fully in place even in Germany, though we're moving towards it. The other target should be patients in the first week of symptoms, especially elderly patients who tend to come to hospital too late at the moment... And we need some kind of sentinel surveillance system, to sample the population regularly and follow the development of the reproduction number" (\citet{guardian_german_expert}).}  \\ \vspace{-2mm}

{\color{blue} Modeling heterogeneity in the population}. It is now clear Covid-19 is much more lethal to the elderly and people with certain specific co-morbidities. One way to model this richness is to create separate compartments for different sections of the population, define system of equations for the evolution of disease dynamics in each compartment with matching functions within and across them. Work along these lines has been done by \citet{farhi_sir} and  \citet{dream_team_SIRmodel} amongst others. Marrying features of behavioral response to these frameworks or introducing age dependent compartments in our model remains an open question.\\

{\color{blue} Modeling economic activity and lockdown.} We modeled lockdown as as a time dependent policy that shuts down a fraction of economic activity. Total output in turn is modeled simply as the total labor supply. These are of course abstractions that do capture the gist of the problem, but miss out some important ingredients. First, lockdown can be selective across various regions of a country, and in varying intensities; see \citet{Schaal_pandemic} and \citet*{elisa_covid_paper} on how this matters for relaxing lockdown policies over time. Second, drop in output itself can kill lives, especially in developing countries where daily wage workers are abundant and social safety nets are weak, see for example \citet{covid_developing} for a discussion. This would, amongst other things, change the way we choose a Pareto-weight for optimal policy. Third, a richer model of production could be interesting to examine. How capital and supply chains are impacted by the pandemic and governmental response to it are salient too. \\ \vspace{-2mm}

{\color{blue} Modeling the end of the pandemic.}  The end of the pandemic is modeled as a negative binomial distribution. The parameters are chosen so that the pandemic could not have ended 360 days after its inception and it almost surely ends 720 days after with an average duration of 540 days. In reality it is difficult to define what the "end of the pandemic" means. We gave it the interpretation of the arrival of a vaccine. To that end, a more careful modeling of the roll out of the vaccine and related factors such as vaccine hesitancy, and mutations to the virus also seems important.\footnote{In a different approach \citet*{flavio_paper4} model the disease dynamics as SIRS which includes the possibility of reinfection, and the pandemic there can last for a century where dynamics of infections are characterized by dampened oscillations.}\\ \vspace{-2mm}

{\color{blue} On the Pareto frontier and value of life.} Putting a Pareto-weight on the tradeoff between economic output (or consumption) and rates of mortality is a deep philosophical question perennially relevant to how we structure society and institutions, see \citet{schelling} for a classic reference. More recently \citet*{jones_pareto} have taken up the question in the context of the recent Covid-19 crises. For our analysis, we were reluctant to pick a particular number and thus eventually provided an entire frontier. A potentially interesting question here is to measure how much societies value life through a revealed preference argument with the pandemic as a natural experiment. \\ \vspace{-2mm}

{\color{blue} On pandemic fatigues, second waves and new variants}. A salient feature of the pandemic across countries has been the rise of infection in waves. While our model is flexible enough to match the data in the US which went through at least two waves of the pandemic, it does not explicitly model the underlying behavioral reasons for it. \citet{atkenson_SIR} takes first steps in this direction by introducing ``a one-time reduction in the semi-elasticity of the transmission rate to the daily death rate late in the year," as pandemic fatigue. Much more work needs to be done in understanding behavioral underpinnings of the pandemic fatigue and how optimal policy should take it into account. Moreover, the successive waves of the pandemic have been associated with the rise of new "variants". At the time of the writing of this draft the Delta variant of Covid has created much havoc and the Omicron variant is on the rise. Incorporating the ebb and flow of variants precisely because of the global nature of the spread of the virus is another challenge for modeling. \\ \vspace{-2mm}

{\color{blue} On mental health consequences.} We have abstracted here from the mental heath consequences of lockdown and to some extent of social distancing as well. There is widespread evidence of a spurt in mental health troubles during the course of Covid-19, see for example \citet{mental_health_lancent} and \citet{mental_health_mayo}. To the best of our knowledge, these haven't been incorporated in the economic models of the pandemic yet. It presents an important challenge for policymakers for blanket lockdowns policies may have other perverse consequences not immediately observable, which need to be incorporated in trade-offs of policymaking. \\ \vspace{-2mm}

In conclusion, the most important caveat is that while we hope this paper helps economists and maybe perhaps even epidemiologists and public health researchers think carefully about the ways in which to incorporate behavioral response in models of disease dynamics and evaluate optimal policy, great care should be taken to interpret any of these insights to inform actual policy. As social scientists, we are in the long-haul of understanding socio-politico-economic contexts of this deep health crisis.


\section{Appendix}

\subsection{Derivation of the agent's optimal behavior}

\label{agentapp}

The agent's optimization problem is described in details in Section \ref{agent_opt}: Taking the aggregate variables $(S_t,I_t,\widehat S_t,\widehat I_t)$ and $(\bl_t,\;\bt_t)$ as given, the agent solves
\[
\max\limits_{(\alpha_t^k)_{k \leqslant t},\pi_t} \quad Y_A-C + (1-\da)c\kappa P \quad \text{subject to}\quad \text{\hyperref[eqpi A]{$(\Delta \pi_t)$}}.
\]

We now set up the Lagrangian for the agent's problem denoting the undiscounted dual variables associated with Equations \hyperref[eqpi A]{$(\Delta \pi_t)$} by $\Upsilon_t=\big(\s^k_t)_{k \leqslant t+1},(\i^k_t)_{k \leqslant t},\i\t_t,(\r^k_t)_{k \leqslant t},\r\t_t,\mathbbm{H}_t\big)$. The Lagrangian can be unpacked as follows:
\begin{scriptsize}
\begin{align*}
&\underbrace{(1-\da)\sum_{t =1}^{\infty} \da^{t-1}\bigg(y_tq_t + \frac{\da}{1-\da}(1-d_{t+1}) p_t\bigg)}_{Y_A}
-\underbracea{
(1-\da)\frac{c}{2}\sum_{t =1}^{\infty}\da^{t-1}\bigg((it_t+h_t+d_t)q_t+\frac{\da}{1-\da}d_{t+1}p_t + \sum_{k\leqslant t}(1-\alpha_t^k)^2(s_t^k+i_t^k+r_t^k)q_t \bigg)}- 
\\
&-\underbracebd{(1-\da)\sum_{t =1}^{\infty}\da^{t-1}\bigg(\phi^+(\beta_w\bl_t^2s_t{I}_t+\beta_s\widehat s_t{\widehat I}_t)+\phi^-(\beta_w\bl_t^2{S}_ti_t+\beta_s{\widehat S}_t \widehat i_t)\bigg)q_t}_{C}
+(1-\da)c\kappa\underbrace{\sum_{t=1}^{\infty}\da^{t-1}\sum_{k \leqslant t}\big(\ln \alpha^k_t+\ln \big(1-\alpha^k_t\big)\big)q_t}_{P}+
\\
&+\sum_{t=1}^{\infty}\sum_{k \leqslant t}\s^k_t\bigg(-\underbrace{s^k_{t+1} + (1-\bt_t\gamma)\bigg(s^k_t-\beta_w\bl^2_ts^k_tI_t - \beta_s\alpha^k_ts^k_t\widehat{I_t}\bigg)}_{\Delta s_t^k}\bigg)
+\sum_{t=1}^{\infty}\s^{t+1}_t\bigg(-\underbrace{s^{t+1}_{t+1} + \bt_t\gamma\bigg(s_t -\beta_w\bl^2_ts_tI_t - \beta_s\widehat{s_t}\widehat{I_t}\bigg)}_{\Delta s_t^{t+1}}\bigg)+
\\
&+\sum_{t=1}^{\infty}\sum_{k \leqslant t}\i^k_t\bigg(-\underbrace{i^k_{t+1} + (1-\bt_t)\bigg(\bigg(1-\frac{1}{t_i}\bigg)i^k_t + \beta_w\bl^2_ts^k_tI_t + \beta_s\alpha^k_ts^k_t\widehat{I_t}\bigg)}_{\Delta i_t^k}\bigg)+
\sum_{t=1}^{\infty}\i\t_t\bigg(-\underbrace{it_{t+1} + \bigg(\bigg(1-\frac{1}{t_i}\bigg)it_t + \bt_t\bigg(\bigg(1-\frac{1}{t_i}\bigg)i_t + \beta_w\bl^2_ts_tI_t + \beta_s\widehat{s_t}\widehat{I_t}\bigg)\bigg)}_{\Delta it_t}\bigg)+
\\
&+\sum_{t=1}^{\infty}\sum_{k \leqslant t}\r^k_t \bigg(-\underbrace{r^k_{t+1} + \bigg((1-\bt_t\gamma)\bigg(r^k_t +\frac{1-m_i}{t_i}i^k_t\bigg)\bigg)}_{\Delta r_t^k}\bigg)
+\sum_{t=1}^{\infty}\r\t_t\bigg(-\underbrace{rt_{t+1} + \bigg(rt_t + \frac{1-m_i}{t_i}it_t +\frac{1-m_h}{h_t}h_t + \bt_t\gamma\bigg(r_t+\frac{1-m_i}{t_i}i_t\bigg)\bigg)}_{=\Delta rt_t}\bigg)+
\\
&+\sum_{t=1}^{\infty}\h_t\bigg(-\underbrace{h_{t+1} + \bigg(\bigg(1-\frac{1}{t_h}\bigg)h_t + \frac{m_i}{t_i}(i_t+it_t)\bigg)}_{\Delta h_t}\bigg),
\end{align*}
\end{scriptsize}
where we used our standard shorthand notations: $s_t = \sum_{k \leqslant t}s_t^k$, $i_t = \sum_{k \leqslant t}i_t^k$, $r_t = \sum_{k \leqslant t}r_t^k$, $\widehat s_t = \sum_{k \leqslant t}\alpha_t^ks_t^k$, $\widehat i_t = \sum_{k \leqslant t}\alpha_t^ki_t^k$, $\widehat r_t = \sum_{k \leqslant t}\alpha_t^kr_t^k$, $y_t = \bl_t(s_t+i_r+r_t+rt_t)$ and $d_t=1-(s_t+i_t+it_t+r_t+rt_t+h_t)$.
\\

The optimal level of social distancing is characterized by the necessary first-order condition (\ref{agent_foc}). For completeness, we copy the condition:
\begin{scriptsize}
\begin{align*}
1-\alpha^k_t + \frac{\kappa}{s_t^k+i_t^k+r_t^k}\bigg(\frac{1}{\alpha^k_t}-\frac{1}{1-\alpha^k_t}\bigg)
&= \frac{\phi^+}{c}\frac{s^k_t}{s^k_t+i^k_t+r^k_t}\beta_s{\widehat I_t} + \frac{\phi^-}{c}\frac{i^k_t}{s^k_t+i^k_t+r^k_t}\beta_s{\widehat S_t} +\frac{\Delta_t^k}{c(1-\da)\da^{t-1}q_t}\frac{s_t^k}{s_t^k+i_t^k+r_t^k}\beta_s {\widehat I_t},
\nonumber
\end{align*}
\end{scriptsize}
where $\Delta_t^k = (1-\bt_t\gamma)\s^k_t +\bt_t\gamma\s^{t+1}_t -(1-\bt_t)\i^k_t -\bt_t\i\t_t$. It is easy to see that the equation above admits a unique solution for small enough $\kappa > 0$.\footnote{In fact, Equation (\ref{agent_foc}) can be rewritten as $\alpha_t^k = A_t^k + \frac{\kappa}{s_t^k+i_t^k+r_t^k}\bigg(\frac{1}{\alpha_t^k}-\frac{1}{1-\alpha_t^k}\bigg)$ for a certain number $A_t^k$, which is independent of $\kappa$. The equation defines a polynomial of degree 3. Then, it is possible to show that the solution in trigonometric functions can be obtained using the Viete's formula, and that $\alpha_t^k$ converges to $\max\{0,\min\{1,A_t^k\}\}$ as $\kappa \downarrow 0$.} 

The remaining first-order conditions with respect to the vector of state variables are as follows:
\begin{scriptsize}
\begin{align}
\label{adjs}
\s^k_{t} &= (1-\bt_{t+1}\gamma) \s^k_{t+1}  + \bt_{t+1}\gamma \s^{t+2}_{t+1} - \Delta^k_{t+1}\left(\beta_w\bl^2_{t+1}I_{t+1} + \beta_s\widehat{I}_{t+1}\alpha^k_{t+1}\right) +\da^t\bigg(1+\frac{c}{2}\bigg)p_t + \nonumber
\\
\tag{$\Delta \s_t^k$}
&+(1-\da)\da^t\bigg(\bl_{t+1} + \frac{c}{2}\left(1-(1-\alpha^k_{t+1})^2\right)-\phi^+\left(\beta_w\bl^2_{t+1}I_{t+1} + \beta_s\alpha^k_{t+1}\widehat{I}_{t+1}\right)\bigg)q_{t+1},\;\; \tiny{k=1,\ldots, t+1}
\\
\label{adji}
\i^k_t &= \bigg(1-\frac{1}{t_i}\bigg)\bigg((1-\bt_{t+1})\i^k_{t+1}+\bt_{t+1}\i\t_{t+1}\bigg)  + \frac{1-m_i}{t_i}\bigg((1-\bt_{t+1}\gamma) \r^k_{t+1} + \bt_{t+1}\gamma\r\t_{t+1}\bigg) + \frac{m_i}{t_i}\h_{t+1} + \da^t\bigg(1+\frac{c}{2}\bigg)p_t +
\nonumber
\\
\tag{$\Delta \i_t^k$}
&+(1-\da)\da^t\bigg(\bl_{t+1} + \frac{c}{2}\left(1-(1-\alpha^k_{t+1})^2\right)-\phi^-\left(\beta_w\bl^2_{t+1}S_{t+1} + \beta_s\alpha^k_{t+1}\widehat{S}_{t+1}\right)\bigg)q_{t+1},\;\; \tiny{k=1,\ldots, t}
\\
\label{adjit}
\tag{$\Delta \i\t_t$}
\i\t_t &=\bigg(1-\frac{1}{t_i}\bigg)\i\t_{t+1} + \frac{1-m_i}{t_i}\r\t_{t+1} + \frac{m_i}{t_i}\h_{t+1} + \da^t \bigg(1 +\frac{c}{2}\bigg)p_t,
\\
\label{adjr}
\tag{$\Delta \r_t^k$}
\r^k_{t} &= (1-\bt_{t+1}\gamma)\r^k_{t+1} + \bt_{t+1}\gamma\r\t_{t+1} + \da^t\bigg(1+\frac{c}{2}\bigg)p_t + (1-\da)\da^t\bigg(\bl_{t+1} + \frac{c}{2}\bigg(1-(1-\alpha^k_{t+1})^2\bigg)\bigg)q_{t+1} ,\;\;\tiny{k=1,\ldots,t} 
\\
\label{adjrt}
\tag{$\Delta \r\t_t$}
\r\t_{t} &=\r\t_{t+1} + (1-\da)\da^t\bigg(\bl_{t+1} + \frac{c}{2}\bigg)q_{t+1} + \da^t\bigg(1+\frac{c}{2}\bigg)p_t,
\\
\label{adjh}
\tag{$\Delta \h_t$}
\h_t &=\bigg(1-\frac{1}{t_h}\bigg)\h_{t+1}  + \frac{1-m_h}{t_h}\r\t_{t+1} + \da^t\bigg(1 +\frac{c}{2}\bigg)p_t. 
\end{align}
\end{scriptsize}
As discussed in Section \ref{agent_opt}, the system of adjoint equations is backwards with the following boundary condition:
\begin{small}
\[
\lim_{t \to \infty} \frac{\s^k_t}{p_t\da^{t-1}} = \lim_{t \to \infty} \frac{\i^k_t}{p_t\da^{t-1}} = \lim_{t \to \infty} \frac{\r^k_t}{p_t\da^{t-1}} = \lim_{t \to \infty} \frac{\i\t_t}{p_t\da^{t-1}} = \lim_{t \to \infty} \frac{\r\t_t}{p_t\da^{t-1}} = \lim_{t \to \infty} \frac{\h_t}{p_t\da^{t-1}} = \da\left(1+\frac{c}{2}\right) \; \forall k.
\]
\end{small}
As usual, we write $\Delta \Upsilon_t := \big((\Delta \s_t^k)_{k \leqslant t+1},(\Delta \i_t^k)_{k \leqslant t},\Delta \i\t_t,(\Delta \r_t^k)_{k \leqslant t},\Delta \r\t_t,\Delta \h_t\big) \label{equps A}$ for the system of adjoint equations. For  the fixed aggregates $\big(S_t,I_t,\widehat{S}_t,\widehat{I}_t\big)$ and policies $(\lambda_t,\tau_t)$, Equations  (\hyperref[eqpi A]{$\Delta \pi_t$}), (\hyperref[equps A]{$\Delta \Upsilon_t$}) and (\ref{agent_foc}) determine a representative agent's best response, that is her rate of social distancing  $(\alpha_t^k)_{k \leqslant t}$ as well as the individual state and adjoint vectors, $\pi_t$ and $\Upsilon_t$, respectively.  Since the agents are ex-ante identical, the Law of large numbers implies that in equilibrium the individual state variables match exactly the aggregate state variables, that is $\pi_t = \Pi_t$, in particular: $\big(s_t,i_t,\widehat{s}_t,\widehat{i}_t\big)=\big(S_t,I_t,\widehat{S}_t,\widehat{I}_t\big)$.

\subsection{Derivation of the government's optimal policy}

\label{govtapp}

The government's optimization problem is described in details in Section \ref{govt_opt}: The government chooses the rates of lockdown, social distancing and testing as well as the aggregate state variables and agent's adjoint variables to maximize her own payoff. In this task the government is constrained by the fact every agent must indeed find it in her interest to follow the designated social distancing policy, that is the agent's optimality conditions must be respected. In addition, the total test count must be exactly equal to the capacity, i.e., $\overline{X}_t$. Thus, we arrive at the following problem:
\begin{align*}
&\max\limits_{\lambda_t,(\alpha_t^k)_{k \leqslant t},\bt_t,\Pi_t,\Upsilon_t} \quad Y_G + \xi \cdot (1-D)  \quad \text{subject to}\quad \text{\hyperref[eqpi]{$(\Delta \Pi_t)$}},\;\text{\hyperref[equps A]{$(\Delta \Upsilon_t)$}},\;\text{(\ref{agent_foc})}\;\text{and}\;\text{(\ref{eq testing capacity})}.
\end{align*}

We now set up the Lagrangian for the problem denoting the undiscounted dual variables associated with Equations \hyperref[eqpi]{$(\Delta \Pi_t)$} by $\overline{\Upsilon}_t=\big((\gs^k_t)_{k \leqslant t+1},(\gi^k_t)_{k \leqslant t},\git_t,(\gr^k_t)_{k \leqslant t},\grt_t,\gh_t\big)$ and those associated with Equations \hyperref[equps A]{$(\Delta \Upsilon_t)$} by $\big((\gas_t^k)_{k \leqslant t},(\gai_t^k)_{k \leqslant t},\gait_t,(\gar_t^k)_{k \leqslant t},\gart_t,\gah_t\big)$. In addition, we let $(\eta_t^k)_{k \leqslant t}$ and $\chi_t$ be the Lagrange multipliers on (\ref{agent_foc}) and (\ref{eq testing capacity}), respectively. Taking all pieces together, the Lagrangian can be written as follows:
\begin{tiny}
\begin{align*}
&\underbrace{(1-\delta_G)\sum_{t=1}^{\infty}\dg^{t-1}\bigg(Y_tq_t + \frac{\dg+\xi/\dg^{t-1}}{1-\delta_G}(1-D_{t+1}) p_t\bigg)}_{Y_G}
+\sum_{t=1}^{\infty}\chi_t\bigg(\underbrace{-\overline X_t+\tau_t\bigg(\gamma\bigg(S_t-\beta_w\lambda^2_tS_tI_t-\beta_s\widehat{S}_t\widehat{I}_t\bigg)+\gamma\bigg(R_t+\frac{1-m_i}{t_i}I_t\bigg)+\bigg(1-\frac{1}{t_i}\bigg)I_t + \beta_w\lambda^2_tS_tI_t+\beta_s\widehat{S}_t\widehat{I}_t\bigg)}_{F-\bt_t}\bigg)+
\\+
&\sum_{t=1}^{\infty}\sum_{k \leqslant t}\eta_t^k 
\underbrace{\bigg((1-\da)c\da^{t-1}\bigg((1-\alpha_t^k)(S_t^k+I_t^k+R_t^k)+\kappa\bigg(\frac{1}{\alpha^k_t}-\frac{1}{1-\alpha^k_t}\bigg)\bigg)q_t-(1-\da)\da^{t-1}\beta_s\left(\phi^+S_t^k\widehat I_t+\phi^-I_t^k\widehat S_t\right)-\Delta_t^k\beta_sS_t^k\widehat I_t\bigg)}_{FOC-\alpha_t^k}+
\\
&+\sum_{t=1}^{\infty}\sum_{k \leqslant t}\gs^k_t\bigg(-\underbrace{S^k_{t+1} + (1-\bt_t\gamma)\bigg(S^k_t-\beta_w\bl^2_tS^k_tI_t - \beta_s\alpha^k_tS^k_t\widehat{I_t}\bigg)}_{\Delta S_t^k}\bigg)
+\sum_{t=1}^{\infty}\gs^{t+1}_t\bigg(-\underbrace{S^{t+1}_{t+1} + \bt_t\gamma\bigg(S_t -\beta_w\bl^2_tS_tI_t - \beta_s\widehat{S_t}\widehat{I_t}\bigg)}_{\Delta S_t^{t+1}}\bigg)+
\\
&+\sum_{t=1}^{\infty}\sum_{k \leqslant t}\gi^k_t\bigg(-\underbrace{I^k_{t+1} + (1-\bt_t)\bigg(\bigg(1-\frac{1}{t_i}\bigg)I^k_t + \beta_w\bl^2_tS^k_tI_t + \beta_s\alpha^k_tS^k_t\widehat{I_t}\bigg)}_{\Delta I_t^k}\bigg)+
\sum_{t=1}^{\infty}\git_t\bigg(-\underbrace{IT_{t+1} + \bigg(\bigg(1-\frac{1}{t_i}\bigg)IT_t + \bt_t\bigg(\bigg(1-\frac{1}{t_i}\bigg)I_t + \beta_w\bl^2_tS_tI_t + \beta_s\widehat{S_t}\widehat{I_t}\bigg)\bigg)}_{\Delta IT_t}\bigg)+
\\
&+\sum_{t=1}^{\infty}\sum_{k \leqslant t}\gr^k_t \bigg(-\underbrace{R^k_{t+1} +\bigg((1-\bt_t\gamma)\bigg(R^k_t +\frac{1-m_i}{t_i}I^k_t\bigg)\bigg)}_{\Delta R_t^k}\bigg)
+\sum_{t=1}^{\infty}\grt_t\bigg(-\underbrace{RT_{t+1} + \bigg(RT_t + \frac{1-m_i}{t_i}IT_t +\frac{1-m_h}{h_t}H_t + \bt_t\gamma\bigg(R_t+\frac{1-m_i}{t_i}I_t\bigg)\bigg)}_{=\Delta RT_t}\bigg)+
\\
&+\sum_{t=1}^{\infty}\gh_t\bigg(-\underbrace{H_{t+1} + \bigg(\bigg(1-\frac{1}{t_h}\bigg)H_t + \frac{m_i}{t_i}(I_t+IT_t)\bigg)}_{\Delta H_t}\bigg)+\sum_{t=1}^{\infty}\sum_{k \leqslant t+1}\gas_{t+1}^k\bigg(\underbracea{-\s^k_{t} + \bigg((1-\bt_{t+1}\gamma) \s^k_{t+1}  + \bt_{t+1}\gamma \s^{t+2}_{t+1} - \Delta^k_{t+1}\left(\beta_w\bl^2_{t+1}I_{t+1} + \beta_s\widehat{I}_{t+1}\alpha^k_{t+1}\right) +\da^t\bigg(1+\frac{c}{2}\bigg)p_t}+
\\
&+\underbracebd{(1-\da)\da^t\bigg(\bl_{t+1} + \frac{c}{2}\left(1-(1-\alpha^k_{t+1})^2\right)-\phi^+\left(\beta_w\bl^2_{t+1}I_{t+1} + \beta_s\alpha^k_{t+1}\widehat{I}_{t+1}\right)\bigg)q_{t+1}\bigg)}_{\Delta \s_t^k}\bigg)+
\\
&+\sum_{t=1}^{\infty}\sum_{k \leqslant t+1}\gai_{t+1}^k\bigg(\underbracea{-\i^k_{t} + \bigg(\bigg(1-\frac{1}{t_i}\bigg)\bigg((1-\bt_{t+1})\i^k_{t+1}+\bt_{t+1}\i\t_{t+1}\bigg)  + \frac{1-m_i}{t_i}\bigg((1-\bt_{t+1}\gamma) \r^k_{t+1} + \bt_{t+1}\gamma\r\t_{t+1}\bigg) + \frac{m_i}{t_i}\h_{t+1} + \da^t\bigg(1+\frac{c}{2}\bigg)p_t}+
\\
&+\underbracebd{(1-\da)\da^t\bigg(\bl_{t+1} + \frac{c}{2}\left(1-(1-\alpha^k_{t+1})^2\right)-\phi^-\left(\beta_w\bl^2_{t+1}S_{t+1} + \beta_s\alpha^k_{t+1}\widehat{S}_{t+1}\right)\bigg)q_{t+1}\bigg)}_{\Delta \i_t^k}\bigg)+\sum_{t=1}^{\infty}\gait_{t+1}\bigg(\underbrace{-\i\t_{t}+\bigg(\bigg(1-\frac{1}{t_i}\bigg)\i\t_{t+1} + \frac{1-m_i}{t_i}\r\t_{t+1} + \frac{m_i}{t_i}\h_{t+1} + \da^t \bigg(1 +\frac{c}{2}\bigg)p_t\bigg)}_{\Delta \i\t_t}\bigg)+
\\
&+\sum_{t=1}^{\infty}\sum_{k \leqslant t+1}\gar_{t+1}^k\bigg(\underbrace{-\r^k_{t} + \bigg((1-\bt_{t+1}\gamma)\r^k_{t+1} + \bt_{t+1}\gamma\r\t_{t+1} + \da^t\bigg(1+\frac{c}{2}\bigg)p_t + (1-\da)\da^t\bigg(\bl_{t+1} + \frac{c}{2}\bigg(1-(1-\alpha^k_{t+1})^2\bigg)\bigg)q_{t+1}\bigg)}_{\Delta \r_t^k}\bigg)+
\\
&+\sum_{t=1}^{\infty}\gart_{t+1}\bigg(\underbrace{-\r\t_{t}+\bigg(\r\t_{t+1} + (1-\da)\da^t\bigg(\bl_{t+1} + \frac{c}{2}\bigg)q_{t+1} + \da^t\bigg(1+\frac{c}{2}\bigg)p_t\bigg)}_{\Delta \r\t_t}\bigg)+\sum_{t=1}^{\infty}\gah_{t+1}\bigg(\underbrace{-\h_{t}+\bigg(\bigg(1-\frac{1}{t_h}\bigg)\h_{t+1}  + \frac{1-m_h}{t_h}\r\t_{t+1} + \da^t\bigg(1 +\frac{c}{2}\bigg)p_t\bigg)}_{\Delta \h_t}\bigg),
\end{align*}
\end{tiny}
In the problem above the following short notations are used: $S_t = \sum_{k \leqslant t}S_t^k$, $I_t = \sum_{k \leqslant t}I_t^k$, $R_t = \sum_{k \leqslant t}R_t^k$, $\widehat S_t = \sum_{k \leqslant t}\alpha_t^kS_t^k$, $\widehat I_t = \sum_{k \leqslant t}\alpha_t^kI_t^k$, $\widehat r_t = \sum_{k \leqslant t}\alpha_t^kR_t^k$, $Y_t = \bl_t(S_t+I_r+R_t+RT_t)$, $\overline{Y}_t = \bl_t(\gas_t+\gai_t+\gar_t+\gart_t)$ and $D_t=1-(S_t+I_t+IT_t+R_t+RT_t+H_t)$. In addition, we let $\Delta_t^k = (1-\bt_t\gamma)\s^k_t +\bt_t\gamma\s^{t+1}_t -(1-\bt_t)\i^k_t -\bt_t\i\t_t$ and $\overline{\Delta}_t^k = (1-\bt_t\gamma)\gs^k_t +\bt_t\gamma\gs^{t+1}_t -(1-\bt_t)\gi^k_t -\bt_t\git_t$.
\\

The optimal lockdown is a unique maximizer of Equation (\ref{lambda_foc}), which is given by
\begin{small}
\begin{align*}
\tag{OPT-$\bl_t$}
\lambda_t 
&= \arg\max\limits_{\bl \in [\overline{\bl},1]} \; \bigg\{\underbrace{\bigg((1-\dg)\dg^{t-1}Y_tq_t+(1-\da)\da^{t-1} \overline{Y}_tq_t\bigg)}_{=:a_t}\bl - \\
&- \underbrace{\bigg(\beta_w\sum_{k \leqslant t} \bigg(\overline{\Delta}_t^k S_t^k I_t+\Delta_t^k\gas_t^k I_t\bigg)+\beta_w\chi_t\bt_t(1-\gamma)S_tI_t+\beta_w(1-\da)\da^{t-1}\bigg(\phi^+\gas_tI_t+\phi^-S_t\gai_t\bigg)q_t
\bigg)}_{=:b_t}\bl^2\bigg\}.\nonumber
\end{align*}
\end{small}
Denote by $a_t$ and $b_t$ the first and second coefficients in Equation (\ref{lambda_foc}), respectively. Then, the optimal value of $\bl_t$ can be succinctly written as
\begin{small}
\[
\lambda_t = 
\begin{cases}
1 &\text{if}\; b_t \leqslant 0,
\\
\min\left\{1,\max\left\{0,\overline{\lambda},\frac{a_t}{2b_2}\right\}\right\} &\text{if}\; b_t > 0.
\end{cases}  
\] 
\end{small}
\vspace{-2mm}

As for the other remaining control variables: The first-order conditions with respect to the testing rate $\tau_t$ and social distancing $(\alpha_t^k)_{k \leqslant t}$ pin down the Lagrange multipliers on the corresponding constraints, $\chi_t$ and $(\eta_t^k)_{k \leqslant t}$, respectively. The former is given by the following:
\begin{footnotesize}
\begin{align}
\label{tau_foc}
\tag{OPT-$\bt_t$}
&\chi_t\bigg(\gamma S_t +\bigg(1-\frac{1}{t_i}\bigg)I_t + \gamma\bigg(R_t+\frac{1-m_i}{t_i}I_t\bigg) + (1-\gamma)\left(\beta_w\lambda^2_tS_tI_t+\beta_s\widehat{S}_t\widehat{I}_t\right)\bigg)=\\
&=\bigg(\sum_{k \leqslant t}\bigg(\gamma\left(\gs_t^k-\gs_{t}^{t+1}\right)S_t^k+\bigg(1-\frac{1}{t_i}\bigg)\left(\gi_t^k-\git_{t}\right)I_t^k+\gamma\left(\gr_t^k-\grt_{t}\right)\bigg(R_t^k+\bigg(1-\frac{1-m_i}{t_i}\bigg)I_t^k\bigg)\bigg)+\nonumber\\
&+\sum_{k \leqslant t}\bigg(\gamma\left(\s_t^k-\s_{t}^{t+1}\right)\gas_t^k+\bigg(1-\frac{1}{t_i}\bigg)\left(\i_t^k-\i\t_{t}\right)\gai_t^k+\gamma\left(\r_t^k-\r\t_{t}\right)\bigg(\gar_t^k+\bigg(1-\frac{1-m_i}{t_i}\bigg)\gai_t^k\bigg)\bigg)+\nonumber\\
&+\left(\gi_t^k-\git_{t}-\gamma\left(\gs_t^k-\gs_{t}^{t+1}\right)\right)\left(\beta_w\bl^2S_t^kI_t+\beta_s\alpha_t^k+\eta_t^k)S_t^k\widehat I_t\right)
+\left(\i_t^k-\i\t_{t}-\gamma\left(\s_t^k-\s_{t}^{t+1}\right)\right)\left(\beta_w\bl^2\gas_t^kI_t+\beta_s(\alpha_t^k+\eta_t^k)k\gas_t^k\widehat I_t\right)\bigg).
\nonumber
\end{align}
\end{footnotesize}
We note that Equation (\ref{tau_foc}) uniquely determines the Lagrange multiplier $\chi_t$ for the interior choice of $\bt_t \in (0,1)$. Recall that the testing rate is determined by Equation (\ref{eq testing capacity}) in a way that exactly $\overline{X}_t$ are conducted at date $t$. As a result, as long as $\overline{X}_t$ is positive and not too large, the implied testing rate is indeed interior. 
\\

The first-order condition of the government's problem with respect to the rate of social distancing is as follows:
\begin{footnotesize}
\begin{align}
&\eta_t^k(1-\da)\da^{t-1}c\bigg(S_t^k+I_t^k+R_t^k+\kappa\bigg(\frac{1}{(\alpha_t^k)^2}+\frac{1}{(1-\alpha_t^k)^2}\bigg)\bigg)q_t + \sum_{m \leqslant t}\eta_t^m\bigg((1-\da)\da^{t-1}\beta_s\left(\phi^+S_t^mI_t^k+\phi^-I_t^mS_t^k\right)q_t-\Delta_t^m\beta_sS_t^mI_t^k\bigg)
=\nonumber
\\
&=\chi_t\bt_t(1-\gamma)\beta_s\left(S_t^k\widehat I_t+I_t^k \widehat S_t\right) - \beta_s\left(\overline{\Delta}_t^kS_t^k+\Delta_t^k\gas_t^k\right)\widehat{I}_t - \beta_s\sum_{m \leqslant t}\left(\overline{\Delta}_t^mS_t^m+\Delta_t^m\gas_t^m\right)\widehat{I}_t^k + (1-\da)\da^{t-1}c(1-\alpha_t^k)\left(\gas_t^k+\gai_t^k+\gar_t^k\right)q_t-\nonumber
\\
\label{govt_foc}
\tag{OPT-$\alpha_t^k$}
&-(1-\da)\da^{t-1}\beta_s\bigg(\phi^+\gas_t^k\widehat{I}_t+\phi^-\gai_t^k\widehat{S}_t+\phi^+\widehat{\gas_t}I_t^k+\phi^-\widehat{\gai_t}S_t^k\bigg)q_t.
\end{align}
\end{footnotesize}
Equation (\ref{govt_foc}) can be used to solve for the Lagrange variable $(\eta_t^k)_{k \leqslant t}$. According to the aforementioned equation, $(\eta_t^k)_{k \leqslant t}$ must be a solution to the linear system, i.e., $\mathcal{A}_t\vec{\eta_t} = \mathcal{B}_t$ for the certain $t \times t$ matrix $\mathcal{A}_t$ and  $t \times 1$ vector $\mathcal{B}_t$. Thus, as long as the matrix $\mathcal{A}_t$ is invertible the solution is unique, and it is given by 
\begin{small}
\[
\vec{\eta_t} = \mathcal{A}_t^{-1}\mathcal{B}_t.  
\]
\end{small}
\vspace{-8mm}

Moving on, recall that the vector of agent's adjoints $\Upsilon_t=\big((\s^k_t)_{k \leqslant t+1},(\i^k_t)_{k \leqslant t},\i\t_t,(\r^k_t)_{k \leqslant t},\r\t_t,\mathbbm{H}_t\big)$ is associated to the Lagrange variables $\overline{\Delta\Pi}_t = \big((\gas_t^k)_{k \leqslant t},(\gai_t^k)_{k \leqslant t},\gait_t,(\gar_t^k)_{k \leqslant t},\gart_t,\gah_t\big)$. The first-order condition with respect to $\Upsilon_t$ specifies the forward system of equations (see \hyperref[eqpi bar]{$(\Delta \overline{\Pi}_t)$}) describing the dynamics of $\overline{\Pi}_t$. For completeness, we copy the system from  the main text:
\begin{small}
\begin{align}
\tag{$\Delta \gas_t^k$}
\overline{S}^k_{t+1} &= (1-\bt_t\gamma)\bigg(\overline{S}^k_t-\beta_w\bl^2_t\overline{S}^k_tI_t - \beta_s\bigg(\alpha^k_t\overline{S}^k_t-\eta^k_t{S}^k_t\bigg)\widehat{I}_t\bigg)\;\text{for } k=1,\ldots, t, 
\\
\tag{$\Delta \gas_t^{t+1}$}
\overline{S}^{t+1}_{t+1}&=\bt_t\gamma\bigg(\overline{S}_t -\beta_w\bl^2_t\overline{S}_tI_t - \beta_s\bigg(\widehat{\overline{S}}_t-\sum_{k \leqslant t}\eta_t^kS_t^k\bigg)\widehat{I}_t\bigg),
\\
\tag{$\Delta \gai_t^k$}
\overline{I}^k_{t+1} &= (1-\bt_t)\bigg(\bigg(1-\frac{1}{t_i}\bigg)\overline{I}^k_t + \beta_w\bl^2_t\overline{S}^k_tI_t +\beta_s\bigg(\alpha^k_t\overline{S}^k_t-\eta^k_t{S}^k_t\bigg)\widehat{I}_t\bigg)\;\text{for } k=1,\ldots, t, 
\\
\tag{$\Delta \gait_t$}
\overline{IT}_{t+1} &=\bigg(1-\frac{1}{t_i}\bigg)\overline{IT}_t + \bt_t\bigg(\bigg(1-\frac{1}{t_i}\bigg)\overline{I}_t + \beta_w\bl^2_t\overline{S}_tI_t +\beta_s\bigg(\widehat{\overline{S}}_t-\sum_{k \leqslant t}\eta_t^kS_t^k\bigg)\widehat{I}_t\bigg),
\\
\tag{$\Delta \gar_t^k$}
\overline{R}^k_{t+1} &=(1-\bt_t\gamma)\bigg(\overline{R}^k_t +\frac{1-m_i}{t_i}\overline{I}^k_t\bigg),\;\;\;\text{for } k=1,\ldots, t, 
\\
\tag{$\Delta \gart_t$}
\overline{RT}_{t+1} &= \overline{RT}_t + \frac{1-m_i}{t_i}\overline{IT}_t +\frac{1-m_h}{h_t}\overline{H}_t + \bt_t\gamma\bigg(\overline{R}_t+\frac{1-m_i}{t_i}\overline{I}_t\bigg),
\\
\tag{$\Delta \gah_t$}
\overline{H}_{t+1} &= \bigg(1-\frac{1}{t_h}\bigg)\overline{H}_t + \frac{m_i}{t_i}(\overline{I}_t+\overline{IT}_t),
\end{align}
\end{small}
and $\overline{I}_{t+1}^{t+1}=\overline{R}_{t+1}^{t+1}$ at all dates. The initial condition is that $\overline{\Pi}_1 = 0$.
\\

It remains only to identify the first-order condition with respect to the state variables. The vector of state variables $\Pi_t = \big((S_t^k)_{k \leqslant t},(I_t^k)_{k \leqslant t},IT_t,(R_t^k)_{k \leqslant t},RT_t,H_t\big)$ is associated to the vector of Lagrange multipliers $\overline \Upsilon_t$, which is $\overline \Upsilon_t=\big((\gs^k_t)_{k \leqslant t+1},(\gi^k_t)_{k \leqslant t},\git_t,(\gr^k_t)_{k \leqslant t},\grt_t,\gh_t\big)$. Thus, the first-order conditions with respect to $\Pi_t$ outputs the backward system of equations that describes the dynamics of $\overline{\Upsilon}_t$. The system is as follows:
\begin{footnotesize}
\begin{align}
\label{adjs govt}
\tag{$\gs_t^k$}
\gs_t^k &= \left((1-\bt_{t+1}\gamma)\gs_{t+1}^{k}+\bt_{t+1}\gamma\gs_{t+1}^{t+2}\right)+\chi_{t+1}\tau_{t+1}\bigg(\gamma+(1-\gamma)\left(\beta_w\bl^2I_{t+1}+\beta_s\widehat{I}_{t+1}\alpha_{t+1}^k\right)\bigg) + 
\\
&+ (1-\dg)\dg^{t}\bl_{t+1}q_{t+1}  + (\dg^t+\xi)p_t +\eta_{t+1}^k\bigg((1-\da)\da^tc(1-\alpha_{t+1}^k)q_{t+1}-(1-\da)\da^t\beta_s\phi^+\widehat{I}_{t+1}q_{t+1}-\Delta_{t+1}^k\beta_s\widehat{I}_{t+1}\bigg)-\nonumber\\
&-\sum_{m \leqslant t+1}\eta_{t+1}^m(1-\da)\da^t\beta_s\phi^-I_{t+1}^m\alpha_{t+1}^k-\overline{\Delta}_{t+1}^k\bigg(\beta_w\bl^2_{t+1}I_{t+1}^k+\beta\alpha_{t+1}^m\widehat{I}_{t+1}\alpha_{t+1}^k\bigg)-\nonumber\\
&-(1-\da)\da^t\sum_{m \leqslant t+1}\gai_{t+1}^m\phi^-\left(\beta_w\bl_{t+1}^2+\beta_s\alpha_{t+1}^m\right)q_{t+1},\nonumber
\\
\label{adji govt}
\tag{$\gi_t^k$}
\gi_t^k &= \bigg(1-\frac{1}{t_i}\bigg)\left((1-\bt_{t+1})\gi_{t+1}^{k}+\bt_{t+1}\git_{t+1}\right)+\frac{1-m_i}{t_i}\left((1-\bt_{t+1}\gamma)\gr_{t+1}^{k}+\bt_{t+1}\gamma\grt_{t+1}\right)+\frac{m_i}{t_i}\gh_{t+1}+\\
&+\chi_{t+1}\tau_{t+1}\bigg(\gamma\frac{1-m_i}{t_i}+1-\frac{1}{t_i}+(1-\gamma)\left(\beta_w\bl^2S_{t+1}+\beta_s\widehat{S}_{t+1}\alpha_{t+1}^k\right)\bigg) + (1-\dg)\dg^{t}\bl_{t+1}q_{t+1}  + (\dg^t+\xi)p_t+\nonumber
\\
&+\eta_{t+1}^k(1-\da)\da^t\bigg(c(1-\alpha_{t+1}^k)-\beta_s\phi^-\widehat{S}_{t+1}\bigg)q_{t+1}-\sum_{m \leqslant t+1}\eta_{t+1}^m(1-\da)\da^t\beta_s\bigg(\phi^++\Delta^m_{t+1}\bigg)S^m_{t+1}\alpha_{t+1}^k-\nonumber
\\
&-\sum_{m \leqslant t+1}\Delta_{t+1}^m\bigg(\beta_w\bl^2_{t+1}\gas_{t+1}^m+\beta\alpha_{t+1}^m\gas_{t+1}^m\alpha_{t+1}^k\bigg)-\sum_{m \leqslant t+1}\overline{\Delta}_{t+1}^m\bigg(\beta_w\bl^2_{t+1}S_{t+1}^m+\beta\alpha_{t+1}^mS_{t+1}^m\alpha_{t+1}^k\bigg)-\nonumber
\\
&-(1-\da)\da^t\sum_{m \leqslant t+1}\gas_{t+1}^m\phi^+\left(\beta_w\bl_{t+1}^2+\beta_s\alpha_{t+1}^m\right)q_{t+1},\nonumber
\\
\label{adjit govt}
\tag{$\git_t$}
\git_t 
&= \bigg(1-\frac{1}{t_i}\bigg)\git_{t+1} + \frac{1-m_i}{t_i}\grt_{t+1} + \frac{m_i}{t_i}\gh_{t+1} + (\dg^t+\xi)p_t,
\\
\label{adsr govt}
\tag{$\gr_t^k$}
\gr^k_t 
&= (1-\bt_{t+1}\gamma)\gr_{t+1}^k+\bt_{t+1}\gamma\grt_{t+1}+\chi_{t+1}\bt_{t+1}\gamma+\eta_{t+1}^k(1-\da)c\da^{t}(1-\alpha_{t+1}^k)q_{t+1} + (1-\dg)\dg^{t}\bl_{t+1}q_{t+1}  + (\dg^t+\xi)p_t\nonumber
\\
\label{adjrt govt}
\tag{$\Delta \grt_t$}
\grt_t 
&= \grt_{t+1} + (1-\dg)\dg^{t}\bl_{t+1}q_{t+1}  + (\dg^t+\xi)p_t,
\\
\label{adjh govt}
\tag{$\Delta \gh_t$}
\gh_t 
&= \bigg(1-\frac{1}{t_h}\bigg)\gh_{t+1} + \frac{1-m_h}{t_h}\grt_{t+1} + (\dg^t+\xi)p_t.
\end{align}
\end{footnotesize}
As already standard, we write $\overline{\Delta \Upsilon}_t := \big((\Delta \gs_t^k)_{k \leqslant t+1},(\Delta \gi_t^k)_{k \leqslant t},\Delta \git_t,(\Delta \gr_t^k)_{k \leqslant t},\Delta \grt_t,\Delta \gh_t\big) \label{equps G}$ for the system of adjoint equations.  We note that this system, as one for $\Upsilon_t$, is backwards with the following ``initial" condition:
\begin{small}
\begin{align*}
\lim_{t \to \infty} \frac{\gs^k_t}{p_t}-\dg^{t} = \lim_{t \to \infty} \frac{\gi^k_t}{p_t}-\dg^{t} = \lim_{t \to \infty} \frac{\gr^k_t}{p_t}-\dg^{t} = \lim_{t \to \infty}
\frac{\git_t}{p_t}-\dg^{t} = \lim_{t \to \infty} \frac{\grt_t}{p_t}-\dg^{t} = \lim_{t \to \infty} \frac{\gh_t}{p_t}-\dg^{t} = \xi \; \forall k.
\end{align*}
\end{small}

\subsection{Calibration}

In this section we discuss in details the rationale behind the assumed government policies and the set of fixed parameters which are used to calibrate the parameters of interest.

\subsubsection{Lockdown function}

We start with the approximation of actual lockdown (see Figure \ref{fig: lock}) which is assumed to take the following parametric form:
\begin{small}
\begin{align*}
\bl_t & = 
\begin{cases}
1 & t = 1,\ldots, 16, \\
 \sqrt{1- \frac{1-\overline{\bl}}{\big(\frac{32}{17} - 1\big)^{\sigma_1}}\big(\frac{t}{55} - 1\big)^{\sigma_1}} & t = 17, \ldots, 32, \\
\overline{\bl} & t = 33,\ldots, 103, \\
\sqrt{\overline{\bl}^2 \sigma_2^{t-1-103} + (1-\sigma_2)^{t-1-103}} & t = 104,\ldots, T,
\end{cases}
\end{align*}
\end{small}
where $1-\overline{\bl}$ reflects the maximum level of lockdown that can be imposed by the government, while $\sigma_1$ and $\sigma_2$ controls its severity. The assumed lockdown function has four broad phases: (i) between Feb 29, 2020 ($t=1$) and Mar 15, 2020 ($t=16$), the US government did not impose any lockdown and the economy worked at full capacity, (ii) severe lockdown including bans on international travel were imposed between Mar 16 ($t=17$) and Mar 31, 2020 ($t=32$), reflected by the steep decline in the lockdown function.\footnote{This was the period in which the US government formally acknowledged Covid-19 to be a national emergency. \href{https://www.federalregister.gov/documents/2020/03/18/2020-05794/declaring-a-national-emergency-concerning-the-novel-coronavirus-disease-covid-19-outbreak}{Link} to Proclamation 9994 by President Donald J. Trump.} 
The flat line (phase (iii)) between March 31 ($t=32$) to June 5 ($t=103$)  reflects  the severe lockdown imposed in majority of the states and cities. \citet{moreland2020timing} provides a comprehensive analysis of lockdowns imposed by various states in the US. They show that about 80\% of the states had mandatory stay-at-home orders in some form by March 31st, 2020. Most of these orders were gradually rescinded or allowed to exhaust from June 6 ($t=103$) onwards. This is reflected in the gradual rise in phase (iv) of the assumed lockdown function.

The choice of $\sigma_1$ and $\sigma_2$ reflects the curvature of the lockdown policy. After the proclamation of national emergency, there was a flurry of measures taken to curb down international and interstate travel, as well as closure of most offices. This is denoted by a steep curvature in phase (ii), reflected by a high value of $\sigma_1 = 2$ ensuring approximately the maximum level of lockdown at the start of phase (iii).  On the other hand, the relaxation of lockdown has been much slower, reflected in a gradual increase in the lockdown function at the fixed rate of $\sigma_2$ in phase (iv). We set $\sigma_2 = 0.9995$, so that $60\%$ of the economy is functioning by Fall, 2021. 

\subsubsection{Test count}
\label{section tests}

As explained above in Section \ref{section calibration}, measuring the total number of tests and the number of positive and negative cases has been a tough task for researchers. The difficulty is driven by the fact that in the US there is no Federal standard for reporting testing data, and each individual state can decide to report a number of tests in terms of a) people tested at least once (unique people), b) samples collected and c) testing encounters. The "\href{https://covidtracking.com/}{COVID Tracking Project}" aggregates the state level data as follows: for each state, a total number of tests equals to a number of testing encounters if it is reported, otherwise it equals to a number of unique people if it is reported, otherwise it equals to a number of samples collected. 

In order to calibrate and estimate the model, we aggregate the testing data by imputing a number of unique people reported by each state. For each state $s$, let $\overline{X}_{s,t}$ be the number of unique tested at time $t$ if it is reported by this state; otherwise, let $\overline{X}_{s,t}$ be the total number of test in state $s$ at time $t$ multiplied by the average ratio of the total number of tests to the number of unique people at time $t$ across states which reported both. Then, we simple set 
\[
\overline{X}_t := \sum_{s \in \text{states}}\overline{X}_{s,t}.
\]

\begin{figure*}[h]
\centering
\begin{subfigure}[b]{0.48\textwidth}    
\includegraphics[scale=0.28]{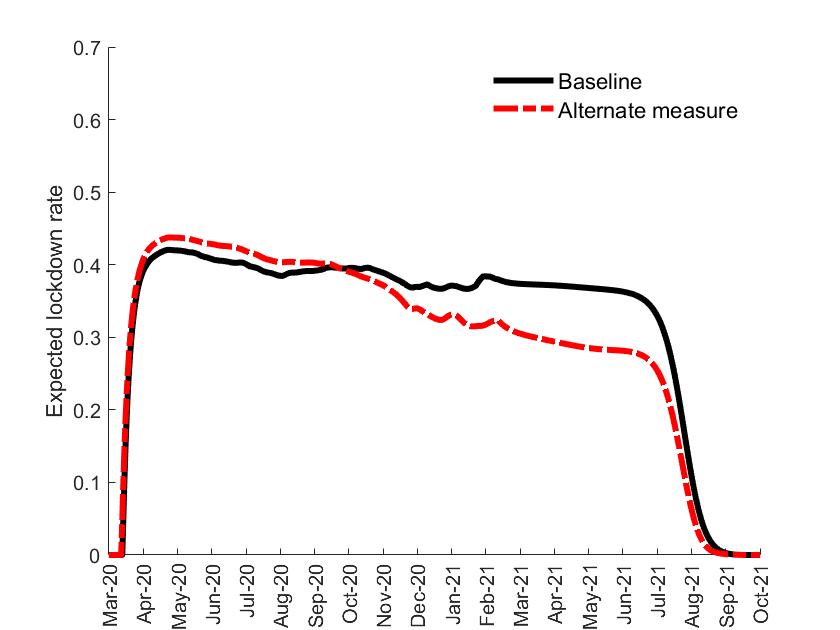}
\caption{Lockdown}
\label{fig: lockdown different test}
\end{subfigure}
\begin{subfigure}[b]{0.48\textwidth}    
\includegraphics[scale=0.28]{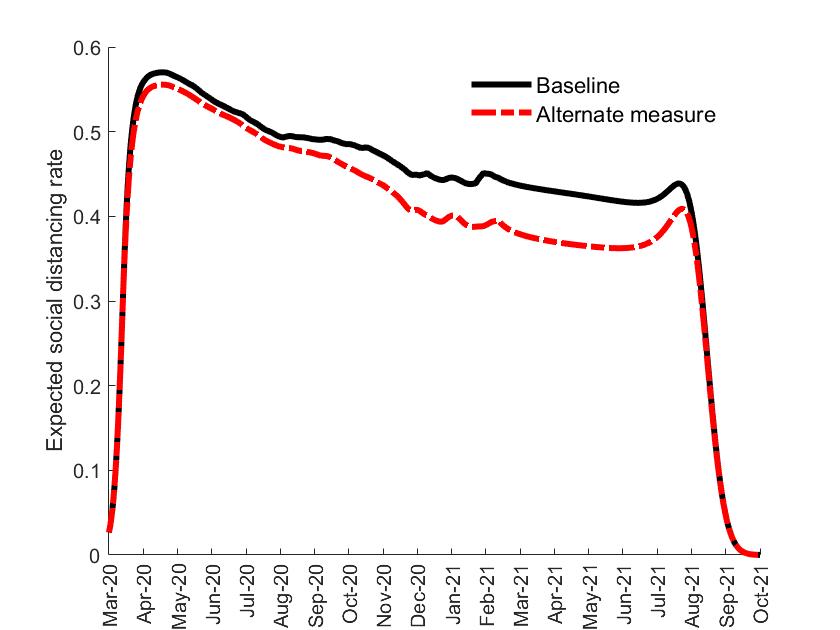}
\caption{Social Distancing}
\label{fig: sd different test}
\end{subfigure}\\
\begin{subfigure}[b]{0.48\textwidth}    
\includegraphics[scale=0.28]{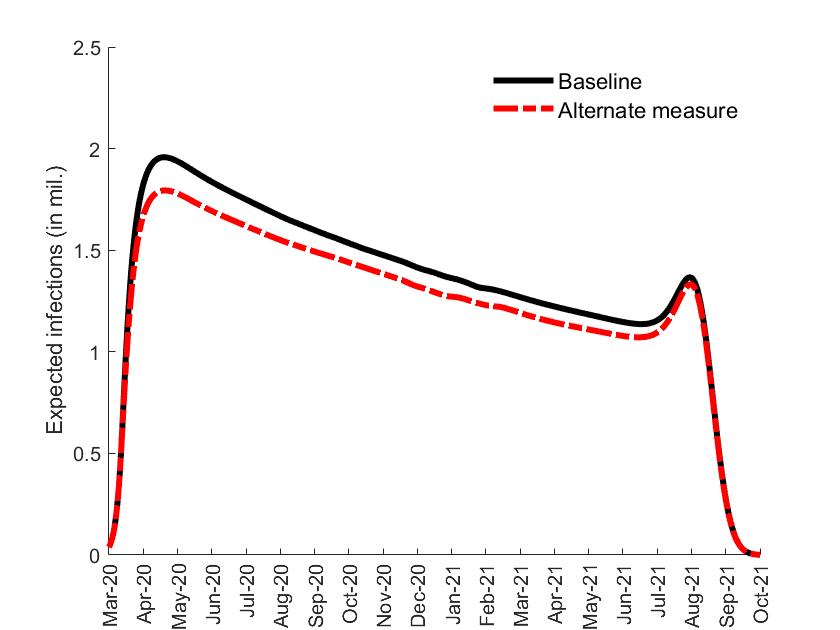}
\caption{Flow of infections}
\label{fig: infections different test}
\end{subfigure}
\begin{subfigure}[b]{0.48\textwidth}    
\includegraphics[scale=0.28]{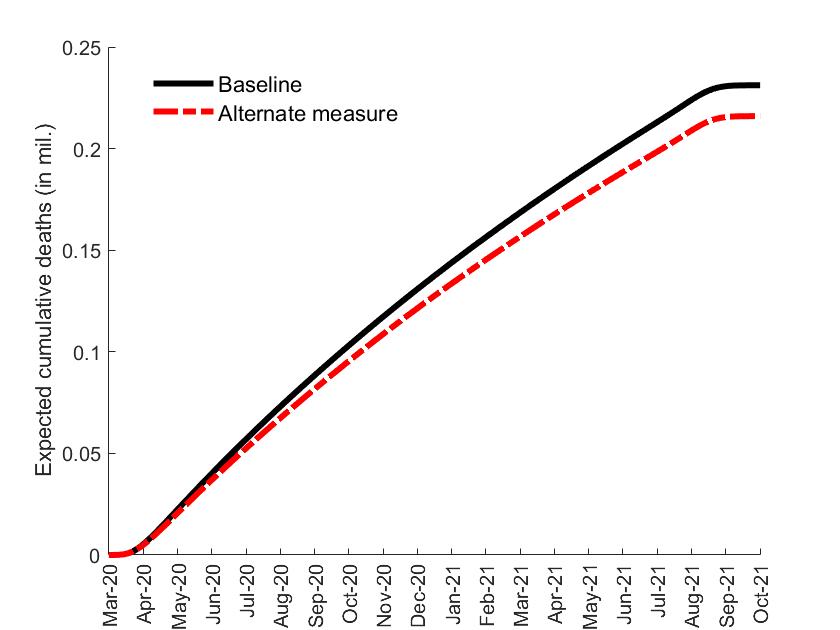}
\caption{Cumulative deaths}
\label{fig: deaths different test}
\end{subfigure}
\caption{Optimal policy when the total number of tests is measured differently} 
\label{fig: different test}
\end{figure*}

The main reason for working with unique people is that this metric is reported by most states in our dataset, thus its imputation is more reliable; in contrast, testing encounters were used only by a few smaller states. We do however re-calibrate and re-estimate the model using the aggregated testing data provided by the  "\href{https://covidtracking.com/}{COVID Tracking Project}". It turns out that both approaches yield very similar results. Instead of reporting all the estimation and optimal policy results again, we simply plot four time series of lockdown, social distancing, infections and deaths for these two specifications of deaths---see Figure \ref{fig: different test}.

The total number of tests available almost doubles in comparison to the baseline. However, in order to fit the data, the efficacy of tracing $\gamma$ must become worse. So the gains in the total number of deaths in non-trivial but not large.

\subsubsection{Fixed parameters}

We now move to the fixed parameters shown in Table \ref{table parameters}. First of all, we utilize the collective wisdom of various medical studies to calibrate the medical parameters. The average number of days of infection $t_{i}$  is set to $10$ days which includes the incubation and latency period. Conditional on developing further severe symptoms, the average number of days of hospitalization (or at home critical care) is assumed to be $t_{h}=7$. Our main basis for these numbers come from the \href{https://www.cdc.gov/coronavirus/2019-ncov/hcp/planning-scenarios.html}{CDC Planning Scenario} dated March 19, 2021. For instance, their estimate for the mean time from exposure of the disease to the onset of symptoms is around 8 days. Similarly, the population weighted mean time between the onset of symptoms to hospitalization is around 9 days. The average hospitalization period is calculated similarly. 

Since the time series of data on hospitalizations and daily deaths is available in the United States, and majority of Covid-19 related deaths have taken place at hospitals, we do a simple OLS regression between the two variables to calculate the rate at which those who develop severe symptoms are likely to die, which turns out to be $m_{h} = 0.1705$. This basically means that everything else being equal, $17\%$ of  all people who develop a severe case of Covid-19 will end up dying. Technically this parameter is not fixed because we estimate it from the data we have, but once estimated it will be fixed for the grand-estimation procedure. 

We calculate the rate at which an infected person can develop severe symptoms and require extra care, that is the rate of transition from the state $(\bf I)$ to the state $(\bf H)$, to be $m_{i} = 0.0176$ using the following argument: First, the IFR or the infection fatality rate is taken to be $0.003$. This means, ceteris paribus, there is a $0.3\%$ chance of death upon contracting the virus. This corroborates with estimates obtained in some of the medical literature (see for instance \citet{ioannidis2021infection}). 
Note that the value of IFR is by no means settled in the medical literature. For instance, the CDC Planning Scenario of 19th March, 2021 pegs IFR to be around 0.004. \citet{lockdown_SIR} uses a value of 0.0005, while \citet{farboodietal_covid} uses an IFR of 0.0062. If anything, our choice of $0.003$ may turn out to be on the higher side when the dust settles and one widely agreed upon number emerges. Once IFR is fixed, $m_i$ is calculated using the identity $m_i \times m_H = IFR$.

The arrival of vaccine and more broadly vaccine dissemination that culminates in the date when the pandemic can be declared to have ended is modelled through a negative binomial distribution with mean 540 days and variance 180 days (see the entries for $\mathbb{E}T$ and $\mathbb{V}T$ in Table \ref{table parameters}). This ensures that the pandemic will surely not end till around 404 days and it will almost certainly end in 593 days from the first day of the virus spread. This roughly squares with the prevailing wisdom, aggregated well in \citet{vaccine_nyt}, which had pegged the best case scenario for ending the pandemic to be the summer of 2021. Since the arrival of vaccine in our model coincides with the end of the pandemic, a mean of 540 days implies that the pandemic is expected to end in the Fall of 2021 (around August or September, 2021). At the current pace of vaccination in the US, it is expected that around 80\% of the US population will be vaccinated by September, 2021 (see \href{https://www.nytimes.com/interactive/2020/us/covid-19-vaccine-doses.html}{New York Times [2020]}).

The maximum amount of lockdown the government can impose is capped at 70 percent. So $\overline{\bl}=0.3$ means that thirty percent of economic activities need to happen for the basic ingredients of society to keep functioning. We have in mind  health, retail, government, utilities, and food manufacturing. This is the number used, for example, by \citet{lockdown_SIR} and the uniform lockdown policy used in \citet{dream_team_SIRmodel}. Then, as standard in the literature, the government is forward-looking and discounts the future in a way that the annual interest rate equals to $5$ percent. Our model allows for the flexibility of choosing any level of discounting by the agent, we calibrate the model for $\delta_A = \delta_G$.

Finally, we note that there is some indeterminacy in jointly identifying the total prevalence and the initial number of infected agent, see \citet{sir_unident}. The standard approach in the literature is either to fix both or fix one of these parameters and calibrate the other. We decided to calibrate the total prevalence by taking $e_{0}=0.007$ percent, i.e., there were $23000$ infected people at the end of February (roughly one month since the first positive case was reported in the US). It must be noted that the choice of the initial seed varies greatly in the literature, depending on the time frame of the analysis. For instance, \citet{ert_covid} works with an initial seed as high as 328,000 while \citet{farboodietal_covid} work with an initial seed of roughly 17,000. 

\subsubsection{Fixing $\delta_{A}=0$}

The agents are modeled to be as forward looking as the government. \citet{mcadams_survey} notes that some studies in the literature fix $\delta_{A}=0$, that is take the agents to be myopic optimizers, both for tractability and as thumb rule which approximates some reality. To assess robustness of our results and make them comparable to other studies, we fix $\delta_{A}=0$, re-estimate the model and then solve for the optimal policy. All the other fixed parameters are kept the same as before. 
\begin{figure*}[h]
\centering
\begin{subfigure}[b]{0.48\textwidth}    
\includegraphics[scale=0.28]{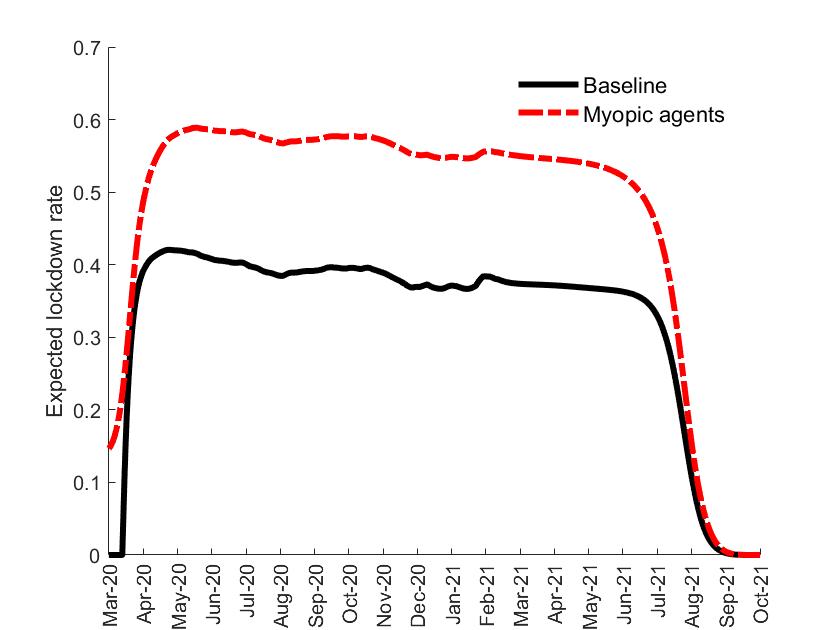}
\caption{Lockdown}
\label{fig: lockdown d zero}
\end{subfigure}
\begin{subfigure}[b]{0.48\textwidth}    
\includegraphics[scale=0.28]{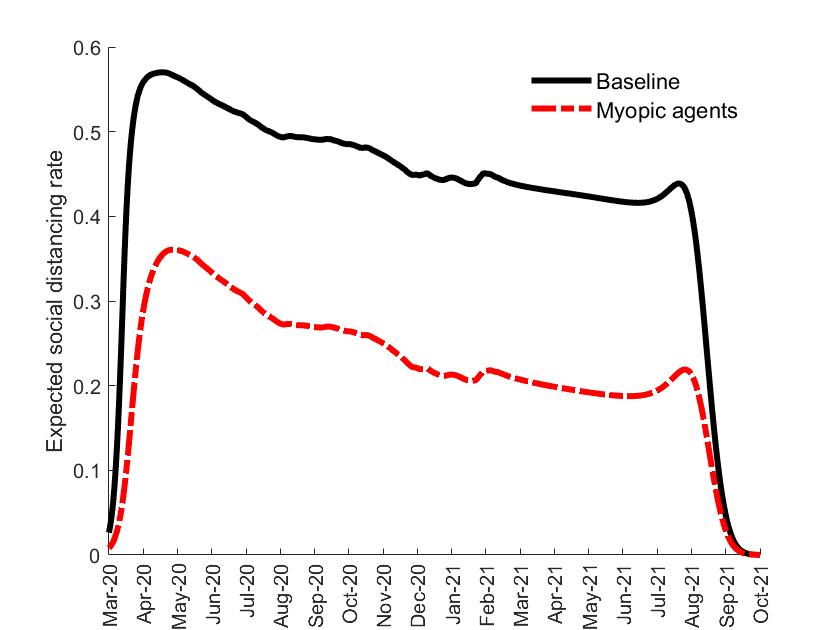}
\caption{Social Distancing}
\label{fig: sd d zero}
\end{subfigure}\\
\begin{subfigure}[b]{0.48\textwidth}    
\includegraphics[scale=0.28]{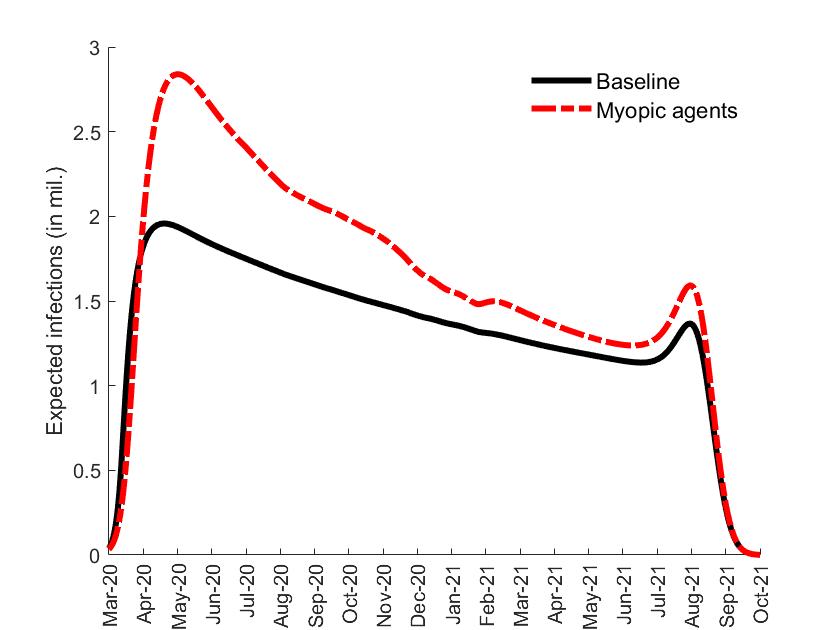}
\caption{Flow of infections}
\label{fig: infections d zero}
\end{subfigure}
\begin{subfigure}[b]{0.48\textwidth}    
\includegraphics[scale=0.28]{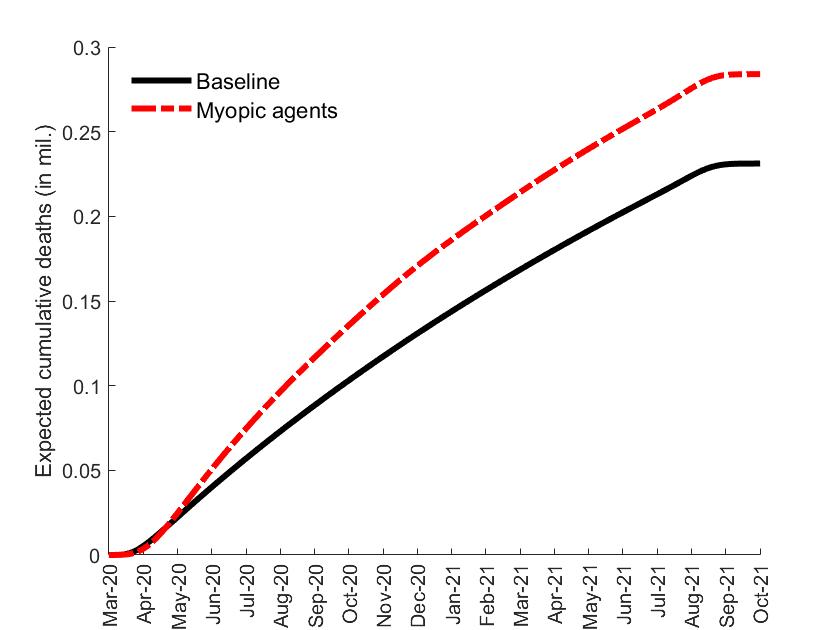}
\caption{Cumulative deaths}
\label{fig: deaths d zero}
\end{subfigure}
\caption{Results for myopic agents, $\delta_{A}=0$} 
\label{fig: counterfactuals d zero}
\end{figure*}

Since the agents are myopic, they internalize the costs of getting infected to a lesser extent and social distance less (Figure \ref{fig: sd d zero}). Expecting the reduced social distancing from the agents, the government thus locks down more (Figure \ref{fig: lockdown d zero}). Both social distancing and lockdown have the similar shape as the forward looking model --- broadly persistent with a gradual decrease during the course of the pandemic, with a sharp decline towards the end. As a consequence of the agents' myopia, the number of infections and deaths is larger (Figures \ref{fig: infections d zero} and \ref{fig: deaths d zero}). While the greater lockdown picks up some of the slack from reduced social distancing, the government still prefers not to completely lockdown at the maximal amount for a prolonged period for that would choke the economy. As before the optimal policy aims at reducing the effective reproductive number to one through the course of the pandemic.


\bibliographystyle{abbrvnat}
\bibliography{PandemicBehaviorPolicy_08Feb2022}

\begin{thebibliography}{53}
\providecommand{\natexlab}[1]{#1}
\providecommand{\url}[1]{\texttt{#1}}
\expandafter\ifx\csname urlstyle\endcsname\relax
  \providecommand{\doi}[1]{doi: #1}\else
  \providecommand{\doi}{doi: \begingroup \urlstyle{rm}\Url}\fi

\bibitem[Acemoglu et~al.(2021)Acemoglu, Chernozhukov, Werning, and
  Whinston]{dream_team_SIRmodel}
D.~Acemoglu, V.~Chernozhukov, I.~Werning, and M.~Whinston.
\newblock Optimal targeted lockdowns in a multi-group sir model.
\newblock {\it American Economic Review: Insights}, forthcoming, 2021.

\bibitem[Alvarez et~al.(2021)Alvarez, Argente, and Lippi]{lockdown_SIR}
F.~Alvarez, D.~Argente, and F.~Lippi.
\newblock A simple planning problem for {COVID-19} lock-down, testing, and
  tracing.
\newblock \emph{American Economic Review: Insights}, 3\penalty0 (3):\penalty0
  367--82, 2021.

\bibitem[Andreoni(1989)]{warm_glow}
J.~Andreoni.
\newblock Giving with impure altruism: Applications to charity and ricardian
  equivalence.
\newblock \emph{Journal of Political Economy}, 97\penalty0 (6):\penalty0
  1447--1458, 1989.

\bibitem[Atkeson(2021{\natexlab{a}})]{atkenson_SIR}
A.~Atkeson.
\newblock A parsimonious behavioral {SEIR} model of the {2020 COVID} epidemic
  in the {United States and the United Kingdom}.
\newblock UCLA, 2021{\natexlab{a}}.

\bibitem[Atkeson(2021{\natexlab{b}})]{atkeson_brookings}
A.~Atkeson.
\newblock Behavior and the dynamics of epidemics.
\newblock UCLA, 2021{\natexlab{b}}.

\bibitem[Bandyopadhyay et~al.(2021)Bandyopadhyay, Chatterjee, Das, and
  Roy]{kalyan_da_paper}
S.~Bandyopadhyay, K.~Chatterjee, K.~Das, and J.~Roy.
\newblock Learning versus habit formation: Optimal timing of lockdown for
  disease containment.
\newblock \emph{Journal of Mathematical Economics}, 93\penalty0 (102452), 2021.
\newblock \doi{https://doi.org/10.1016/j.jmateco.2020.11.008}.

\bibitem[Baqaee et~al.(2020)Baqaee, Farhi, Mina, and Stock]{farhi_sir}
D.~Baqaee, E.~Farhi, M.~Mina, and J.~H. Stock.
\newblock Policies for a second wave.
\newblock \emph{Brookings Papers on Economic Activity, Summer Special ed.}, 51,
  2020.

\bibitem[Barnett-Howell and Mobarak(2020)]{covid_developing}
Z.~Barnett-Howell and A.~M. Mobarak.
\newblock Should low-income countries impose the same social distancing
  guidelines as {Europe and North America} to halt the spread of {COVID-19?}
\newblock Yale University, 2020.

\bibitem[Barro et~al.(2020)Barro, Urs\'{u}a, and Weng]{barro_flu}
R.~J. Barro, J.~F. Urs\'{u}a, and J.~Weng.
\newblock The coronavirus and the great influenza pandemic: Lessons from the
  {"Spanish Flu"} for the coronavirus's potential effects on mortality and
  economic activity.
\newblock NBER, 2020.

\bibitem[Bergera et~al.(2021)Bergera, Herkenhoff, Huang, and
  Mongey]{testing_SIR}
D.~Bergera, K.~Herkenhoff, C.~Huang, and S.~Mongey.
\newblock Testing and reopening in an {SEIR} model.
\newblock \emph{Review of Economic Dynamics}, 2021.

\bibitem[Brodeur et~al.(2021)Brodeur, Gray, Islam, and Bhuiyan]{covid_survey}
A.~Brodeur, D.~Gray, A.~Islam, and S.~Bhuiyan.
\newblock A literature review of the economics of {COVID-19}.
\newblock \emph{Journal of Economic Surveys}, 35\penalty0 (4):\penalty0
  1007--1044, 2021.

\bibitem[Caballero and Simsek(2021)]{caba_simsek_assets}
R.~J. Caballero and A.~Simsek.
\newblock A model of endogenous risk intolerance and {LSAPs}: Asset prices and
  aggregate demand in a {"COVID-19"} shock.
\newblock \emph{The Review of Financial Studies}, 34\penalty0 (11):\penalty0
  5522--5580, 2021.

\bibitem[Chari et~al.(2021)Chari, Kirpalani, and Phelan]{rishabh_da_paper}
V.~V. Chari, R.~Kirpalani, and C.~Phelan.
\newblock The hammer and the scalpel: On the economics of indiscriminate versus
  targeted isolation policies during pandemics.
\newblock \emph{Review of Economic Dynamics}, 42:\penalty0 1--14, 2021.

\bibitem[Correia et~al.(2020)Correia, Luck, and Verner]{spanish_flu}
S.~Correia, S.~Luck, and E.~Verner.
\newblock Pandemics depress the economy, public health interventions do not:
  Evidence from the 1918 flu.
\newblock The Federal Reserve and MIT, 2020.

\bibitem[{COVID-19 Mental Disorders
  Collaborators}(2021)]{mental_health_lancent}
{COVID-19 Mental Disorders Collaborators}.
\newblock Global prevalence and burden of depressive and anxiety disorders in
  204 countries and territories in 2020 due to the {COVID-19} pandemic.
\newblock \emph{The Lancent}, 398\penalty0 (10312):\penalty0 1700--1712, 2021.
\newblock \doi{https://doi.org/10.1016/S0140-6736(21)02143-7}.

\bibitem[Dasaratha(2020)]{dasratha_SIR}
K.~Dasaratha.
\newblock Virus dynamics with behavioral responses.
\newblock Harvard University, 2020.

\bibitem[{Deutsche Welle}(2020)]{Merkel_speech}
{Deutsche Welle}.
\newblock {Merkel: Coronavirus is Germany's greatest challenge since World War
  II}.
\newblock
  \href{https://www.dw.com/en/merkel-coronavirus-is-germanys-greatest-challenge-since-world-war-ii/a-52830797}{Weblink},
  March 18, 2020.

\bibitem[Eichenbaum et~al.(2021{\natexlab{a}})Eichenbaum, Rebelo, and
  Trabandt]{ert_covid}
M.~Eichenbaum, S.~Rebelo, and M.~Trabandt.
\newblock The macroeconomics of epidemics.
\newblock \emph{The Review of Financial Studies}, 34\penalty0 (11):\penalty0
  5149--5187, 2021{\natexlab{a}}.

\bibitem[Eichenbaum et~al.(2021{\natexlab{b}})Eichenbaum, Rebelo, and
  Trabandt]{macro_testing_quaran}
M.~S. Eichenbaum, S.~Rebelo, and M.~Trabandt.
\newblock The macroeconomics of testing and quarantining.
\newblock {\it NBER} Working Paper No. 27104, 2021{\natexlab{b}}.

\bibitem[Engle et~al.(2021)Engle, Keppo, Kudlyak, Quercioli, Smith, and
  Wilson]{lones_SIR}
S.~Engle, J.~Keppo, M.~Kudlyak, E.~Quercioli, L.~Smith, and A.~Wilson.
\newblock The behavioral sir model, with applications to the swine flu and
  {COVID-19} pandemics.
\newblock University of Wisconsin-Madison, 2021.

\bibitem[Fajgelbaum et~al.(2020)Fajgelbaum, Khandelwal, Kim, Mantovani, and
  Schaal]{Schaal_pandemic}
P.~D. Fajgelbaum, A.~Khandelwal, W.~Kim, C.~Mantovani, and E.~Schaal.
\newblock Optimal lockdown in a commuting network.
\newblock American Economic Review: Insights, forthcoming, 2020.

\bibitem[Farboodi et~al.(2021)Farboodi, Jarosch, and
  Shimer]{farboodietal_covid}
M.~Farboodi, G.~Jarosch, and R.~Shimer.
\newblock Internal and external effects of social distancing in a pandemic.
\newblock {\it Journal of Economic Theory}, forthcoming, 2021.

\bibitem[Fenichel et~al.(2011)Fenichel, Castillo-Chavez, Ceddia, Chowell,
  Parra, Hickling, Holloway, Horan, Morin, Perrings, Springborn, Velazquez, and
  Villalobos]{behavior_pnas}
E.~P. Fenichel, C.~Castillo-Chavez, M.~G. Ceddia, G.~Chowell, P.~A.~G. Parra,
  G.~J. Hickling, G.~Holloway, R.~Horan, B.~Morin, C.~Perrings, M.~Springborn,
  L.~Velazquez, and C.~Villalobos.
\newblock Adaptive human behavior in epidemiological models.
\newblock \emph{Proceedings of the National Academy of Sciences of the United
  States}, 108\penalty0 (15):\penalty0 6306--6311, 2011.

\bibitem[Fern\'{a}ndez-Villaverde and Jones(2020)]{jesus_jones}
J.~Fern\'{a}ndez-Villaverde and C.~I. Jones.
\newblock Estimating and simulating a {SIRD} model of {COVID-19} for many
  countries, states, and cities.
\newblock University of Pennsylvania and Stanford University, 2020.

\bibitem[{Financial Times}(2020)]{FT_UK}
{Financial Times}.
\newblock {Covid-19 restrictions not affecting social distancing, says ONS}.
\newblock
  \href{https://www.ft.com/content/8b7d688d-7153-43f3-9539-56db62bcc8a7}{Weblink},
  September 30, 2020.

\bibitem[{Financial Times}(2021{\natexlab{a}})]{FT_Sweden}
{Financial Times}.
\newblock {Sweden no longer stands out in pandemic, says architect of ‘no
  lockdown’ policy}.
\newblock
  \href{https://www.ft.com/content/0c07de5f-e852-4c23-823b-5f8f7d18ebef}{Weblink},
  November 12, 2021{\natexlab{a}}.

\bibitem[{Financial Times}(2021{\natexlab{b}})]{FT_Vietnam}
{Financial Times}.
\newblock {Vietnam abandons zero-Covid strategy after record drop in GDP}.
\newblock
  \href{https://www.ft.com/content/37f7f400-20aa-4e52-8f3b-f9359fa73fe8}{Weblink},
  September 30, 2021{\natexlab{b}}.

\bibitem[Giannitsarou et~al.(2021)Giannitsarou, Kissler, and
  Toxvaerd]{flavio_paper4}
C.~Giannitsarou, S.~Kissler, and F.~Toxvaerd.
\newblock Waning immunity and the second wave: Some projections for
  {SARS-CoV-2}.
\newblock \emph{American Economic Review: Insights}, 3\penalty0 (3):\penalty0
  321--338, 2021.

\bibitem[Giannone et~al.(2020)Giannone, Paixao, and Pang]{elisa_covid_paper}
E.~Giannone, N.~Paixao, and X.~Pang.
\newblock Pandemic in an interregional model-- staggered restart.
\newblock Penn State University and Bank of Canada, 2020.

\bibitem[Glover et~al.(2021)Glover, Heathcote, Krueger, and
  Rios-Rull]{krueger_victor_SIR}
A.~Glover, J.~Heathcote, D.~Krueger, and J.-V. Rios-Rull.
\newblock Health versus wealth: On the distributional effects of controlling a
  pandemic.
\newblock Federal Reserve Bank and University of Pennsylvania, 2021.

\bibitem[Guerrieri et~al.(2021)Guerrieri, Lorenzoni, Straub, and
  Werning]{glsw_covid}
V.~Guerrieri, G.~Lorenzoni, L.~Straub, and I.~Werning.
\newblock Macroeconomic implications of covid-19: Can negative supply shocks
  cause demand shortages?
\newblock {\it American Economic Review}, forthcoming, 2021.

\bibitem[Hale et~al.(2021)Hale, Angrist, Goldszmidt, Kira, Petherick, Phillips,
  Webster, Cameron-Blake, Hallas, Majumdar, et~al.]{hale2021global}
T.~Hale, N.~Angrist, R.~Goldszmidt, B.~Kira, A.~Petherick, T.~Phillips,
  S.~Webster, E.~Cameron-Blake, L.~Hallas, S.~Majumdar, et~al.
\newblock A global panel database of pandemic policies (oxford covid-19
  government response tracker).
\newblock \emph{Nature Human Behaviour}, 5\penalty0 (4):\penalty0 529--538,
  2021.

\bibitem[Hall et~al.(2020)Hall, Jones, and Klenow]{jones_pareto}
R.~E. Hall, C.~I. Jones, and P.~J. Klenow.
\newblock Trading off consumption and covid-19 deaths.
\newblock Stanford University, 2020.

\bibitem[Ioannidis(2021)]{ioannidis2021infection}
J.~P. Ioannidis.
\newblock Infection fatality rate of covid-19 inferred from seroprevalence
  data.
\newblock \emph{Bulletin of the World Health Organization}, 99\penalty0
  (1):\penalty0 19, 2021.

\bibitem[Jones et~al.(2021)Jones, Philippon, and Venkateswaran]{jpv_covid}
C.~Jones, T.~Philippon, and V.~Venkateswaran.
\newblock Optimal mitigation policies in a pandemic: Social distancing and
  working from home.
\newblock \emph{The Review of Financial Studies}, 34\penalty0 (11):\penalty0
  5188--5223, 2021.

\bibitem[Kaplan et~al.(2020)Kaplan, Moll, and Violante]{moll_SIR}
G.~Kaplan, B.~Moll, and G.~L. Violante.
\newblock The great lockdown and the big stimulus: Tracing the pandemic
  possibility frontier for the u.s.
\newblock University of Chicago, LSE and Princeton University, 2020.

\bibitem[Kermack and McKendrick(1927)]{epidemic_first}
W.~O. Kermack and A.~G. McKendrick.
\newblock A contribution to the mathematical theory of epidemics.
\newblock \emph{Proceedings of the royal society of London Series A},
  115\penalty0 (772):\penalty0 700--721, 1927.

\bibitem[Korolev(2021)]{sir_unident}
I.~Korolev.
\newblock Identification and estimation of the {SEIRD} epidemic model for
  {COVID-19}.
\newblock \emph{Journal of Econometrics}, 220\penalty0 (1):\penalty0 63--85,
  2021.

\bibitem[Krueger et~al.(2020)Krueger, Uhlig, and Xie]{kh_covid}
D.~Krueger, H.~Uhlig, and T.~Xie.
\newblock Macroeconomic dynamics and reallocation in an epidemic.
\newblock University of Pennsylvania, University of Chicago and National
  University of Singapore, 2020.

\bibitem[{Mayo Cinic}(2021)]{mental_health_mayo}
{Mayo Cinic}.
\newblock {COVID-19 and your mental health}.
\newblock
  \href{https://www.mayoclinic.org/diseases-conditions/coronavirus/in-depth/mental-health-covid-19/art-20482731}{Weblink},
  November 21, 2021.

\bibitem[McAdams(2021)]{mcadams_survey}
D.~McAdams.
\newblock The blossoming of economic epidemiology.
\newblock \emph{Bulletin of the World Health Organization}, 13:\penalty0
  539--570, 2021.

\bibitem[Moreland et~al.(2020)Moreland, Herlihy, Tynan, Sunshine, McCord,
  Hilton, Poovey, Werner, Jones, Fulmer, et~al.]{moreland2020timing}
A.~Moreland, C.~Herlihy, M.~A. Tynan, G.~Sunshine, R.~F. McCord, C.~Hilton,
  J.~Poovey, A.~K. Werner, C.~D. Jones, E.~B. Fulmer, et~al.
\newblock Timing of state and territorial covid-19 stay-at-home orders and
  changes in population movement—united states, march 1--may 31, 2020.
\newblock \emph{Morbidity and Mortality Weekly Report}, 69\penalty0
  (35):\penalty0 1198, 2020.

\bibitem[Narita and Sudo(2021)]{narita_democracy}
Y.~Narita and A.~Sudo.
\newblock Curse of democracy: Evidence from the 21st century.
\newblock Yale University, 2021.

\bibitem[Neher et~al.(2020)Neher, Aksamentov, Noll, Albert, and
  Dyrdak]{covid_basel}
R.~Neher, I.~Aksamentov, N.~Noll, J.~Albert, and R.~Dyrdak.
\newblock Covid-19 scenarios.
\newblock University of Basel,
  \href{https://covid19-scenarios.org/about}{https://covid19-scenarios.org/about},
  2020.

\bibitem[{New York Times}(2020)]{vaccine_nyt}
{New York Times}.
\newblock How long will a vaccine really take?
\newblock
  \href{https://www.nytimes.com/interactive/2020/04/30/opinion/coronavirus-covid-vaccine.html}{Weblink},
  April 30, 2020.

\bibitem[\`{O}scar. et~al.(2020)\`{O}scar., Singh, and Taylor]{gabru_ka_paper}
J.~\`{O}scar., S.~R. Singh, and A.~M. Taylor.
\newblock Longer-run economic consequences of pandemics.
\newblock University of California, Davis, 2020.

\bibitem[Perra et~al.(2011)Perra, Balcan, Gonçalves, and
  Vespignani]{behavior_plos}
N.~Perra, D.~Balcan, B.~Gonçalves, and A.~Vespignani.
\newblock Towards a characterization of behavior-disease models.
\newblock \emph{PLoS One}, 6\penalty0 (8):\penalty0 PMC3149628, 2011.

\bibitem[Rizzo et~al.(2014)Rizzo, Frasca, and Porfiri]{behavior_physics}
A.~Rizzo, M.~Frasca, and M.~Porfiri.
\newblock Effect of individual behavior on epidemic spreading in
  activity-driven networks.
\newblock \emph{Physical Review E}, 90:\penalty0 042801, 2014.

\bibitem[Rowthorn and Toxvaerd(2020)]{flavio_paper3}
R.~Rowthorn and F.~Toxvaerd.
\newblock The optimal control of infectious diseases via prevention and
  treatment.
\newblock Cambridge-INET Working Paper Series No: 2020/13, 2020.

\bibitem[Schelling(1968)]{schelling}
T.~C. Schelling.
\newblock The life you save may be your own.
\newblock In S.~B. Chase~Jr., editor, \emph{Problems in Public Expenditure
  Analysis}, pages 127--161. Brookings Institution, Washington D.C., 1968.

\bibitem[{The Gaurdian}(2020)]{guardian_german_expert}
{The Gaurdian}.
\newblock {Germany's Covid-19 expert: 'For many, I'm the evil guy crippling the
  economy'}.
\newblock
  \href{https://www.theguardian.com/world/2020/apr/26/virologist-christian-drosten-germany-coronavirus-expert-interview?CMP=share_btn_tw}{Weblink},
  April 26, 2020.

\bibitem[Toxvaerd(2019)]{flavio_ier}
F.~Toxvaerd.
\newblock Rational disinhibition and externalities in prevention.
\newblock \emph{International Economic Review}, 60\penalty0 (4):\penalty0
  1737--1755, 2019.

\bibitem[Toxvaerd(2020)]{flavio_covid}
F.~Toxvaerd.
\newblock Equilibrium social distancing.
\newblock University of Cambridge, 2020.

\end{thebibliography}

\end{document}